\documentclass[useAMS,usenatbib,referee]{biom}

\usepackage{amsmath,mathtools,amsfonts,mathabx}
\usepackage{url}
\usepackage{natbib}

\usepackage{xr} 
\usepackage{graphicx}
\graphicspath{{./figures/}}

\DeclareMathOperator*{\argmax}{arg\,max}

\newcommand{\trans}{\text{T}}

\newcommand{\mbf}[1]{\boldsymbol{#1}}

\allowdisplaybreaks

\usepackage{xcolor}

\newcommand{\revise}{\textcolor{black}}

\title[Prediction-Oriented Transfer Learning]{Prediction-Oriented Transfer Learning for Survival Analysis}

\author{Yu Gu$^{1,*}$\email{yugu@hku.hk}, 
Donglin Zeng$^{2}$, and D. Y. Lin$^{3}$ \\
$^{1}$Department of Statistics and Actuarial Science, University of Hong Kong, Pokfulam Road, Hong Kong \\
$^{2}$Department of Biostatistics, University of Michigan, Ann Arbor, Michigan, U.S.A. \\
$^{3}$Department of Biostatistics, University of North Carolina at Chapel Hill, Chapel Hill, North Carolina, U.S.A.}

\begin{document}


\date{{\it Received October} 2007. {\it Revised February} 2008.  {\it
Accepted March} 2008.}



\pagerange{\pageref{firstpage}--\pageref{lastpage}} 
\volume{64}
\pubyear{2008}
\artmonth{December}


\doi{10.1111/j.1541-0420.2005.00454.x}


\label{firstpage}


\begin{abstract}
Transfer learning is beneficial for survival analysis, especially when the target study has a limited number of events.
However, existing transfer learning methods rely on the restrictive assumption that the target and source studies share similar parameters under Cox models, and most require access to individual-level source data.
In this article, we propose a novel transfer learning framework that enhances model-based survival prediction by transferring predictive rather than distributional knowledge from source studies. 
Our approach employs flexible semiparametric transformation models for the target data while eliminating the need to model or share the source data. 
The ingeniously designed penalty enables simple and stable computation via an EM algorithm.
We rigorously establish the asymptotic properties of the proposed estimator and show that it achieves a faster convergence rate than the target-only estimator when source knowledge is sufficiently accurate.
We demonstrate the advantages of our methods through extensive simulation studies and an application to two major breast cancer studies, with the objective of improving survival prediction in a target cohort using predictive information from an external study.

\end{abstract}

%



\maketitle


%

\section{Introduction}
\label{sec:intro}
Transfer learning has been extensively studied in computer science over the past decade. 
By leveraging knowledge from related source studies, transfer learning can address challenges posed by insufficient samples and improve learning performance in the target study. 
It has been successfully applied to a wide range of real-world problems, such as natural language processing, image processing, speech recognition, biomedical research, and recommender systems \citep{zhuang2020comprehensive}. 

Recently, transfer learning has gained considerable attention within the statistics community. 
Various methods have been developed to tackle statistical problems such as high-dimensional linear regression \citep{bastani2021predicting,li2022transfer,gu2024robust}, high-dimensional generalized linear models \citep{tian2023transfer,li2024estimation}, functional linear regression models \citep{lin2022hypothesis}, and Cox regression models \citep{li2023accommodating,xie2024transfer}. 
Most existing methods incorporate source information into the target learning procedure via a penalty that encourages similarity between the modeled target and source distributions, with the specific penalty determined by the underlying similarity assumption. 
In regression models, a typical assumption is that the $L_r$-distance ($r\in[0,2]$) between the target and source parameters is small, motivating the use of the $L_r$-penalty \citep{bastani2021predicting,li2022transfer,tian2023transfer,li2023accommodating,xie2024transfer}. 
Alternatively, \citet{gu2024robust} proposed an angle-based penalty assuming the angle distance between the target and source parameters is small. 
These methods fall into the category of distance-based transfer learning. 
Another category, representation transfer learning, assumes shared or similar low-dimensional representations between the target and source parameters \citep{tian2023learning,he2024representation}.

In this article, we explore transfer learning in the model-based prediction context of survival analysis, which holds important implications for medicine and public health. 
For example, studies of chronic diseases often face limitations due to a small number of events resulting from short study periods or low disease incidence rates. 
Relying solely on data from a single study can lead to unsatisfactory performance in risk evaluation and survival prediction. 
Transfer learning can enhance performance by using auxiliary knowledge from related studies with higher event rates or longer study periods. 
Another example is survival analysis for underrepresented populations, such as racial and ethnic minorities and children, where transferring knowledge from majority populations can improve estimation and prediction performance on the target population.

Despite its potential, the literature on transfer learning for survival analysis is quite limited.
All existing methods are established under the Cox regression model, and none have theoretical justifications.
\citet{li2016transfer} used an $L_{2,1}$-penalty to learn a shared representation between the target and source studies.
\citet{li2023accommodating} and \citet{xie2024transfer} developed distance-based transfer learning methods for right- and interval-censored data, respectively, with $L_0$- or $L_1$-penalties applied to both the regression coefficients and the cumulative baseline hazard function.
\citet{lu2026adaptive} addressed covariate shift using density ratio weighting and aggregated the weighted partial likelihood functions from the target and source studies as the objective function.
All these methods specify Cox models for both the target and source data and require the two studies to share the same set of covariates.
Thus, they are susceptible to model misspecification and can be suboptimal when the target and source studies collect different covariates, since only the shared covariates can be used.
Moreover, these methods either discard source information about the baseline hazard function, which may limit improvement in prediction, or impose the restrictive assumption that the target and source studies have similar cumulative baseline hazard functions, which can be easily violated in practice due to heterogeneities in study populations.
Last but not least, the methods of \citet{li2016transfer} and \citet{lu2026adaptive} require sharing individual-level source data, which is often infeasible due to privacy concerns.
For instance, individual-level data from large biobanks, 
major cohort studies, 
and electronic health records are generally not shareable due to regulatory and institutional restrictions.

To overcome the limitations of existing transfer learning methods for survival analysis, we propose a novel framework called Prediction-Oriented Transfer Learning (POTL). 
POTL aims to enhance model-based prediction by directly integrating predictive knowledge from source studies.
Unlike existing approaches, which focus on transferring knowledge of model parameters, POTL is fully prediction-driven: the entire transfer learning procedure operates on survival predictions rather than on parameter sharing. 
This prediction-oriented feature distinguishes POTL from all existing methods.
Moreover, transferring predictive knowledge requires imposing penalties directly on survival probabilities, which is computationally far more challenging than penalizing individual model parameters. 
For example, even under simple Cox models, applying conventional $L_r$-penalties ($r\in[0,2]$) to survival probabilities leads to intractable computation. 
We overcome this difficulty in POTL through a new penalization strategy and an efficient optimization algorithm.

Our approach has several major advantages. 
First, unlike existing methods that require similar distributions or model parameters between the target and source studies, we impose a much weaker assumption that the studies share similar survival predictions within the target-relevant domain. Thus, our approach remains applicable even when the source models differ substantially from the target model, such as when the source model is non-Cox beyond the target study period, and direct parameter transfer is infeasible or inappropriate.
Second, we consider a broad class of semiparametric transformation models with potentially time-dependent covariates for the target study, while leaving the models for all source studies completely unrestricted---they can be traditional regression models, machine learning models, or artificial intelligence (AI) models such as large language models. This flexibility makes POTL highly robust to model misspecification and allows it to accommodate a wide variety of source models.
Third, POTL does not require sharing individual-level source data, as only summary-level prediction information is transferred.
Finally, we introduce a novel cross-entropy-type penalty that is computationally much easier than traditional $L_r$ penalties. 

We establish a rigorous asymptotic theory for transfer learning in survival analysis through the novel use of empirical process theory. 
Specifically, we show that the proposed estimator for the survival function converges to the true survival function at an optimal rate that is no slower than the standard $n^{1/2}$ rate based on target data only. 
This theoretical foundation ensures that our POTL approach effectively leverages source information to enhance prediction performance in the target study. 
We evaluate the operating characteristics of our approach through extensive simulation studies and provide an application to the TCGA--BRCA \citep{cancer2012comprehensive} and METABRIC \citep{curtis2012genomic} studies on breast cancer.

\section{Methods}
\label{sec:methods}
\subsection{Transformation Models for the Target Study}
\label{sub:models_data_and_likelihood}
For the target study, let $T$ denote the failure time and $\mbf{X}(\cdot)$ denote a $p$-vector of potentially time-dependent covariates.
We consider a broad class of semiparametric transformation models for the conditional cumulative hazard function of $T$ given $\mbf{X}$ in the form of
\begin{equation} \label{trans_model}
\Lambda(t | \mbf{X}) = G\left[\int_0^t\exp\{\mbf{\beta}^{\trans}\mbf{X}(s)\}d\Lambda(s)\right], 
\end{equation}
where $G(\cdot)$ is a strictly increasing transformation function, $\mbf{\beta}$ is a $p$-vector of unknown regression parameters, and $\Lambda(\cdot)$ is an unknown increasing function with $\Lambda(0)=0$. 
Let $C$ denote the censoring time, which is assumed to be independent of $T$ conditional on $\mbf{X}$.
We observe $Y = \min(T, C)$ and $\Delta = I(T\le C)$.
Thus, the data from $n$ independent subjects consist of $\{Y_i, \Delta_i, \mbf{X}_i(t): t\in[0,\tau]\}$ ($i=1,\dots,n$), where $\tau$ is the end time of the target study. 
The likelihood function concerning $(\mbf{\beta}, \Lambda)$ is 
\[
L_n(\mbf{\beta}, \Lambda) =
\prod_{i=1}^{n} \left[\Lambda^{\prime}(Y_{i})e^{\mbf{\beta}^{\trans}\mbf{X}_{i}(Y_{i})}G^{\prime}\left\{\int_0^{Y_{i}}e^{\mbf{\beta}^{\trans}\mbf{X}_{i}(s)}d\Lambda(s)\right\}\right]^{\Delta_{i}}  
\exp\left[-G\left\{\int_0^{Y_{i}}e^{\mbf{\beta}^{\trans}\mbf{X}_{i}(s)}d\Lambda(s)\right\}\right],
\]
where we use $f^{\prime}(x)$ to denote the first-order derivative of function $f(x)$ with respect to $x$.
It is useful to consider the log-Laplace transformation 
\[
G(x) = -\log \int_0^{\infty} e^{-xz}f(z)dz,
\] 
where $f(z)$ is the density function of a frailty variable with support $[0,\infty)$.
The choice of the gamma density with mean 1 and variance $r>0$ for $f(z)$ yields the class of logarithmic transformations $G(x) = r^{-1}\log(1+rx)$, which includes the proportional odds model ($r=1$) and the proportional hazards model ($r = 0$).
Under the log-Laplace transformation, the likelihood can be rewritten as
\[
  L_n(\mbf{\beta}, \Lambda) = 
\prod_{i=1}^{n} \int_0^{\infty}\left\{z_i\Lambda^{\prime}(Y_{i})e^{\mbf{\beta}^{\trans}\mbf{X}_{i}(Y_{i})}\right\}^{\Delta_{i}} 
\exp\left\{-z_i\int_0^{Y_{i}}e^{\mbf{\beta}^{\trans}\mbf{X}_{i}(s)}d\Lambda(s)\right\}f(z_i)dz_i,
\]
which is the likelihood of the proportional hazards frailty model with frailty $z_i$.

\subsection{Prediction-Oriented Transfer Learning (POTL)}
\label{sub:POTL}
Let $S(t|\mbf{X}; \mbf{\beta}, \Lambda) = \exp[-G\{\int_0^{t}e^{\mbf{\beta}^{\trans}\mbf{X}(s)}d\Lambda(s)\}]$ denote the covariate-specific survival function in the target study.
Our goal is to improve the estimation of $S(t|\mbf{X})$ by incorporating predictive information from a source study.
Let $\widecheck{S}(t|\mbf{X})$ denote the survival function predictor obtained from the source study.
This predictor can be constructed using any survival analysis method, such as Cox regression, transformation models similar to \eqref{trans_model}, or machine-learning approaches.
We require only that the covariate information used by the source predictor be fully determined by the covariates in the target model \eqref{trans_model}; the source and target models are not required to use identical covariate sets.
For instance, the covariates used in $\widecheck{S}(t|\mbf{X})$ may be a subset of $\mbf{X}$ or, more generally, a lower-dimensional function of $\mbf{X}$. 
We define the similarity metric between $S(t|\mbf{X})$ and $\widecheck{S}(t|\mbf{X})$ by 
\begin{equation} \label{similarity_metric}
\psi_m(\mbf{\beta}, \Lambda) = 
m^{-1}\sum_{i=1}^{m}\Bigl[\widecheck{S}(\widetilde{Y}_{i}|\widetilde{\mbf{X}}_{i})\log S(\widetilde{Y}_{i}|\widetilde{\mbf{X}}_{i}) 
+ \left\{1-\widecheck{S}(\widetilde{Y}_{i}|\widetilde{\mbf{X}}_{i})\right\}\log\left\{1-S(\widetilde{Y}_{i}|\widetilde{\mbf{X}}_{i})\right\}\Bigr],
\end{equation}
where $m>0$ is a sufficiently large integer, $\{\widetilde{Y}_{i}: i=1,\dots,m\}$ and $\{\widetilde{\mbf{X}}_{i}: i=1,\dots,m\}$ are independent copies of $Y$ and $\mbf{X}$, respectively.
A simple and practical choice is to set $m=n$ and $(\widetilde{Y}_{i}, \widetilde{\mbf{X}}_{i}) = (Y_{i}, \mbf{X}_{i})$ for $i=1,\dots,n$, which achieves satisfactory performance with low computational cost in our numerical studies.

\begin{remark}
Although our asymptotic theory in Section~\ref{sec:theory} requires $m$ to be sufficiently large in order to achieve the optimal convergence rate, we find that setting $m = n$ is generally adequate to achieve satisfactory performance in finite-sample settings. Our methods are flexible and allow for many other choices of $(\widetilde{Y}_{i}, \widetilde{\mbf{X}}_{i})$. For example, 
$\widetilde{\mbf{X}}_{i}$ can be generated from the empirical distribution of $\mbf{X}$ based on the target samples, and $\widetilde{Y}_{i}$ can be generated from $\text{Unif}(0,\tau)$. Based on our numerical experience, the performance of the proposed methods is insensitive to these choices.
In some cases, the source study may provide $\widecheck{S}(t^*|\mbf{X})$ only at a fixed time point $t^*$, such as a 5-year survival prediction.
Our methods remain applicable in this setting by taking $\widetilde{Y}_{i} = t^*$ for $i = 1,\dots,m$.
\end{remark}

The similarity metric in \eqref{similarity_metric} is reminiscent of the negative cross-entropy loss and can serve as a penalty to encourage similarity between target and source predictions, thereby facilitating knowledge transfer.
In light of this, we estimate $(\mbf{\beta}, \Lambda)$ by solving the optimization problem
\begin{equation} \label{obj_target}
\begin{aligned}
(\widehat{\mbf{\beta}}, \widehat{\Lambda}) ={} & \argmax_{(\mbf{\beta}, \Lambda)} n^{-1}\log L_n(\mbf{\beta}, \Lambda) + \xi_n \psi_m(\mbf{\beta}, \Lambda) \\
={} & 
\begin{multlined}[t]
\argmax_{(\mbf{\beta}, \Lambda)}
  n^{-1} \sum_{i=1}^n \log\left[\int_0^{\infty}\left\{z_i\Lambda^{\prime}(Y_{i})e^{\mbf{\beta}^{\trans}\mbf{X}_{i}(Y_{i})}\right\}^{\Delta_{i}} 
\exp\left\{-z_i\int_0^{Y_{i}}e^{\mbf{\beta}^{\trans}\mbf{X}_{i}(s)}d\Lambda(s)\right\}f(z_i)dz_i\right] \\
+\xi_n m^{-1}\sum_{i=1}^{m}\Bigl[\widecheck{S}(\widetilde{Y}_{i}|\widetilde{\mbf{X}}_{i})\log S(\widetilde{Y}_{i}|\widetilde{\mbf{X}}_{i}) + \left\{1-\widecheck{S}(\widetilde{Y}_{i}|\widetilde{\mbf{X}}_{i})\right\}\log\left\{1-S(\widetilde{Y}_{i}|\widetilde{\mbf{X}}_{i})\right\}\Bigr],
\end{multlined}
\end{aligned}
\end{equation}
where $\xi_n\ge0$ is a tuning parameter that controls the degree of knowledge transfer and may depend on $n$, and 
\[
  S(\widetilde{Y}_{i}|\widetilde{\mbf{X}}_{i}) = \int_0^{\infty} \exp\biggl\{-\widetilde{z}_i\int_0^{\widetilde{Y}_{i}}e^{\mbf{\beta}^{\trans}\widetilde{\mbf{X}}_{i}(s)}d\Lambda(s)\biggr\}f(\widetilde{z}_i)d\widetilde{z}_i, \quad\text{ for } i=1,\dots,m,
\] 
with $\widetilde{z}_i$ being the frailty variable associated with the generated sample $(\widetilde{Y}_{i}, \widetilde{\mbf{X}}_{i})$. 
In practice, $\xi_n$ can be selected using standard data-driven procedures such as cross-validation. When the assumed similarity in survival predictions between the target and source studies is weak or violated, cross-validation tends to choose a small $\xi_n$, so that our approach effectively reduces to the target-only method.

Compared to conventional $L_r$-penalties on survival probabilities, a major advantage of our cross-entropy-type penalty $\psi_m(\mbf{\beta}, \Lambda)$ is that it leads to more tractable optimization within the maximum likelihood framework, owing to its connection with the likelihood for current status data.
To see this, consider $m$ subjects with current status data $\{\widetilde{Y}_{i}, \delta_{i} = 0, \widetilde{\mbf{X}}_{i}(t): t\in[0,\tau]\}$ ($i=1,\dots,m$), and another $m$ subjects with current status data $\{\widetilde{Y}_{i}, \delta_{i} = 1, \widetilde{\mbf{X}}_{i}(t): t\in[0,\tau]\}$ ($i=1,\dots,m$).
Then the penalty term $\xi_n\psi_m(\mbf{\beta}, \Lambda)$ is exactly the weighted log-likelihood for these current status data, with weights $\xi_nm^{-1}\widecheck{S}(\widetilde{Y}_{i}|\widetilde{\mbf{X}}_{i})$ ($i=1,\dots,m$) for the first $m$ subjects and weights $\xi_nm^{-1}\{1-\widecheck{S}(\widetilde{Y}_{i}|\widetilde{\mbf{X}}_{i})\}$ ($i=1,\dots,m$) for the second $m$ subjects. 
Thus, solving the optimization problem in \eqref{obj_target} is equivalent to maximizing the weighted log-likelihood arising from a mixture of right-censored and current status data. 
This can be done by adapting the nonparametric likelihood strategy to accommodate weighted likelihood contributions.

\subsection{Estimation Procedure}
\label{sub:computational_algorithm}
We adopt the nonparametric maximum likelihood estimation approach by treating $\Lambda$ as a step function with nonnegative jumps at $t_1<t_2<\cdots<t_L$, which are the unique values of $Y_{i}$ ($i=1,\dots,n$) and $\widetilde{Y}_i$ ($i=1,\dots,m$).
For $l=1,\dots,L$, let $\lambda_l$ denote the jump size of $\Lambda$ at time $t_l$. 
Then the objective function in \eqref{obj_target} becomes
\begin{equation} \label{discrete_obj_target}
\begin{split}
  n^{-1} \sum_{i=1}^n \log\left[\int_0^{\infty}\left\{z_i\Lambda\{Y_{i}\}e^{\mbf{\beta}^{\trans}\mbf{X}_{i}(Y_{i})}\right\}^{\Delta_{i}} 
\exp\Bigl(-\sum_{l: t_l\le Y_{i}}z_i\lambda_le^{\mbf{\beta}^{\trans}\mbf{X}_{il}}\Bigr)f(z_i)dz_i\right] \\
+\xi_n m^{-1}\sum_{i=1}^{m}\widecheck{S}(\widetilde{Y}_{i}|\widetilde{\mbf{X}}_{i})\log\Biggl[\int_0^{\infty} \exp\Bigl(-\sum_{l: t_l\le \widetilde{Y}_{i}}\widetilde{z}_i\lambda_{l}e^{\mbf{\beta}^{\trans}\widetilde{\mbf{X}}_{il}}\Bigr) f(\widetilde{z}_i)d\widetilde{z}_i\Biggr] \\
+\xi_n m^{-1}\sum_{i=1}^{m}\left\{1-\widecheck{S}(\widetilde{Y}_{i}|\widetilde{\mbf{X}}_{i})\right\}\log\Biggl[\int_0^{\infty} \biggl\{1-\exp\Bigl(-\sum_{l: t_l\le \widetilde{Y}_{i}}\widetilde{z}_i\lambda_{l}e^{\mbf{\beta}^{\trans}\widetilde{\mbf{X}}_{il}}\Bigr)\biggr\} f(\widetilde{z}_i)d\widetilde{z}_i\Biggr], 
\end{split}
\end{equation}
where $\Lambda\{t\}$ denotes the jump size of $\Lambda$ at time $t$, $\mbf{X}_{il} = \mbf{X}_{i}(t_l)$, and $\widetilde{\mbf{X}}_{il} = \widetilde{\mbf{X}}_{i}(t_l)$.
Direct maximization of \eqref{discrete_obj_target} over $\mbf{\beta}$ and $\lambda_l$'s is challenging because the $\lambda_l$'s are high-dimensional parameters and do not admit analytical expressions. 
To address this challenge, we introduce independent Poisson random variables $W_{0il}$ and $W_{1il}$ ($i=1,\dots,m; \, l=1,\dots,L$), both with means $\widetilde{z}_i\lambda_{l}e^{\mbf{\beta}^{\trans}\widetilde{\mbf{X}}_{il}}$.
It is easy to see that, conditional on the frailty variable $\widetilde{z}_i$, the likelihood contributions from the current status data $(\widetilde{Y}_{i}, \delta_{i} = 0, \widetilde{\mbf{X}}_{i})$ and $(\widetilde{Y}_{i}, \delta_{i} = 1, \widetilde{\mbf{X}}_{i})$ are equivalent to the likelihoods of the events $V_{0i}:= \{\sum_{l: t_l\le \widetilde{Y}_{i}}W_{0il} = 0\}$ and $V_{1i}:= \{\sum_{l: t_l\le \widetilde{Y}_{i}}W_{1il} > 0\}$, respectively. 
Thus, maximizing the objective function in \eqref{discrete_obj_target} is equivalent to maximizing the weighted log-likelihood based on the right-censored data $(Y_i, \Delta_i, \mbf{X}_i)$ $(i=1,\dots,n)$ and the events $V_{0i}$ and $V_{1i}$ ($i=1,\dots,m$).

We develop an EM algorithm to maximize the equivalent objective function based on the right-censored data and the Poisson variables, treating the frailty variables $z_i$ ($i=1,\dots,n$) and $\widetilde{z}_i$ ($i=1,\dots,m$), and the Poisson variables $W_{0il}$ and $W_{1il}$ ($i=1,\dots,m$ and $l=1,\dots,L$) as missing data. 
The complete-data weighted log-likelihood is
\begin{multline*}
n^{-1}\sum_{i=1}^{n} \biggl[\Delta_{i}\left\{\log z_{i} +\log \Lambda\{Y_{i}\} + \mbf{\beta}^{\trans}\mbf{X}_{i}(Y_{i})\right\} 
-\sum_{l: t_l\le Y_{i}}z_{i}\lambda_le^{\mbf{\beta}^{\trans}\mbf{X}_{il}} + \log f(z_{i})\biggr] \\
+\xi_n m^{-1}\sum_{i=1}^{m}\widecheck{S}(\widetilde{Y}_{i}|\widetilde{\mbf{X}}_{i})\sum_{l: t_l\le \widetilde{Y}_{i}}\biggl[-\widetilde{z}_i\lambda_le^{\mbf{\beta}^{\trans}\widetilde{\mbf{X}}_{il}} + W_{0il}\left\{\log \widetilde{z}_i+\log\lambda_l+\mbf{\beta}^{\trans}\widetilde{\mbf{X}}_{il}\right\}-\log(W_{0il}!)\biggr] \\
+\xi_n m^{-1}\sum_{i=1}^{m}\left\{1-\widecheck{S}(\widetilde{Y}_{i}|\widetilde{\mbf{X}}_{i})\right\}\sum_{l: t_l\le \widetilde{Y}_{i}}\biggl[-\widetilde{z}_i\lambda_le^{\mbf{\beta}^{\trans}\widetilde{\mbf{X}}_{il}} + W_{1il}\left\{\log \widetilde{z}_i+\log\lambda_l+\mbf{\beta}^{\trans}\widetilde{\mbf{X}}_{il}\right\}-\log(W_{1il}!)\biggr] \\
+\xi_n m^{-1}\sum_{i=1}^{m} \log f(\widetilde{z}_{i}).
\end{multline*}
Let $\widetilde{E}(\cdot)$ denote the conditional expectation given the observed data. 
In the E-step, we only need to evaluate $\widetilde{E}(z_{i})$, $\widetilde{E}(\widetilde{z}_{i})$, $\widetilde{E}(W_{0il})$, and $\widetilde{E}(W_{1il})$.
The conditional density of $z_{i}$ given the observed data is proportional to $z_i^{\Delta_{i}} \exp(-\sum_{l: t_l\le Y_{i}}z_i\lambda_le^{\mbf{\beta}^{\trans}\mbf{X}_{il}})f(z_i)$,
while the conditional density of $\widetilde{z}_{i}$ given the observed data is proportional to 
\[
  \biggl\{\exp\Bigl(-\sum_{l: t_l\le \widetilde{Y}_{i}}\widetilde{z}_i\lambda_{l}e^{\mbf{\beta}^{\trans}\widetilde{\mbf{X}}_{il}}\Bigr)\biggr\}^{\widecheck{S}(\widetilde{Y}_{i}|\widetilde{\mbf{X}}_{i})} \\
\times \biggl\{1-\exp\Bigl(-\sum_{l: t_l\le \widetilde{Y}_{i}}\widetilde{z}_i\lambda_{l}e^{\mbf{\beta}^{\trans}\widetilde{\mbf{X}}_{il}}\Bigr)\biggr\}^{1-\widecheck{S}(\widetilde{Y}_{i}|\widetilde{\mbf{X}}_{i})}f(\widetilde{z}_i).
\]
It follows directly from the definition of $V_{0i}$ that $E(W_{0il}|V_{0i}, \widetilde{z}_{i}) = 0$.
Meanwhile, a simple calculation gives, for $t_l \le \widetilde{Y}_i$,
\[
E(W_{1il}|V_{1i}, \widetilde{z}_{i}) = \frac{\widetilde{z}_i\lambda_le^{\mbf{\beta}^{\trans}\widetilde{\mbf{X}}_{il}}}{1-\exp\left(-\sum_{l': t_{l'}\le \widetilde{Y}_{i}}\widetilde{z}_i\lambda_{l'}e^{\mbf{\beta}^{\trans}\widetilde{\mbf{X}}_{il'}}\right)}.
\]
We use Gauss--Laguerre quadrature to approximate all integrals with respect to $z_i$ and $\widetilde{z}_i$.

Define the functions 
\begin{align*}
s^{(0)}(t,\mbf{\beta}) ={} & n^{-1}\sum_{i=1}^{n}I(t\le Y_{i})\widetilde{E}(z_{i})e^{\mbf{\beta}^{\trans}\mbf{X}_{i}(t)}+\xi_nm^{-1}\sum_{i=1}^m I(t\le \widetilde{Y}_{i})\widetilde{E}(\widetilde{z}_{i})e^{\mbf{\beta}^{\trans}\widetilde{\mbf{X}}_{i}(t)}, \\
s^{(1)}(t,\mbf{\beta}) ={} & n^{-1}\sum_{i=1}^{n}I(t\le Y_{i})\widetilde{E}(z_{i})e^{\mbf{\beta}^{\trans}\mbf{X}_{i}(t)}\mbf{X}_{i}(t)+\xi_nm^{-1}\sum_{i=1}^m I(t\le \widetilde{Y}_{i})\widetilde{E}(\widetilde{z}_{i})e^{\mbf{\beta}^{\trans}\widetilde{\mbf{X}}_{i}(t)}\widetilde{\mbf{X}}_{i}(t).
\end{align*}
In the M-step, we first update $\lambda_l$ ($l=1,\dots,L$) by 
\[
\lambda_l = \frac{n^{-1}\sum_{i=1}^{n}\Delta_{i}I(Y_{i} = t_l)+\xi_nm^{-1}\sum_{i=1}^m I(t_l\le \widetilde{Y}_{i})\left\{1-\widecheck{S}(\widetilde{Y}_{i}|\widetilde{\mbf{X}}_{i})\right\}\widetilde{E}(W_{1il})}{s^{(0)}(t_l,\mbf{\beta})}.
\]
We then update $\mbf{\beta}$ by solving the following equation using the one-step Newton--Raphson method:
\[
\mbf{0} = 
\begin{multlined}[t]
n^{-1}\sum_{i=1}^{n}\Delta_{i}\left\{\mbf{X}_{i}(Y_{i})-\frac{s^{(1)}(Y_i,\mbf{\beta})}{s^{(0)}(Y_i,\mbf{\beta})}\right\} \\
+ \xi_nm^{-1}\sum_{i=1}^m \sum_{l: t_l\le \widetilde{Y}_{i}}\left\{1-\widecheck{S}(\widetilde{Y}_{i}|\widetilde{\mbf{X}}_{i})\right\}\widetilde{E}(W_{1il})\left\{\widetilde{\mbf{X}}_{il}-\frac{s^{(1)}(t_l,\mbf{\beta})}{s^{(0)}(t_l,\mbf{\beta})}\right\}.
\end{multlined}
\]

We set the initial value of $\mbf{\beta}$ to $\mbf{0}$ and the initial value of each $\lambda_l$ to $1/L$.
We iterate between the E- and M-steps until the sum of the $L_2$-norm difference in the $\mbf{\beta}$ estimator and the maximum difference in the $\lambda_l$ estimators ($l=1,\dots,L$) between successive iterations is smaller than $10^{-6}$.
As established in~\ref{s-sec:justification_for_the_em_algorithm}, the observed-data weighted likelihood is guaranteed to increase at each iteration. 
Another advantage of our EM algorithm is that the high-dimensional parameters $\lambda_l$ are updated explicitly in the M-step, eliminating the need to invert large matrices.
Given the estimators $(\widehat{\mbf{\beta}}, \widehat{\Lambda})$, we estimate the survival function for the target study by
\[
  \widehat{S}(t|\mbf{X}) = S(t|\mbf{X}; \widehat{\mbf{\beta}}, \widehat{\Lambda}) = \exp\left[-G\left\{\int_0^{t}e^{\widehat{\mbf{\beta}}^{\trans}\mbf{X}(s)}d\widehat{\Lambda}(s)\right\}\right].
\]
Various predictions for the target population can then be made based on $\widehat{S}(t|\mbf{X})$.

\subsection{Extension to Multiple Sources}
\label{sub:multi-source}
When there are $K$ source studies, each providing a survival function predictor $\widecheck{S}_k(t|\mbf{X})$ ($k=1,\dots,K$), we can extend the proposed POTL method by first constructing a pooled source predictor. Specifically, we define $\widecheck{S}(t|\mbf{X})=\sum_{k=1}^K c_k \widecheck{S}_k(t|\mbf{X})$, where the weights satisfy $c_k\ge 0$ and $\sum_{k=1}^K c_k=1$. The pooled predictor $\widecheck{S}(t|\mbf{X})$ can then be incorporated into the POTL procedure described above in the same way as a single-source predictor. This is equivalent to adding a separate penalty term for each source predictor to the objective function in \eqref{obj_target}.
The source weights $c_k$ may be pre-specified when prior knowledge is available regarding the informativeness or relevance of each source study; otherwise, they can be selected in a data-adaptive manner, such as by cross-validation.

\section{Asymptotic Theory}
\label{sec:theory}
Let $(\mbf{\beta}_0, \Lambda_0)$ denote the true values of $(\mbf{\beta}, \Lambda)$, $S_0(t|\mbf{X}) = S(t|\mbf{X}; \mbf{\beta}_0, \Lambda_0)$ denote the true survival function for the target population, and $S_1(t|\mbf{X})$ denote the true survival function for the source population.
We measure the dissimilarity between $S_0(t|\mbf{X})$ and $S_1(t|\mbf{X})$ by
\[
\eta = E\|S_1(t|\mbf{X})-S_0(t|\mbf{X})\|_{L_2}^2,
\]
where $\|\cdot\|_{L_2}$ denotes the $L_2$-norm with respect to the Lebesgue measure over $[0,\tau]$, and the expectation is taken over the covariates $\mbf{X}$.
A small value of $\eta$ indicates that the target and source studies share similar survival predictions, enabling effective knowledge transfer in the proposed POTL approach.
In addition, we assume that the source predictor $\widecheck{S}(t|\mbf{X})$ is well-defined for any $t\in[0,\tau]$ and $\mbf{X}$ in the target domain, and converges to $S_1(t|\mbf{X})$ at the following rate:
\[
E\|\widecheck{S}(t|\mbf{X})-S_1(t|\mbf{X})\|_{L_2}^2\le O_p(q_N),
\]
where $N$ is the sample size of the source study. 
Typically, a rate of $q_N = N^{-1}$ is achieved when the source predictor is derived from commonly used regression models, such as Cox and transformation models.
Machine learning methods, such as random survival forests and deep neural networks, can also be used for source prediction, with the rate $q_N$ generally slower than $N^{-1}$ and depending on the specific settings \citep{mentch2016quantifying,zhong2021deep,zhong2022deep}.

The following theorem states our main results about the convergence rate for $\widehat{S}(t|\mbf{X})$. 
\begin{theorem} \label{thm:conv_rate}
Suppose that $\xi_n\rightarrow\xi$ as $n\rightarrow\infty$. Under Conditions~\ref{s-true_parameter}--\ref{s-identifiability} in~\ref{s-sec:theory}, we have 
\[
E\|\widehat{S}(t|\mbf{X})-S_0(t|\mbf{X})\|_{L_2}^2 \le 
\begin{cases}
O_p\bigl\{n^{-1} + m^{-2/3}\xi_n^{4/3}+(\xi_n-\xi)^{2}+\xi(\eta+q_N)\bigr\} & \text{ if } \xi<\infty, \\
O_p\bigl(\xi_n^{-2} + m^{-2/3} + \eta+q_N\bigr) & \text{ if } \xi=\infty.
\end{cases}
\]
\end{theorem}

We can select $m$, the number of subjects in the penalty term, to be sufficiently large so that the terms involving $m^{-2/3}$ become negligible. 
We then minimize the upper bound established in Theorem~\ref{thm:conv_rate} to derive the optimal convergence rate.
\begin{corollary} \label{coro:opt_conv_rate}
Under Conditions~\ref{s-true_parameter}--\ref{s-identifiability} in~\ref{s-sec:theory}, the optimal convergence rate of $\widehat{S}(t|\mbf{X})$ is given by 
\[
  E\|\widehat{S}(t|\mbf{X})-S_0(t|\mbf{X})\|_{L_2}^2 \le O_p\left\{n^{-1} \land (\eta+q_N)\right\},
\]
where $x\land y=\min(x,y)$.
\end{corollary} 
Thus, when the source predictor is sufficiently accurate such that $\eta+q_N= o(n^{-1})$, POTL achieves a faster convergence rate for $\widehat{S}(t|\mbf{X})$ compared to the target-only method.  
In the ideal case where the target and source predictions are identical (i.e., $\eta=0$) and $q_N= o(n^{-1})$, we achieve the source-only convergence rate $O_p(q_N)$. 

We also provide in Theorem~\ref{s-thm:opt_xi} of~\ref{s-sec:theory} an explicit expression for the tuning parameter $\xi_n$ that guarantees the optimal convergence rate of $\widehat{S}(t|\mbf{X})$. 
The proofs of all theoretical results are provided in~\ref{s-sec:proof_thm}.


\section{Simulation Studies}
\label{sec:simulation}
We conduct extensive simulation studies to evaluate the performance of the proposed POTL method. 
For the target study, we set the sample size to $n=100$ and the study duration to $\tau=2$. 
We consider two independent covariates for the target population: $X_1\sim\text{Ber}(0.5)$ and $X_2\sim\text{Unif}(0, 1)$. 
The failure time is generated from the Cox regression model:  
\begin{equation} \label{target_model}
\Lambda(t|X_1, X_2) = \Lambda(t)\exp(\beta_1X_1+\beta_2X_2),
\end{equation}
with $\Lambda(t) = \log(1+0.5t)$, $\beta_1 = 0.5$, and $\beta_2 = -0.5$.
The censoring time is generated from $\min\{\text{Unif}(1.5, 4), \tau\}$, resulting in a censoring rate of about 50\%. 

We consider one source study with the sample size of $N=1000$ and the study duration of $\tau_s = 5$.
The covariates are generated in the same way as in the target study. 
We consider the following five Cox-type models for the source study:  
\begin{itemize}
  \item SC1: The source model is identical to model~\eqref{target_model}, with $(\beta_1, \beta_2) = (0.5, -0.5)$ and $\Lambda(t) = \log(1+0.5t)$.
  \item SC2: The source model follows model~\eqref{target_model} with $(\beta_1, \beta_2) = (0.5, -0.5)$ and $\Lambda(t) = 0.4t$.
  \item SC3: The source model follows model~\eqref{target_model} with $(\beta_1, \beta_2) = (0.7, -0.7)$ and $\Lambda(t) = 0.4t$.
  \item SC4: The source model is identical to model~\eqref{target_model} within $[0,\tau]$, but contains a time-covariate interaction beyond $\tau$:
  \[
  \Lambda(t|X_1, X_2) = \int_0^t\exp\bigl\{\beta_1X_1+\beta_2X_2+\beta_3tX_1I(t>\tau)\bigr\}d\Lambda(t),
  \]
  with $(\beta_1, \beta_2, \beta_3) = (0.5, -0.5, -0.3)$ and $\Lambda(t) = \log(1+0.5t)$.
  \item SC5: The source model contains a time-covariate interaction throughout $[0,\tau_s]$: 
  \[
  \Lambda(t|X_1, X_2) = \int_0^t\exp\bigl\{\beta_1X_1+\beta_2X_2+\beta_3tX_1\bigr\}d\Lambda(t),
  \]
  with $(\beta_1, \beta_2, \beta_3) = (0.5, -0.5, -0.3)$ and $\Lambda(t) = \log(1+0.5t)$.
\end{itemize}
We generate the censoring time from $\min\{\text{Unif}(3.5, 7), \tau_s\}$, yielding a censoring rate of approximately 25\% in SC1--SC3 and 35\% in SC4--SC5.

We compare the performance of our POTL method with several existing methods: the target-only method, the TransCox method developed by \citet{li2023accommodating}, the CoxTL method developed by \citet{lu2026adaptive}, and pooled analysis using both the target and source data.
Since CoxTL includes the stratified Cox model as a special case when its tuning parameters are fixed at certain values, we do not separately compare with the stratified Cox model.
\revise{To obtain the source predictor required by POTL, we fit a standard Cox model of form~\eqref{target_model} in SC1--SC3 and a varying coefficient Cox model in SC4--SC5. For the latter, the time-varying coefficients for $X_1$ and $X_2$ are estimated using cubic splines, with internal knots placed at equal quantiles of the observed source event times and the number of knots selected by the Akaike Information Criterion (AIC).}
The optimal value of the tuning parameter $\xi_n$ in POTL is selected using 5-fold cross-validation. 
Following recommendations in \citet{li2023accommodating} and \citet{lu2026adaptive}, the tuning parameters in TransCox and CoxTL are selected using the Bayesian Information Criterion (BIC) and 5-fold cross-validation, respectively. 

The evaluation is conducted using the following metrics on a large validation set consisting of $n_v = $ 10,000 samples generated from the target population:
\begin{itemize}
  \item L$_2$D: The $L_2$-distance between $\widehat{S}(t|\mbf{X})$ and $S_0(t|\mbf{X})$ over $[0,\tau]$, defined as 
  \[
  	\mathrm{L}_2\mathrm{D} = n_v^{-1}\sum_{i=1}^{n_v}\left\{\int_0^{\tau} \left|\widehat{S}(t|\mbf{X}_i)-S_0(t|\mbf{X}_i)\right|^2dt\right\}^{1/2}.
  \]
  \item D$_\tau$: The absolute difference between $\widehat{S}(\tau|\mbf{X})$ and $S_0(\tau|\mbf{X})$, defined as $n_v^{-1}\sum_{i=1}^{n_v}|\widehat{S}(\tau|\mbf{X}_i)-S_0(\tau|\mbf{X}_i)|$.
  \item C-index \citep{uno2011c}.
  \item IBS: The integrated Brier score \citep{graf1999assessment} over $[0,\tau]$.
  \item RMST: The error in the estimated restricted mean survival time, defined as
  \[
  \frac{\sum_{i=1}^{n_v}\Delta_i\widehat{H}(Y_i)^{-1}|Y_i-\widehat{T}_i|}{\sum_{i=1}^{n_v}\Delta_i\widehat{H}(Y_i)^{-1}},
  \]
  where $\widehat{H}(\cdot)$ is the Kaplan--Merier estimator of the censoring distribution, and $\widehat{T}_i$ is the $i$th subject's estimated restricted mean survival time. 
\end{itemize}
For the C-index, a larger value indicates better performance, whereas for all other metrics, smaller values indicate better performance. 
The last three metrics incorporate inverse probability of censoring weighting, which is unnecessary for the validation sets in this section, as they do not contain any censored subjects. 
However, this weighting becomes important when applying these metrics to the real data in the subsequent section.

\begin{figure}
\centerline{%
\includegraphics[width=\textwidth]{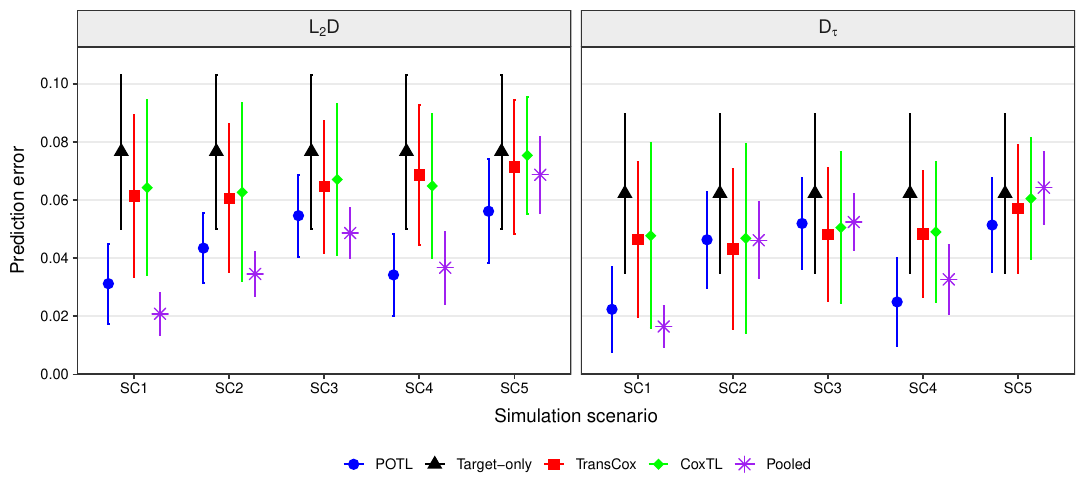}}
\caption{\revise{Simulation results for the L$_2$D and D$_\tau$ metrics. Points represent median over 200 replicates, and vertical bars represent median $\pm$ median absolute deviation.}}
\label{fig:sim_pred_error}
\end{figure}

\revise{Figure~\ref{fig:sim_pred_error} displays the L$_2$D and D$_\tau$ metrics across all scenarios, based on 200 replicates; the full results for all metrics are presented in SC1--SC5 of Web Table~\ref{s-tb:main_sim}. POTL performs comparably to or slightly better than competing methods in the C-index, IBS, and RMST metrics. The relatively small differences among methods in these three metrics are expected because they evaluate predictive performance based on discrimination or prediction accuracy relative to observed outcomes, rather than the direct discrepancy from the true survival function. Therefore, we focus primarily on the L$_2$D and D$_\tau$ metrics, which directly quantify the distance between the estimated and true survival functions. In SC1--SC3, where both the target and source studies follow Cox models, pooled analysis achieves the best performance by leveraging individual-level source data. Nevertheless, POTL performs comparably to pooled analysis while consistently outperforming the other three methods, despite requiring only summary-level source information. In SC4--SC5, where the source model violates the proportional hazards assumption while remaining predictive within the target horizon $[0,\tau]$, POTL achieves substantial improvements over all competing methods. These results demonstrate that POTL can effectively transfer predictive information within the target-relevant time window without requiring similar model structures or parameters between the target and source studies. Therefore, POTL provides a flexible alternative to existing parameter-based transfer learning methods while preserving source data privacy.}

\revise{To assess whether the advantages of POTL in SC4--SC5 are primarily attributable to its flexible source modeling, we let POTL use the same standard Cox model for source prediction as the competing methods, resulting in source model misspecification in SC4--SC5. As shown in SC4$^{\prime}$--SC5$^{\prime}$ of Web Table~\ref{s-tb:main_sim}, POTL still outperforms TransCox and CoxTL in terms of L$_2$D and D$_\tau$ in SC4$^{\prime}$, suggesting that the prediction-oriented transfer mechanism itself contributes to the improvement when source predictions remain reasonably accurate. In SC5$^{\prime}$, however, this advantage is attenuated when the misspecified source model yields substantially biased predictions. Overall, the improvement of POTL in SC4--SC5 reflects both its flexibility in source modeling and the effectiveness of its prediction-oriented transfer mechanism.}

\begin{figure}
\centerline{%
\includegraphics[width=\textwidth]{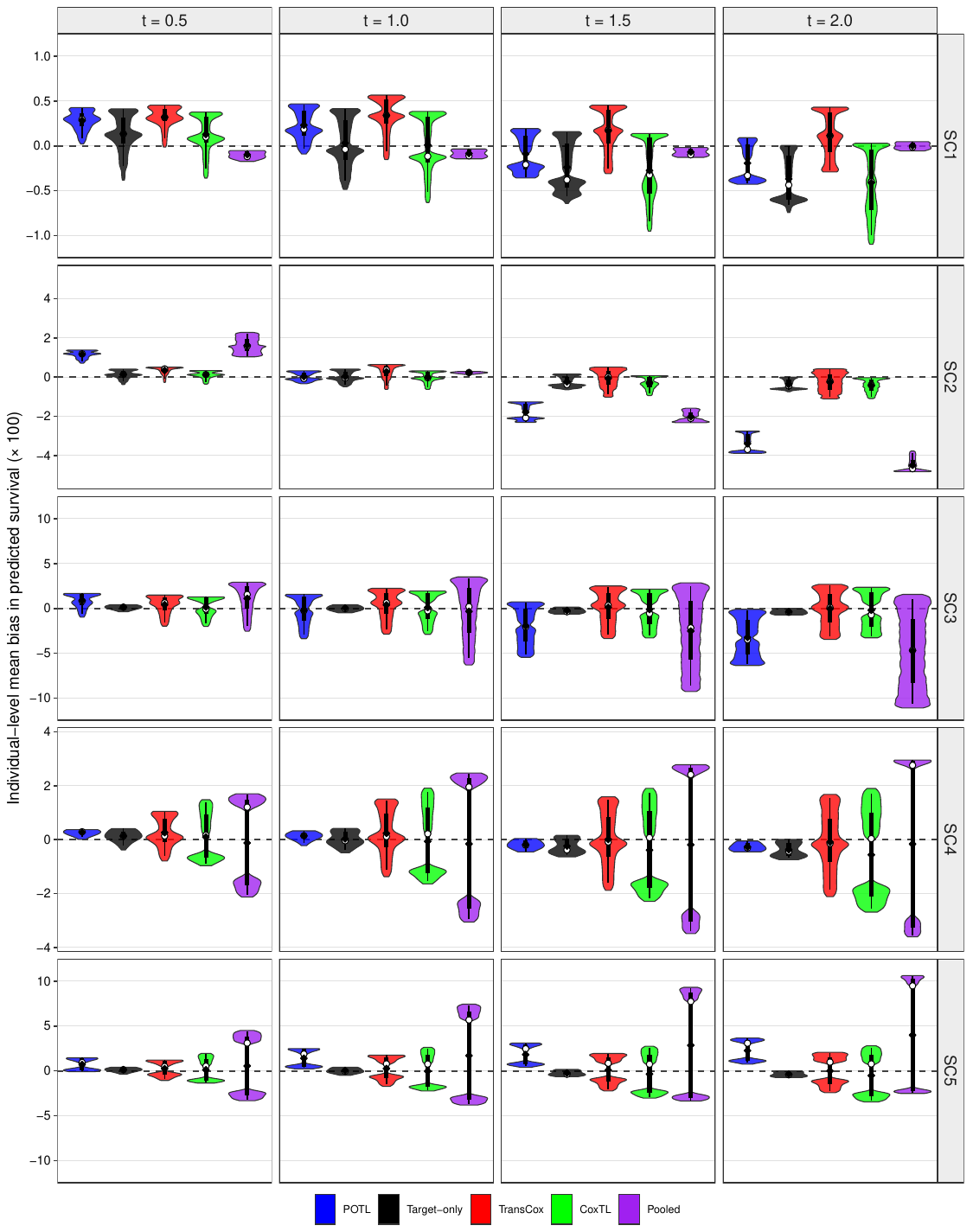}}
\caption{\revise{Violin plots of individual-level mean bias in predicted survival probabilities. For each of 10,000 validation subjects, the mean signed bias is calculated across 200 replicates at prediction times $t=0.5$, $1.0$, $1.5$, and $2.0$. Biases are multiplied by 100 for display.}}
\label{fig:sim_indiv_bias}
\end{figure}

\revise{In addition to the aforementioned population-level metrics, we also evaluate individual-level calibration of the estimated survival functions. Specifically, for each validation subject, we calculate the mean signed bias of the estimated survival probability across 200 replicates. This metric provides a complementary assessment by quantifying systematic over- or under-estimation for individual subjects, whereas L$_2$D and D$_\tau$ summarize survival function estimation accuracy at the population level. Figure~\ref{fig:sim_indiv_bias} displays the distributions of these individual-level mean signed biases at four prediction times across the 10,000 validation subjects. Across scenarios, all methods achieve generally small individual-level mean biases, although systematic deviations may occur at later prediction times. In SC1--SC3, POTL and pooled analysis show similar bias patterns, reflecting their stronger use of source information. In SC4--SC5, POTL maintains stable individual-level bias while achieving smaller L$_2$D and D$_\tau$ values than competing methods. Together, these results indicate that POTL improves overall survival function estimation while maintaining satisfactory individual-level calibration.}

We further evaluate the performance of all five methods under covariate shift between the target and source data, as well as between the training and validation data, and in settings where the covariate sets differ between the target and source studies. The conclusions remain unchanged, demonstrating that POTL is robust to covariate shift and broadly applicable to real-world studies with heterogeneous covariates. Finally, we compare POTL with the target-only, CoxTL, and pooled methods in multi-source transfer learning. The results show that POTL outperforms all competing methods when using either sample-size-based or cross-validation-based source weights. Details of these additional simulation studies are provided in~\ref{s-sec:additional_simulation_studies}.

\section{Application}
\label{sec:application}
We apply the proposed method to breast cancer data from the TCGA--BRCA and METABRIC studies. 
The TCGA--BRCA study, which collected tumor samples from 1,096 patients, is designated as the target study, while the METABRIC study, comprising 2,509 patients, serves as the source study. 
The primary outcome of interest is overall survival.
Clinical variables assessed in both studies include age at diagnosis, number of positive lymph nodes examined, tumor stage, histological type, estrogen receptor (ER) status, progesterone receptor (PR) status, and human epidermal growth factor receptor 2 (HER2) status.
We dichotomize tumor stages into early stage (stages 0, I, and II) versus advanced stage (stages III and IV), and we dichotomize histological types of breast cancer into invasive ductal carcinoma (IDC) versus others.
In addition, we perform principal component analysis on mRNA expression data and include the first 10 principal components as genetic variables.

\begin{figure}
\centerline{%
\includegraphics[width=\textwidth]{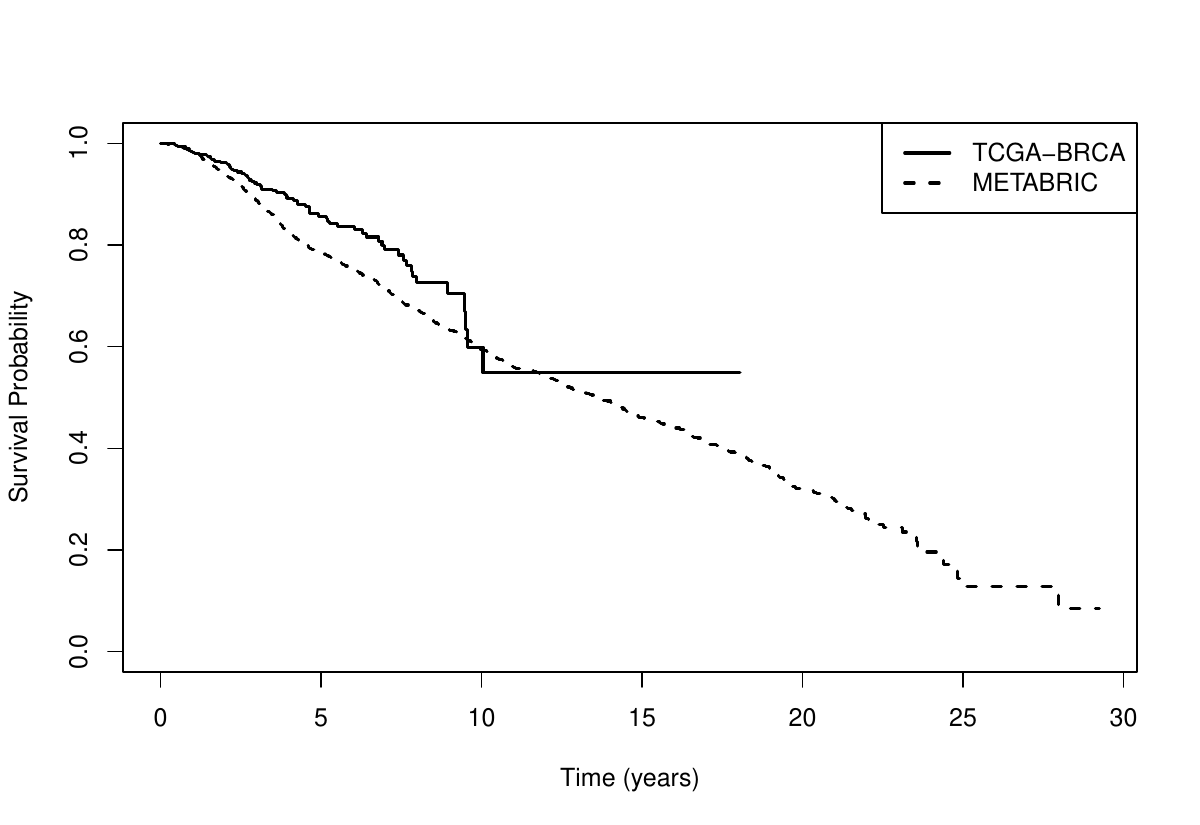}}
\caption{Kaplan--Meier estimators for survival time distributions of the TCGA--BRCA and METABRIC studies.}
\label{fig:app_km}
\end{figure}

After removing patients with missing lymph node and ER/PR/HER2 information, there remain 762 patients in the target study and 1,393 patients in the source study.
The event rate was approximately 10\% in the target study and 56\% in the source study. 
Median follow-up times were approximately 2.4 years for the target study and 9.8 years for the source study, with maximum follow-up times of around 18.1 years in the target study and 29.3 years in the source study.
Figure \ref{fig:app_km} displays the Kaplan--Meier estimators for survival time distributions in the two studies. 
Although overall survival is generally poorer in the source study compared to the target study, the similarities in survival probability during the early and late stages of the target study period support the use of transfer learning methods to improve prediction accuracy for the target population. 
The clinical characteristics of patients in both studies are summarized in Web Table~\ref{s-tb:summary_profile}, which shows broadly similar clinical profiles, with a slight shift in the distribution of tumor stage.    

\begin{figure}
\centerline{%
\includegraphics[width=\textwidth]{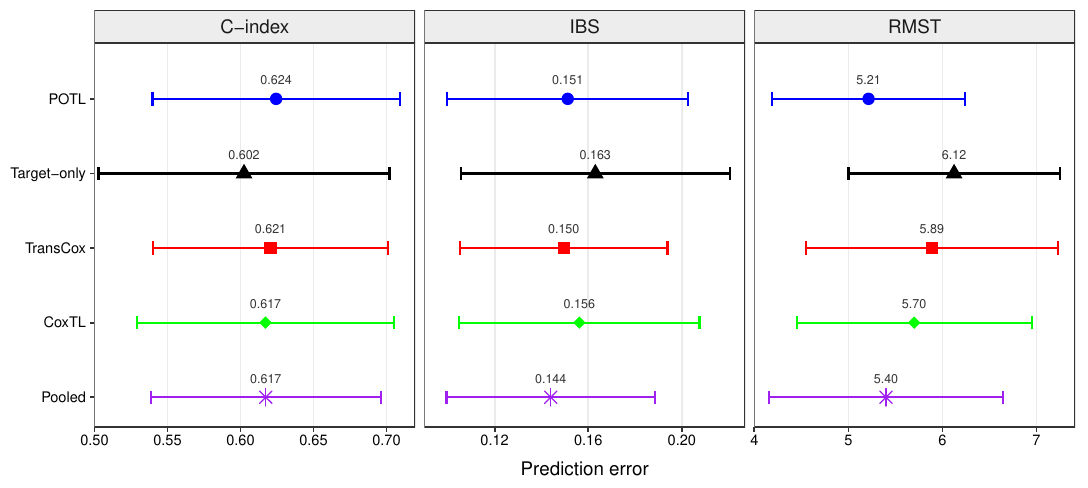}}
\caption{\revise{Prediction performance of all methods in the real data application. Points represent median over 100 replicates, with the corresponding median values labeled above the points; vertical bars represent median $\pm$ median absolute deviation.}}
\label{fig:app_pred_error}
\end{figure}

We assess the performance of POTL, target-only analysis, TransCox, CoxTL, and pooled analysis using the metrics C-index, IBS, and RMST described in Section~\ref{sec:simulation}.
For POTL, target-only, and pooled methods, the optimal transformation models are selected based on AIC, resulting in optimal $r$ values of 1.5, 0.1, and 0.1 for the target, source, and combined data, respectively. 
However, the AIC value does not vary substantially as $r$ increases from 0 to 1.5, suggesting that fitting a Cox model to each dataset would generally be sufficient. 
\revise{The target data are randomly divided into training and validation sets at a 7:3 ratio, and the entire evaluation procedure is repeated 100 times. 
All tuning parameters are selected in the same way as described in Section~\ref{sec:simulation}.
Results from all methods across these 100 replicates are presented in Figure~\ref{fig:app_pred_error}. 
POTL achieves the highest C-index and the lowest RMST error, while achieving an IBS comparable to that of the best-performing method. Despite not using individual-level source data, POTL demonstrates competitive prediction performance compared to existing transfer learning methods and pooled analysis. These results support the accuracy and efficiency of POTL for survival prediction in real-world studies.}

\begin{figure}
\centerline{%
\includegraphics[width=\textwidth]{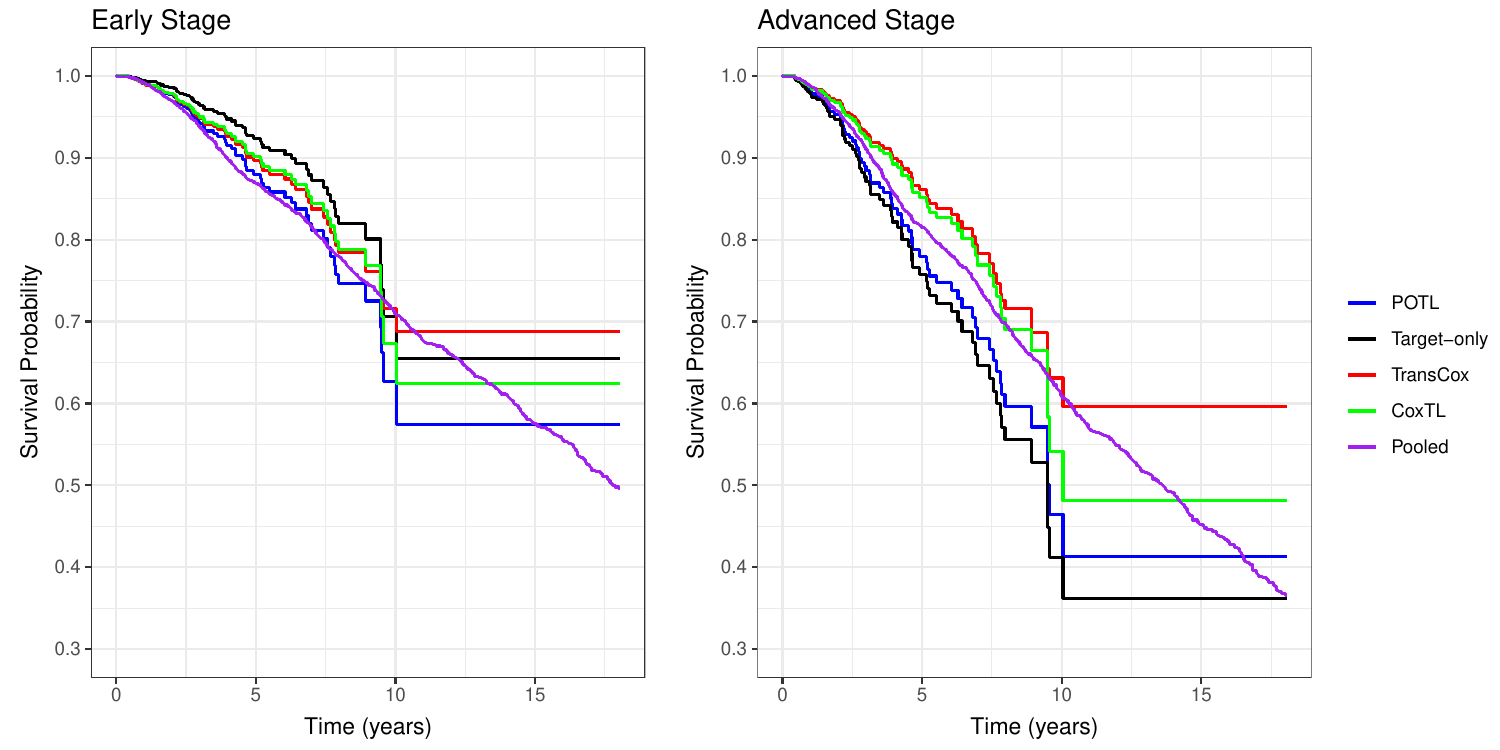}}
\caption{\revise{Predicted survival probabilities for future patients in the TCGA--BRCA study with different tumor stages. All other covariates are fixed at the median values among the current TCGA--BRCA patients.}}
\label{fig:app_surv_pred}
\end{figure}

Finally, we estimate survival probabilities for new patients from the target population. 
We consider two new patients: one with an early-stage tumor and the other with an advanced-stage tumor, while keeping all other covariates fixed at their median values in the target study. 
Figure~\ref{fig:app_surv_pred} illustrates the predicted survival curves for these patients using various methods.
Across all methods, the early-stage patient consistently demonstrates better predicted survival outcomes than the advanced-stage patient. 
This result is consistent with the well-established association between advanced tumor stage and increased mortality risk \citep{edge2010american}.

\section{Discussion}
\label{sec:discussion}
In this article, we introduce a novel, prediction-driven transfer learning approach for survival analysis, which leverages predictive knowledge from source studies to enhance survival prediction in the target study. 
Our approach stands out from existing methods by relying on a less stringent similarity assumption and accommodating various types of source predictors, without the need to model source distributions or share individual-level source data.
We provide theoretical guarantees by showing that the optimal convergence rate of our proposed estimator is faster than the target-only rate when the $L_2$-distance between the target and source survival functions is sufficiently small and the convergence rate of the source predictor is not too slow.
Our approach has shown satisfactory performance in both the simulation studies and the real data application.

The similarity metric $\psi_m(\mbf{\beta}, \Lambda)$ defined in~\eqref{similarity_metric} can be generalized to a weighted average of the pairwise similarities between $\widecheck{S}(\widetilde{Y}_{i}|\widetilde{\mbf{X}}_{i})$ and $S(\widetilde{Y}_{i}|\widetilde{\mbf{X}}_{i})$, where the weights reflect the uncertainty or relevance of the source prediction $\widecheck{S}(\widetilde{Y}_{i}|\widetilde{\mbf{X}}_{i})$, when such information is available. For example, one may use the inverse variance of $\widecheck{S}(\widetilde{Y}_{i}|\widetilde{\mbf{X}}_{i})$ as the weight, so that source predictions with lower uncertainty contribute more strongly to the metric. 
Similarly, source predictions that deviate substantially from their corresponding target-only predictions may be downweighted, thereby reducing their influence in the knowledge transfer process.
Moreover, when prior knowledge suggests that source predictions are more reliable during certain time periods, larger weights can be assigned to generated data points with $\widetilde{Y}_{i}$ falling within those periods.
The asymptotic theory established in Section~\ref{sec:theory} generally continues to hold for the weighted version of $\psi_m(\mbf{\beta}, \Lambda)$, provided that the weights are uniformly bounded away from zero and infinity. 

In Section~\ref{sec:application}, we used a source study (i.e., METABRIC) with available individual-level data to enable direct comparison with CoxTL and pooled analysis, both of which require access to individual-level source data. However, our approach is broadly applicable to privacy-sensitive data sources that do not permit sharing of individual-level information, such as large biobanks, electronic health records, and major cohort studies.
Furthermore, online risk calculators are increasingly used in clinical and research settings. These tools typically estimate 5- or 10-year risk of certain disease using models developed from one or more large cohorts. Examples include the widely used FRAX tool for estimating fracture risk, the National Cancer Institute's Breast Cancer Risk Assessment Tool (Gail model), and the American Heart Association's recently developed PREVENT calculator for estimating cardiovascular disease risk. Our approach can be used to transfer risk estimates from such online calculators, thereby leveraging information from large external cohorts to improve prediction performance in target studies with limited sample sizes.

The proposed POTL method may be suboptimal when multiple source studies involve heterogeneous populations, because each source predictor may be informative only for a subset of the target population. Inspired by \citet{xiong2023distributionally} and \citet{zhan2024transfer} on source-mixing structures in linear regression models, a potential extension is to represent the target population as a mixture of the source populations. Consequently, survival prediction in the target population could be approximated by a mixture of source predictions.
For each target subject, one could identify the most relevant source population and transfer information only from the corresponding matched source predictor, rather than borrowing from all source predictors.
This direction is particularly meaningful for cancer studies, where the target cohort may contain latent demographic or biological subpopulations, and more individualized transfer learning may be needed to optimize survival prediction performance.

\backmatter

\section*{Acknowledgements}
The authors thank the Editor, Associate Editor, and anonymous referees for their valuable comments.
This research was supported by the Hong Kong Research Grant Council grant 27303624 and research funds from the University of Hong Kong.
\vspace*{-8pt}

\section*{Supplementary Materials}
Web Appendices, Tables, and Figures referenced in Sections~\ref{sec:methods}--\ref{sec:application}, and a .zip file containing the code and data needed to reproduce the numerical results, are available with this paper at the Biometrics website on Oxford Academic.
The code and data are also available on GitHub at \url{https://github.com/yugu-stat/POTL}.
\vspace*{-8pt}

\section*{Data Availability}
Data used in this paper to support our findings are available in the following repositories: Broad Institute GDAC (\url{https://gdac.broadinstitute.org/}) and cBioPortal for Cancer Genomics (\url{https://www.cbioportal.org/study/summary?id=brca_metabric}).
\vspace*{-8pt}


%
 \bibliographystyle{biom} 
\bibliography{paper-ref}










\label{lastpage}

\end{document}



\date{{\it Received October} 2007. {\it Revised February} 2008.  {\it
Accepted March} 2008.}



\pagerange{\pageref{firstpage}--\pageref{lastpage}} 
\volume{64}
\pubyear{2008}
\artmonth{December}


\doi{10.1111/j.1541-0420.2005.00454.x}


\label{firstpage}


\begin{abstract}
This Supplementary Material provides additional details on the asymptotic theory presented in~\ref{sec:theory}, with proofs given in~\ref{sec:proof_thm}. Auxiliary lemmas and their proofs are provided in~\ref{sec:useful_lemmas}, and technical details of the EM algorithm are included in~\ref{sec:justification_for_the_em_algorithm}. Additional simulation studies are presented in~\ref{sec:additional_simulation_studies}, followed by all Web Figures and Tables.
\end{abstract}

%



\maketitle


%

\section{Details on Asymptotic Theory}
\label{sec:theory}

\subsection{Regularity Conditions}
\label{sub:regularity_conditions}
\begin{condition} \label{true_parameter}
The true parameter $\mbf{\beta}_0$ lies in the interior of a known compact set $\mathcal{B}\subset\mathbb{R}^p$. The true parameter $\Lambda_0(\cdot)$ is continuously differentiable with a positive derivative in $[0,\tau]$.
\end{condition}
\begin{condition} \label{X_cont_diff}
With probability one, the covariates $\mbf{X}(\cdot)$ are continuously differentiable in $[0,\tau]$. In addition, if there exist some constant vector $\mbf{a}_1$ and some deterministic function $a_2(t)$ such that with probability one, $\mbf{a}_1^{\trans}\mbf{X}(t)+a_2(t) = 0$ for any $t\in[0,\tau]$, then $\mbf{a}_1 = \mbf{0}$ and $a_2(t) = 0$ for $t\in[0,\tau]$.
\end{condition}
\begin{condition} \label{prob_Ctau}
The conditional probability $\pr(C=\tau|\mbf{X})$ is strictly positive. 
In addition, the conditional density function of $C$ given $\mbf{X}$ exists and is uniformly bounded almost everywhere in $[0,\tau]$.  
\end{condition} 
\begin{condition} \label{transformation}
The transformation function $G(\cdot)$ is thrice continuously differentiable and strictly increasing with $G(0) = 0$ and $G(\infty) = \infty$. Moreover, there exists a positive constant $\rho$ such that $\limsup_{x\rightarrow\infty}(1+x)^{\rho}e^{-G(x)}<\infty$ and $\limsup_{x\rightarrow\infty}(1+x)^{1+\rho}e^{-G(x)}G^{\prime}(x)<\infty$.
\end{condition}
\begin{condition} \label{Scheck_bounded}
The conditional density function of $\widetilde{Y}$ given $\widetilde{\mbf{X}}$ is uniformly bounded almost everywhere in $[0,\tau]$.
In addition, with probability one, $\widecheck{S}(\widetilde{Y}|\widetilde{\mbf{X}})$ is bounded away from 0 and 1.
\end{condition}
\begin{condition} \label{identifiability}
For any $\xi\in(0,\infty]$, the function $\xi^{-1}\mathbb{P}\ell(\mbf{\beta}, \Lambda)+\widetilde{\mathbb{P}}\psi(\mbf{\beta}, \Lambda)$ has a unique maximizer $(\mbf{\beta}^*, \Lambda^*)$ such that $\mbf{\beta}^*$ lies in the interior of $\mathcal{B}$ and $\Lambda^*$ is continuously differentiable with positive derivatives in $[0,\tau]$.
\end{condition}
\begin{remark}
Conditions \ref{true_parameter}--\ref{transformation} are standard for semiparametric transformation models with right-censored data \citep{zeng2007maximum}. 
Condition \ref{Scheck_bounded} is easily satisfied with the data generation methods discussed in Section~\ref{m-sub:POTL} of the main manuscript. 
Condition \ref{identifiability} pertains to parameter identifiability and is consistent with those found in the literature on semiparametric regression models (e.g., \citet{zeng2010general}). 
\end{remark}

\subsection{Optimal Tuning Parameter}
\label{sub:optimal_tuning_parameter}
The tuning parameter $\xi_n$ plays an important role in controlling the degree of knowledge transfer from the source studies to the target study.  
Let $\widehat{S}^{\text{tar}}(t|\mbf{X})$ denote the survival function estimator based on the target-only method.
Intuitively, $\xi_n$ should be set to a large value if the source predictor $\widecheck{S}(t|\mbf{X})$ is close to $\widehat{S}^{\text{tar}}(t|\mbf{X})$, and $\xi_n$ should be zero if there is a considerable discrepancy between the two. 
The following theorem gives an explicit choice of $\xi_n$ that ensures the optimal convergence rate of $\widehat{S}(t|\mbf{X})$.

\setcounter{theorem}{1}
\begin{theorem} \label{thm:opt_xi}
Let $\widetilde{\mathbb{P}}_m$ denote the empirical probability measure from $m$ independent observations of $(\widetilde{Y}, \widetilde{\mbf{X}})$
The optimal value of $\xi_n$ can be chosen as 
\[
  \xi_n^{\text{opt}} = \left\{\widetilde{\mathbb{P}}_m\|\widecheck{S}(t|\mbf{X})-\widehat{S}^{\text{tar}}(t|\mbf{X})\|_{L_2}^{2}\right\}^{-\nu/2} I\left\{\widetilde{\mathbb{P}}_m\|\widecheck{S}(t|\mbf{X})-\widehat{S}^{\text{tar}}(t|\mbf{X})\|_{L_2}^{2} +q_N \le cn^{-1}\right\},
\]
where $\nu$ and $c$ are positive constants satisfying $n^{-\nu}\le O(q_N)$.
Under Conditions~\ref{true_parameter}--\ref{identifiability}, $\xi_n^{\text{opt}}$ ensures that $\widehat{S}(t|\mbf{X})$ achieves the optimal convergence rate as stated in Corollary~\ref{m-coro:opt_conv_rate} of the main manuscript.
\end{theorem}

\section{Proofs of Theorems}
\label{sec:proof_thm}
We prove the theoretical results in~\ref{sec:theory}. The proofs make use of some lemmas, which are stated and proved in~\ref{sec:useful_lemmas}.

We first introduce the notation to be used throughout the proofs.  
Let $\|\cdot\|$ denote the Euclidean norm for vectors, $\|\cdot\|_{L_2}$ denote the $L_2$-norm with respect to the Lebesgue measure over $[0,\tau]$ for functions, and $\|\cdot\|_{\infty}$ denote the supremum norm over $[0,\tau]$ for functions.
For some sufficiently large constant $a>0$, let $\Omega_a = \mathcal{B}\times \mathcal{L}_a$ denote the parameter space of $(\mbf{\beta}, \Lambda)$, where $\mathcal{L}_a$ is the set of non-decreasing functions $\Lambda$ satisfying $\Lambda(0)=0$ and $\Lambda(\tau)\le a$.
For a generic function $f(\mbf{\beta}, \Lambda)$, let $\dot{f}_{\mbf{\beta}}$ denote the derivative of $f$ with respect to $\mbf{\beta}$ and $\dot{f}_{\Lambda}[h_{\Lambda}]$ denote the derivative of $f$ with respect to $\Lambda$ along the direction $h_{\Lambda}$. 
In addition, define the score statistic along the direction $(\mbf{h}_{\mbf{\beta}}, h_{\Lambda})$ as $\dot{f}(\mbf{\beta}, \Lambda)[\mbf{h}_{\mbf{\beta}}, h_{\Lambda}]=\dot{f}_{\mbf{\beta}}(\mbf{\beta}, \Lambda)^{\trans}\mbf{h}_{\mbf{\beta}}+\dot{f}_{\Lambda}(\mbf{\beta}, \Lambda)[h_{\Lambda}]$.

We also introduce the following empirical process notation to simplify the presentation.
Let $\mathbb{P}_n$ denote the empirical probability measure from $n$ independent observations of $(Y, \Delta, \mbf{X})$, and $\mathbb{P}$ denote the corresponding true probability measure.
For the penalty term, let $\widetilde{\mathbb{P}}_m$ denote the empirical probability measure from $m$ independent observations of $(\widetilde{Y}, \widetilde{\mbf{X}})$, and $\widetilde{\mathbb{P}}$ denote the corresponding true probability measure.

\subsection{Proof of Theorem 1}
\label{sub:proof_of_theorem_1}
Define the objective function as
\[
\mathbb{M}_{n,m}(\mbf{\beta}, \Lambda) = 
\begin{cases}
\mathbb{P}_n\ell(\mbf{\beta}, \Lambda)+\xi_n\widetilde{\mathbb{P}}_m \psi(\mbf{\beta}, \Lambda) & \text{ if } \xi<\infty, \\
\xi_n^{-1}\mathbb{P}_n\ell(\mbf{\beta}, \Lambda)+\widetilde{\mathbb{P}}_m \psi(\mbf{\beta}, \Lambda) & \text{ if } \xi=\infty;
\end{cases}
\]
and define the limit objective function as 
\[
\mathbb{M}(\mbf{\beta}, \Lambda) = 
\begin{cases}
\mathbb{P}\ell(\mbf{\beta}, \Lambda)+\xi\widetilde{\mathbb{P}} \psi(\mbf{\beta}, \Lambda) & \text{ if } \xi<\infty, \\
\widetilde{\mathbb{P}} \psi(\mbf{\beta}, \Lambda) & \text{ if } \xi=\infty.
\end{cases}
\] 
Here, $\ell(\mbf{\beta}, \Lambda)$ and $\psi(\mbf{\beta}, \Lambda)$ are the log-likelihood function and the penalty based on one single observation, respectively: 
\begin{gather*}
\ell(\mbf{\beta}, \Lambda) = \Delta\left[\log\Lambda^{\prime}(Y)+\mbf{\beta}^{\trans}\mbf{X}(Y)+\log G^{\prime}\left\{\int_0^Ye^{\mbf{\beta}^{\trans}\mbf{X}(s)}d\Lambda(s)\right\}\right]-G\left\{\int_0^Ye^{\mbf{\beta}^{\trans}\mbf{X}(s)}d\Lambda(s)\right\}, \\
\psi(\mbf{\beta}, \Lambda) = \widecheck{S}(\widetilde{Y}|\widetilde{\mbf{X}})\log S(\widetilde{Y}|\widetilde{\mbf{X}}) 
+ \left\{1-\widecheck{S}(\widetilde{Y}|\widetilde{\mbf{X}})\right\}\log\left\{1-S(\widetilde{Y}|\widetilde{\mbf{X}})\right\}.
\end{gather*}

Define $(\mbf{\beta}^*, \Lambda^*) = \argmax_{(\mbf{\beta}, \Lambda)\in\Omega_a} \mathbb{M}(\mbf{\beta}, \Lambda)$.
The outline of the proof is as follows: First, we show that $(\widehat{\mbf{\beta}}, \widehat{\Lambda})$ converge to $(\mbf{\beta}^*, \Lambda^*)$. 
Then, we study the upper bound for $E\|\widehat{S}(t|\mbf{X})-S^*(t|\mbf{X})\|_{L_2}^2$, where $S^*(t|\mbf{X}) = S(t|\mbf{X};\mbf{\beta}^*, \Lambda^*)$. 
Finally, we establish the convergence rate for $\widehat{S}(t|\mbf{X})$ by incorporating the distance between $S^*(t|\mbf{X})$ and $S_0(t|\mbf{X})$. 
It is necessary to address the cases with $\xi<\infty$ and $\xi=\infty$ separately.

\medskip
\begin{flushleft}
{\bf \underline{Case 1: $\xi<\infty$}}
\end{flushleft}

{\bf Step 1.} We first construct a step function $\overline{\Lambda}$ with jumps only at those $Y_i$ with $\Delta_i = 1$ ($i=1,\dots,n$), which converges uniformly to $\Lambda^*$ over $[0,\tau]$.
By differentiating the objective function $\mathbb{M}_{n,m}(\mbf{\beta}, \Lambda)$ with respect to $\Lambda\{Y_i\}$ and setting it to be 0, we can see that $\widehat{\Lambda}\{Y_i\}$ satisfies 
\[
\frac{\Delta_i}{\widehat{\Lambda}\{Y_i\}} = n\mathbb{P}_n R_1(y;\widehat{\mbf{\beta}}, \widehat{\Lambda})
+\xi_n n\widetilde{\mathbb{P}}_m R_2(y;\widehat{\mbf{\beta}}, \widehat{\Lambda})|_{y=Y_i},
\]
where 
\begin{align*}
R_1(t;\mbf{\beta}, \Lambda) ={} & I(Y\ge t)e^{\mbf{\beta}^{\trans}\mbf{X}(t)}\left[-\frac{\Delta G^{\prime\prime}\{\int_0^Y e^{\mbf{\beta}^{\trans}\mbf{X}(s)}d\Lambda(s)\}}{G^{\prime}\{\int_0^Y e^{\mbf{\beta}^{\trans}\mbf{X}(s)}d\Lambda(s)\}}
+G^{\prime}\{\int_0^Y e^{\mbf{\beta}^{\trans}\mbf{X}(s)}d\Lambda(s)\}\right], \\
R_2(t;\mbf{\beta}, \Lambda)={} & 
\begin{multlined}[t]
I(\widetilde{Y}\ge t)e^{\mbf{\beta}^{\trans}\widetilde{\mbf{X}}(t)}\Biggl[\widecheck{S}(\widetilde{Y}|\widetilde{\mbf{X}})G^{\prime}\{\int_0^{\widetilde{Y}} e^{\mbf{\beta}^{\trans}\widetilde{\mbf{X}}(s)}d\Lambda(s)\} \\
-\{1-\widecheck{S}(\widetilde{Y}|\widetilde{\mbf{X}})\}\frac{G^{\prime}\{\int_0^{\widetilde{Y}} e^{\mbf{\beta}^{\trans}\widetilde{\mbf{X}}(s)}d\Lambda(s)\}}{\exp[G^{\prime}\{\int_0^{\widetilde{Y}} e^{\mbf{\beta}^{\trans}\widetilde{\mbf{X}}(s)}d\Lambda(s)\}]-1}\Biggr].
\end{multlined}
\end{align*}
In light of this, we let the jump size of $\overline{\Lambda}$ satisfy
\[
\frac{\Delta_i}{\overline{\Lambda}\{Y_i\}} = n\mathbb{P}_n R_1(y;\mbf{\beta}^*, \Lambda^*)
+\xi_n n\widetilde{\mathbb{P}}_m R_2(y;\mbf{\beta}^*, \Lambda^*)|_{y=Y_i},
\]
and construct $\overline{\Lambda}$ by $\overline{\Lambda}(t) = \sum_{i=1}^n I(Y_i\le t)\overline{\Lambda}\{Y_i\}$.
To verify the uniform convergence of $\overline{\Lambda}$, we observe that 
\begin{equation} \label{convergence_bar_Lambda}
\begin{aligned}
& \left\|\overline{\Lambda}-\mathbb{P}\left\{\frac{\Delta I(Y\le \cdot)}{\mathbb{P} R_1(y;\mbf{\beta}^*, \Lambda^*)
+\xi \widetilde{\mathbb{P}} R_2(y;\mbf{\beta}^*, \Lambda^*)|_{y=Y}}\right\}\right\|_{\infty} \\
={} & \left\|\mathbb{P}_n\left\{\frac{\Delta I(Y\le \cdot)}{\mathbb{P}_n R_1(y;\mbf{\beta}^*, \Lambda^*)
+\xi_n \widetilde{\mathbb{P}}_m R_2(y;\mbf{\beta}^*, \Lambda^*)|_{y=Y}}\right\}-
\mathbb{P}\left\{\frac{\Delta I(Y\le \cdot)}{\mathbb{P} R_1(y;\mbf{\beta}^*, \Lambda^*)
+\xi \widetilde{\mathbb{P}} R_2(y;\mbf{\beta}^*, \Lambda^*)|_{y=Y}}\right\}\right\|_{\infty} \\
\le{} & 
\begin{multlined}[t]
\left\|\mathbb{P}_n\left\{\frac{\Delta I(Y\le \cdot)}{\mathbb{P}_n R_1(y;\mbf{\beta}^*, \Lambda^*)
+\xi_n \widetilde{\mathbb{P}}_m R_2(y;\mbf{\beta}^*, \Lambda^*)|_{y=Y}}-
\frac{\Delta I(Y\le \cdot)}{\mathbb{P} R_1(y;\mbf{\beta}^*, \Lambda^*)
+\xi \widetilde{\mathbb{P}} R_2(y;\mbf{\beta}^*, \Lambda^*)|_{y=Y}}\right\}\right\|_{\infty} \\
+ \left\|(\mathbb{P}_n-\mathbb{P})\left\{\frac{\Delta I(Y\le \cdot)}{\mathbb{P} R_1(y;\mbf{\beta}^*, \Lambda^*)
+\xi \widetilde{\mathbb{P}} R_2(y;\mbf{\beta}^*, \Lambda^*)|_{y=Y}}\right\}\right\|_{\infty} 
\end{multlined} \\
\le{} & 
\begin{multlined}[t]
\left\|\frac{1}{\mathbb{P}_n R_1(\cdot;\mbf{\beta}^*, \Lambda^*)
+\xi_n \widetilde{\mathbb{P}}_m R_2(\cdot;\mbf{\beta}^*, \Lambda^*)}-\frac{1}{\mathbb{P} R_1(\cdot;\mbf{\beta}^*, \Lambda^*)
+\xi \widetilde{\mathbb{P}} R_2(\cdot;\mbf{\beta}^*, \Lambda^*)}\right\|_{\infty} \\
+ \left\|(\mathbb{P}_n-\mathbb{P})\left\{\frac{\Delta I(Y\le \cdot)}{\mathbb{P} R_1(y;\mbf{\beta}^*, \Lambda^*)
+\xi \widetilde{\mathbb{P}} R_2(y;\mbf{\beta}^*, \Lambda^*)|_{y=Y}}\right\}\right\|_{\infty}. 
\end{multlined}
\end{aligned}
\end{equation}
We show in Lemma~\ref{donsker} that 
\[
\mathcal{F}_{R_i} = \left\{R_{i}(t;\mbf{\beta}, \Lambda): \, t\in[0,\tau], (\mbf{\beta}, \Lambda)\in \Omega_a\right\}
\]
are Donsker classes, for $i=1,2$.
The Donsker property and Slutsky's theorem guarantee that the first term on the right-hand side of \eqref{convergence_bar_Lambda} converges to 0.
Likewise, since the function class
\[
\left\{\frac{\Delta I(Y\le t)}{\mathbb{P} R_1(y;\mbf{\beta}^*, \Lambda^*)
+\xi \widetilde{\mathbb{P}} R_2(y;\mbf{\beta}^*, \Lambda^*)|_{y=Y}}:\,t\in[0,\tau]\right\}
\]
is a Glivenko-Cantelli class, the second term of \eqref{convergence_bar_Lambda} also converges to 0. 
Thus, the uniform convergence of $\overline{\Lambda}$ over $[0,\tau]$ follows. 
It remains to show that the limit of $\overline{\Lambda}$ is exactly $\Lambda^*$. 
By the definition of $(\mbf{\beta}^*, \Lambda^*)$, the derivative of $\mathbb{M}(\mbf{\beta}^*, \Lambda)$ with respect
to $\Lambda$ along the submodel $\Lambda^*+\epsilon \int hd\Lambda^*$ is zero, for any $h\in L_2([0,\tau])$.
Thus, simple calculation yields
\begin{align*}
0={} & \mathbb{P}\dot{\ell}_{\Lambda}(\mbf{\beta}^*, \Lambda^*)[\int hd\Lambda^*]+\xi\widetilde{\mathbb{P}}\dot{\psi}_{\Lambda}(\mbf{\beta}^*, \Lambda^*)[\int hd\Lambda^*] \\
={} & \mathbb{P}\left\{\Delta h(Y) - \int_0^{\tau}R_1(t;\mbf{\beta}^*, \Lambda^*)h(t)d\Lambda^*(t)\right\} 
-\xi\widetilde{\mathbb{P}}\int_0^{\tau}R_2(t;\mbf{\beta}^*, \Lambda^*)h(t)d\Lambda^*(t) \\
={} & \int_0^{\tau}\left[\mathbb{P}\Delta \delta(t-Y) - \bigl\{\mathbb{P}R_1(t;\mbf{\beta}^*, \Lambda^*)+\xi\widetilde{\mathbb{P}}R_2(t;\mbf{\beta}^*, \Lambda^*)\bigr\}\lambda^*(t)\right]h(t)dt,
\end{align*} 
where $\delta(\cdot)$ is the Dirac delta function and $\lambda^*(t)$ is the derivative of $\Lambda^*(t)$. 
Since the above equation holds for any $h\in L_2([0,\tau])$, we have 
\begin{equation} \label{lambda_0}
\lambda^*(t) = \mathbb{P}\left\{\frac{\Delta \delta(t-Y)}{\mathbb{P}R_1(t;\mbf{\beta}^*, \Lambda^*)+\xi\widetilde{\mathbb{P}}R_2(t;\mbf{\beta}^*, \Lambda^*)}\right\}, \quad \text{ for } t\in[0,\tau].
\end{equation}
This implies that
\begin{align*}
\Lambda^*(t) ={} & \mathbb{P}\left\{\int_0^{\tau}I(s\le t)\frac{\Delta \delta(s-Y)}{\mathbb{P}R_1(s;\mbf{\beta}^*, \Lambda^*)+\xi\widetilde{\mathbb{P}}R_2(s;\mbf{\beta}^*, \Lambda^*)}ds\right\} \\
={} & \mathbb{P}\left\{\frac{\Delta I(Y\le t)}{\mathbb{P} R_1(s;\mbf{\beta}^*, \Lambda^*)
+\xi \widetilde{\mathbb{P}} R_2(s;\mbf{\beta}^*, \Lambda^*)|_{s=Y}}\right\}.
\end{align*}
The right-hand side is exactly the limit of $\overline{\Lambda}$ as shown in \eqref{convergence_bar_Lambda}.
Thus, we conclude that $\overline{\Lambda}$ converges uniformly to $\Lambda^*$ over $[0,\tau]$.

{\bf Step 2.} We show that $\limsup_n \widehat{\Lambda}(\tau)<\infty$ almost surely. By the definition of $(\widehat{\mbf{\beta}}, \widehat{\Lambda})$, we have $\mathbb{M}_{n,m}(\widehat{\mbf{\beta}}, \widehat{\Lambda})\ge \mathbb{M}_{n,m}(\mbf{\beta}^*, \overline{\Lambda})$. 
Thus, 
\[
\mathbb{P}_n\ell(\widehat{\mbf{\beta}}, \widehat{\Lambda})+\xi_n\widetilde{\mathbb{P}}_m \psi(\widehat{\mbf{\beta}}, \widehat{\Lambda})\ge \mathbb{P}_n\ell(\mbf{\beta}^*, \overline{\Lambda})+\xi_n\widetilde{\mathbb{P}}_m \psi(\mbf{\beta}^*, \overline{\Lambda}).
\]
By the construction of $\overline{\Lambda}$, we add $\mathbb{P}_n\Delta\log n$ to both sides and take the difference to obtain
\begin{equation} \label{bound_Lambda_1}
0\le
\begin{multlined}[t]
n^{-1}\sum_{i=1}^n \Biggl(\Delta_i\left[\log [n\widehat{\Lambda}\{Y_i\}]+\widehat{\mbf{\beta}}^{\trans}\mbf{X}_i(Y_i)+\log G^{\prime}\{\int_0^{Y_i} e^{\widehat{\mbf{\beta}}^{\trans}\mbf{X}_i(s)}d\widehat{\Lambda}(s)\}\right]\Biggr. \\
\Biggl.-G\{\int_0^{Y_i} e^{\widehat{\mbf{\beta}}^{\trans}\mbf{X}_i(s)}d\widehat{\Lambda}(s)\}\Biggr)
+\xi_n\widetilde{\mathbb{P}}_m \psi(\widehat{\mbf{\beta}}, \widehat{\Lambda})+O(1).
\end{multlined}
\end{equation}
By Condition~\ref{X_cont_diff}, there exists a positive constant $b$ such that $|\mbf{\beta}^{\trans}\widetilde{\mbf{X}}(t)|\le b$ for any $\mbf{\beta}\in\mathcal{B}$ and $t\in[0,\tau]$.
Thus, 
\begin{align*}
\xi_n\widetilde{\mathbb{P}}_m \psi(\widehat{\mbf{\beta}}, \widehat{\Lambda})\le{} & -\xi_n\widetilde{\mathbb{P}}_m \widecheck{S}(\widetilde{Y}|\widetilde{\mbf{X}})G\{\int_0^{\widetilde{Y}} e^{\widehat{\mbf{\beta}}^{\trans}\widetilde{\mbf{X}}(s)}d\widehat{\Lambda}(s)\} \\
\le{} & -\xi_n \widetilde{\mathbb{P}}_m \widecheck{S}(\widetilde{Y}|\widetilde{\mbf{X}})G\{e^{-b}\widehat{\Lambda}(\widetilde{Y})\} \\
\le{} & -\xi_n \widetilde{\mathbb{P}}_m \left\{I(\widetilde{Y}=\tau)\widecheck{S}(\tau|\widetilde{\mbf{X}})\right\}G\{e^{-b}\widehat{\Lambda}(\tau)\}.
\end{align*}
When $m$ is sufficiently large, $\widetilde{\mathbb{P}}_m \{I(\widetilde{Y}=\tau)\widecheck{S}(\tau|\widetilde{\mbf{X}})\}=C_1\ge0$ for some constant $C_1$. 
Then by Condition~\ref{transformation}, we can further write \eqref{bound_Lambda_1} as
\begin{equation} \label{bound_Lambda_2}
\begin{aligned}
0\le{} & 
\begin{multlined}[t]
n^{-1}\sum_{i=1}^n \Biggl(\Delta_i\left[\log [n\widehat{\Lambda}\{Y_i\}]+\widehat{\mbf{\beta}}^{\trans}\mbf{X}_i(Y_i)-(1+\rho)\log \{1+\int_0^{Y_i} e^{\widehat{\mbf{\beta}}^{\trans}\mbf{X}_i(s)}d\widehat{\Lambda}(s)\}\right]\Biggr. \\
\Biggl.-(1-\Delta_i)G\{\int_0^{Y_i} e^{\widehat{\mbf{\beta}}^{\trans}\mbf{X}_i(s)}d\widehat{\Lambda}(s)\}\Biggr) 
-\xi_n C_1G\{e^{-b}\widehat{\Lambda}(\tau)\}+O(1) 
\end{multlined} \\
\le{} & 
\begin{multlined}[t]
n^{-1}\sum_{i=1}^n \left[\Delta_i\left\{\log [n\widehat{\Lambda}\{Y_i\}]+\widehat{\mbf{\beta}}^{\trans}\mbf{X}_i(Y_i)\right\}-(\Delta_i+\rho)\log \{1+\int_0^{Y_i} e^{\widehat{\mbf{\beta}}^{\trans}\mbf{X}_i(s)}d\widehat{\Lambda}(s)\}\right] \\
-\xi_n C_1\rho \log\{1+e^{-b}\widehat{\Lambda}(\tau)\}+O(1)
\end{multlined} \\
\le{} & n^{-1}\sum_{i=1}^n \left[\Delta_i\log [n\widehat{\Lambda}\{Y_i\}]-(\Delta_i+\rho)\log \{1+\widehat{\Lambda}(Y_i)\}\right]-\xi_n C_1\rho \log\{1+\widehat{\Lambda}(\tau)\}+O(1).
\end{aligned}
\end{equation}

We mimic the partition arguments of \citet{murphy1994consistency} to show that the right-hand side of \eqref{bound_Lambda_2} will diverge to $-\infty$ if $\limsup_n \widehat{\Lambda}(\tau)=\infty$.
Consider a partition of $[0,\tau]$ denoted by $\tau=s_0>s_1>\cdots>s_R=0$. Let $A_r = [s_{r}, s_{r-1})$, for $r=1,\dots,R$.
Then \eqref{bound_Lambda_2} can be rewritten as
\begin{equation} \label{bound_Lambda_3}
\begin{aligned}
O(1)\le{} & 
\begin{multlined}[t]
n^{-1}\sum_{i=1}^n \Delta_iI\{Y_i\in[s_1, s_0]\}\log [n\widehat{\Lambda}\{Y_i\}] \\
+ \sum_{r=2}^R n^{-1}\sum_{i=1}^n \Delta_iI\{Y_i\in[s_{r}, s_{r-1})\}\log [n\widehat{\Lambda}\{Y_i\}] \\
-n^{-1}\sum_{i=1}^n(\Delta_i+\rho)I(Y_i=\tau)\log \{1+\widehat{\Lambda}(\tau)\} \\
-\sum_{r=1}^Rn^{-1}\sum_{i=1}^n(\Delta_i+\rho)I\{Y_i\in[s_r, s_{r-1})\}\log \{1+\widehat{\Lambda}(Y_i)\} \\
-\xi_n C_1\rho \log\{1+\widehat{\Lambda}(\tau)\}
\end{multlined}
\end{aligned}
\end{equation}
For $r=1,\dots,R$, by the concavity of the log function, 
\begin{align*}
& n^{-1}\sum_{i=1}^n \Delta_iI\{Y_i\in[s_{r}, s_{r-1})\}\log [n\widehat{\Lambda}\{Y_i\}] \\
\le{} & n^{-1}\sum_{i=1}^n \Delta_iI\{Y_i\in[s_{r}, s_{r-1})\}\log\frac{\sum_{i=1}^n \Delta_iI\{Y_i\in[s_{r}, s_{r-1})\}\widehat{\Lambda}\{Y_i\}}{n^{-1}\sum_{i=1}^n \Delta_iI\{Y_i\in[s_{r}, s_{r-1})\}} \\
\le{} & n^{-1}\sum_{i=1}^n \Delta_iI\{Y_i\in[s_{r}, s_{r-1})\}\log\frac{\widehat{\Lambda}(Y_i)}{n^{-1}\sum_{i=1}^n \Delta_iI\{Y_i\in[s_{r}, s_{r-1})\}} \\
\le{} & 
\begin{multlined}[t]
n^{-1}\sum_{i=1}^n \Delta_iI\{Y_i\in[s_{r}, s_{r-1})\}\log\{1+\widehat{\Lambda}(s_{r-1})\} \\
-\mathbb{P}_n\Delta I\{Y\in[s_{r}, s_{r-1})\}\log[\mathbb{P}_n\Delta I\{Y\in[s_{r}, s_{r-1})\}]. 
\end{multlined}
\end{align*}
Under Condition~\ref{prob_Ctau}, $\mathbb{P}_n\Delta I\{Y\in[s_{r}, s_{r-1})\}\rightarrow \pr\{Y\in[s_{r}, s_{r-1}), \Delta = 1\}>0$ as $n\rightarrow\infty$.
Thus, the second term on the right-hand side of the above inequality is $O(1)$ for large $n$.
We apply this result to \eqref{bound_Lambda_3} and obtain
\begin{align*}
O(1) \le{} & \begin{multlined}[t]
n^{-1}\sum_{i=1}^n \Delta_iI\{Y_i\in[s_{1}, s_{0}]\}\log\{1+\widehat{\Lambda}(\tau)\} \\
+\sum_{r=2}^R n^{-1}\sum_{i=1}^n \Delta_iI\{Y_i\in[s_{r}, s_{r-1})\}\log\{1+\widehat{\Lambda}(s_{r-1})\} \\
-n^{-1}\sum_{i=1}^n(\Delta_i+\rho)I(Y_i=\tau)\log \{1+\widehat{\Lambda}(\tau)\} \\
-\sum_{r=1}^Rn^{-1}\sum_{i=1}^n(\Delta_i+\rho)I\{Y_i\in[s_r, s_{r-1})\}\log \{1+\widehat{\Lambda}(s_r)\} \\
-\xi_n C_1\rho \log\{1+\widehat{\Lambda}(\tau)\}
\end{multlined} \\
\le{} & 
\begin{multlined}[t]
-(2n)^{-1}\sum_{i=1}^n(\Delta_i+\rho)I(Y_i=\tau)\log \{1+\widehat{\Lambda}(\tau)\} -\xi_n C_1\rho \log\{1+\widehat{\Lambda}(\tau)\} \\
-\Biggl\{(2n)^{-1}\sum_{i=1}^n(\Delta_i+\rho)I(Y_i=\tau)\log \{1+\widehat{\Lambda}(\tau)\}\Biggr. \\
-\Biggl.n^{-1}\sum_{i=1}^n \Delta_iI\{Y_i\in[s_{1}, s_{0}]\}\log\{1+\widehat{\Lambda}(\tau)\}\Biggr\} \\
-\sum_{r=1}^{R-1}\Biggl\{n^{-1}\sum_{i=1}^n(\Delta_i+\rho)I\{Y_i\in[s_r, s_{r-1})\}\log \{1+\widehat{\Lambda}(s_r)\}\Biggr. \\
\Biggl.-n^{-1}\sum_{i=1}^n \Delta_iI\{Y_i\in[s_{r+1}, s_{r})\}\log\{1+\widehat{\Lambda}(s_{r})\}\Biggr\}.
\end{multlined}
\end{align*}
We can choose $s_0>s_1>\cdots>s_R$ such that the first line on the right-hand side diverges to $-\infty$ as $\widehat{\Lambda}(\tau)$ diverges to $\infty$; and in the remaining terms, the coefficients in front of $\log\{1+\widehat{\Lambda}(s_{r})\}$ are all negative for large $n$. This contradicts the fact that the right-hand side should be bounded. 
Thus, we conclude that $\limsup_n \widehat{\Lambda}(\tau)<\infty$ almost surely.

{\bf Step 3.} We show that $\|\widehat{\mbf{\beta}}-\mbf{\beta}^*\|+ \|\widehat{\Lambda}-\Lambda^*\|_{\infty}\overset{a.s.}{\rightarrow}0$.
We have shown in Step 2 that $\widehat{\Lambda}$ has bounded total variation. By Helly's selection lemma, for any subsequence of $(\widehat{\mbf{\beta}}, \widehat{\Lambda})$, we can choose a further subsequence such that $\widehat{\mbf{\beta}}\rightarrow \mathring{\mbf{\beta}}$ and $\widehat{\Lambda}$ converges pointwise to some increasing function $\mathring{\Lambda}$ in $[0,\tau]$.
It suffices to show that $(\mathring{\mbf{\beta}}, \mathring{\Lambda}) = (\mbf{\beta}^*, \Lambda^*)$.

We first verify the differentiability of $\mathring{\Lambda}$. The construction of $\overline{\Lambda}$ implies that
\begin{equation} \label{relation_hat_tilde_Lambda}
\widehat{\Lambda}(t) = \int_0^t\frac{|\mathbb{P}_n R_1(s;\mbf{\beta}^*, \Lambda^*)
+\xi_n \widetilde{\mathbb{P}}_m R_2(s;\mbf{\beta}^*, \Lambda^*)|}{|\mathbb{P}_n R_1(s;\widehat{\mbf{\beta}}, \widehat{\Lambda})
+\xi_n \widetilde{\mathbb{P}}_m R_2(s;\widehat{\mbf{\beta}}, \widehat{\Lambda})|}d\overline{\Lambda}(s).
\end{equation}
It follows from the Donsker property proved in Lemma~\ref{donsker} and the dominated convergence theorem that 
\begin{gather*}
\sup_{t\in[0,\tau]}\left|\{\mathbb{P}_n R_1(s;\mbf{\beta}^*, \Lambda^*)
+\xi_n \widetilde{\mathbb{P}}_m R_2(s;\mbf{\beta}^*, \Lambda^*)\}-\{\mathbb{P} R_1(s;\mbf{\beta}^*, \Lambda^*)
+\xi \widetilde{\mathbb{P}} R_2(s;\mbf{\beta}^*, \Lambda^*)\}\right| \overset{a.s.}{\rightarrow}0, \\
\sup_{t\in[0,\tau]}\left|\{\mathbb{P}_n R_1(s;\widehat{\mbf{\beta}}, \widehat{\Lambda})
+\xi_n \widetilde{\mathbb{P}}_m R_2(s;\widehat{\mbf{\beta}}, \widehat{\Lambda})\}-\{\mathbb{P} R_1(s;\mathring{\mbf{\beta}}, \mathring{\Lambda})
+\xi \widetilde{\mathbb{P}} R_2(s;\mathring{\mbf{\beta}}, \mathring{\Lambda})\}\right| \overset{a.s.}{\rightarrow}0.
\end{gather*}
We claim that $|\mathbb{P} R_1(t;\mathring{\mbf{\beta}}, \mathring{\Lambda})
+\xi \widetilde{\mathbb{P}} R_2(t;\mathring{\mbf{\beta}}, \mathring{\Lambda})|$ is uniformly bounded away from 0 over $[0,\tau]$. 
Then we can take limits on both sides of \eqref{relation_hat_tilde_Lambda} to obtain
\[
\mathring{\Lambda}(t) = \int_0^t\frac{|\mathbb{P} R_1(s;\mbf{\beta}^*, \Lambda^*)
+\xi \widetilde{\mathbb{P}} R_2(s;\mbf{\beta}^*, \Lambda^*)|}{|\mathbb{P} R_1(s;\mathring{\mbf{\beta}}, \mathring{\Lambda})
+\xi \widetilde{\mathbb{P}} R_2(s;\mathring{\mbf{\beta}}, \mathring{\Lambda})|}d\Lambda^*(s).
\]
We conclude that $\mathring{\Lambda}$ is absolutely continuous with respect to $\Lambda^*$, thus is differentiable with respect to $t$. 
Denote the derivative of $\mathring{\Lambda}$ by $\mathring{\lambda}$.
We also have $\widehat{\Lambda}\{t\}/\overline{\Lambda}\{t\}$ uniformly converges to $\mathring{\lambda}(t)/\lambda^*(t)$ in $[0,\tau]$.

It remains to verify that $\inf_{t\in[0,\tau]}|\mathbb{P} R_1(t;\mathring{\mbf{\beta}}, \mathring{\Lambda})+\xi \widetilde{\mathbb{P}} R_2(t;\mathring{\mbf{\beta}}, \mathring{\Lambda})|>0$.
If this is not true, then there exists some $\mathring{t}\in[0,\tau]$ such that $\mathbb{P} R_1(\mathring{t};\mathring{\mbf{\beta}}, \mathring{\Lambda})
+\xi \widetilde{\mathbb{P}} R_2(\mathring{t};\mathring{\mbf{\beta}}, \mathring{\Lambda}) = 0$.  
Under Conditions~\ref{X_cont_diff}--\ref{transformation}, $\mathbb{P} R_1(t;\mathring{\mbf{\beta}}, \mathring{\Lambda})
+\xi \widetilde{\mathbb{P}} R_2(t;\mathring{\mbf{\beta}}, \mathring{\Lambda})$ is differentiable almost everywhere. 
Thus, there exist some $\delta t>0$ and $C_2>0$ such that for $t\in (\mathring{t}-\delta t, \mathring{t}+\delta t)$,
\begin{align*}
& |\mathbb{P} R_1(t;\mathring{\mbf{\beta}}, \mathring{\Lambda})+\xi \widetilde{\mathbb{P}} R_2(t;\mathring{\mbf{\beta}}, \mathring{\Lambda})| \\
={} & |\{\mathbb{P} R_1(t;\mathring{\mbf{\beta}}, \mathring{\Lambda})+\xi \widetilde{\mathbb{P}} R_2(t;\mathring{\mbf{\beta}}, \mathring{\Lambda})\}-\{\mathbb{P} R_1(\mathring{t};\mathring{\mbf{\beta}}, \mathring{\Lambda})+\xi \widetilde{\mathbb{P}} R_2(\mathring{t};\mathring{\mbf{\beta}}, \mathring{\Lambda})\}| \\
\le{} & C_2|t-\mathring{t}|.
\end{align*}
On the other hand, it follows from \eqref{relation_hat_tilde_Lambda} that for any $\epsilon>0$, 
\[
\widehat{\Lambda}(\tau) = \int_0^{\tau}\frac{|\mathbb{P}_n R_1(s;\mbf{\beta}^*, \Lambda^*)
+\xi_n \widetilde{\mathbb{P}}_m R_2(s;\mbf{\beta}^*, \Lambda^*)|}{|\mathbb{P}_n R_1(s;\widehat{\mbf{\beta}}, \widehat{\Lambda})
+\xi_n \widetilde{\mathbb{P}}_m R_2(s;\widehat{\mbf{\beta}}, \widehat{\Lambda})|+\epsilon}d\overline{\Lambda}(s).
\]
Taking limits on both sides of the above equation yields 
\[
O(1)=\limsup_n \widehat{\Lambda}(\tau) \ge \int_0^{\tau}\frac{|\mathbb{P} R_1(s;\mbf{\beta}^*, \Lambda^*)
+\xi \widetilde{\mathbb{P}} R_2(s;\mbf{\beta}^*, \Lambda^*)|}{|\mathbb{P} R_1(s;\mathring{\mbf{\beta}}, \mathring{\Lambda})
+\xi \widetilde{\mathbb{P}} R_2(s;\mathring{\mbf{\beta}}, \mathring{\Lambda})|+\epsilon}d\Lambda^*(s).
\]
We let $\epsilon\rightarrow0$ and apply the monotone convergence theorem to obtain
\begin{align*}
O(1)\ge{} & \int_0^{\tau}\frac{|\mathbb{P} R_1(s;\mbf{\beta}^*, \Lambda^*)
+\xi \widetilde{\mathbb{P}} R_2(s;\mbf{\beta}^*, \Lambda^*)|}{|\mathbb{P} R_1(s;\mathring{\mbf{\beta}}, \mathring{\Lambda})
+\xi \widetilde{\mathbb{P}} R_2(s;\mathring{\mbf{\beta}}, \mathring{\Lambda})|}d\Lambda^*(s) \\
\ge{} & 
\begin{multlined}[t]
\int_{\mathring{t}}^{\mathring{t}+\delta t}\frac{|\mathbb{P} R_1(s;\mbf{\beta}^*, \Lambda^*)
+\xi \widetilde{\mathbb{P}} R_2(s;\mbf{\beta}^*, \Lambda^*)|}{C_2|s-\mathring{t}|}d\Lambda^*(s) \\
+\int_{\mathring{t}-\delta t}^{\mathring{t}}\frac{|\mathbb{P} R_1(s;\mbf{\beta}^*, \Lambda^*)
+\xi \widetilde{\mathbb{P}} R_2(s;\mbf{\beta}^*, \Lambda^*)|}{C_2|s-\mathring{t}|}d\Lambda^*(s).
\end{multlined}
\end{align*}
This is a contradiction since the right-hand side of the above inequality is infinite. 
Thus, $|\mathbb{P} R_1(t;\mathring{\mbf{\beta}}, \mathring{\Lambda})+\xi \widetilde{\mathbb{P}} R_2(t;\mathring{\mbf{\beta}}, \mathring{\Lambda})|$ is uniformly bounded away from 0. 
Therefore, we complete the proof for the differentiability of $\mathring{\Lambda}$.

To prove the consistency of $(\widehat{\mbf{\beta}}, \widehat{\Lambda})$, we first define the function
\[
\Phi(\mbf{\beta}, \Lambda) = \left[e^{\mbf{\beta}^{\trans}\mbf{X}(Y)}G^{\prime}\left\{\int_0^Ye^{\mbf{\beta}^{\trans}\mbf{X}(s)}d\Lambda(s)\right\}\right]^{\Delta}\exp\left[-G\left\{\int_0^Ye^{\mbf{\beta}^{\trans}\mbf{X}(s)}d\Lambda(s)\right\}\right],
\]
such that the likelihood function $L(\mbf{\beta}, \Lambda) = \Lambda^{\prime}(Y)^{\Delta}\Phi(\mbf{\beta}, \Lambda)$.
Note that 
\begin{equation} \label{diff_Qhat_Q0}
\begin{aligned}
0\le{} & \mathbb{M}_{n,m}(\widehat{\mbf{\beta}}, \widehat{\Lambda})- \mathbb{M}_{n,m}(\mbf{\beta}^*, \overline{\Lambda}) \\
={} & \mathbb{P}_n\left\{\Delta\log\frac{\widehat{\Lambda}\{Y\}}{\overline{\Lambda}\{Y\}}+\log\frac{\Phi(\widehat{\mbf{\beta}}, \widehat{\Lambda})}{\Phi(\mbf{\beta}^*, \overline{\Lambda})}\right\}+\xi_n\widetilde{\mathbb{P}}_m\left\{\psi(\widehat{\mbf{\beta}}, \widehat{\Lambda})-\psi(\mbf{\beta}^*, \overline{\Lambda})\right\}.
\end{aligned}
\end{equation}
By the uniform convergence of $\widehat{\Lambda}\{t\}/\overline{\Lambda}\{t\}$ over $[0,\tau]$ and the strong law of large numbers, we have 
\[
\mathbb{P}_n\Delta\log[\widehat{\Lambda}\{Y\}/\overline{\Lambda}\{Y\}]\overset{a.s.}{\rightarrow} \mathbb{P}\Delta\log\{\mathring{\lambda}(Y)/\lambda^*(Y)\}.
\]
In addition, by the Donsker properties of the relevant function classes proved in Lemma~\ref{donsker} and the dominated convergence theorem, we have 
\begin{gather*}
\mathbb{P}_n\log\{\Phi(\widehat{\mbf{\beta}}, \widehat{\Lambda})/\Phi(\mbf{\beta}^*, \overline{\Lambda})\}\overset{a.s.}{\rightarrow} \mathbb{P}\log\{\Phi(\mathring{\mbf{\beta}}, \mathring{\Lambda})/\Phi(\mbf{\beta}^*, \Lambda^*)\}, \\
\xi_n\widetilde{\mathbb{P}}_m\{\psi(\widehat{\mbf{\beta}}, \widehat{\Lambda})-\psi(\mbf{\beta}^*, \overline{\Lambda})\}\overset{a.s.}{\rightarrow} \xi\widetilde{\mathbb{P}}\{\psi(\mathring{\mbf{\beta}}, \mathring{\Lambda})-\psi(\mbf{\beta}^*, \Lambda^*)\}.
\end{gather*}
Based on the above arguments, we take the limit on the right-hand side of \eqref{diff_Qhat_Q0} to obtain
\begin{align*}
0\le{} & \mathbb{P}\log\frac{\mathring{\lambda}(Y)^{\Delta}\Phi(\mathring{\mbf{\beta}}, \mathring{\Lambda})}{\lambda^*(Y)^{\Delta}\Phi(\mbf{\beta}^*, \Lambda^*)}+\xi\widetilde{\mathbb{P}}\left\{\psi(\mathring{\mbf{\beta}}, \mathring{\Lambda})-\psi(\mbf{\beta}^*, \Lambda^*)\right\} \\
={} & \mathbb{M}(\mathring{\mbf{\beta}}, \mathring{\Lambda})-\mathbb{M}(\mbf{\beta}^*, \Lambda^*).
\end{align*}
By the definition, $(\mbf{\beta}^*, \Lambda^*)$ maximizes $\mathbb{M}(\mbf{\beta}, \Lambda)$.
If $\xi>0$, we have $\mathbb{M}(\mathring{\mbf{\beta}}, \mathring{\Lambda})=\mathbb{M}(\mbf{\beta}^*, \Lambda^*)$, which further implies $(\mathring{\mbf{\beta}}, \mathring{\Lambda}) = (\mbf{\beta}^*, \Lambda^*)$ by Condition~\ref{identifiability}.
If $\xi=0$, we have $\mathbb{P}\ell(\mathring{\mbf{\beta}}, \mathring{\Lambda}) = \mathbb{P}\ell(\mbf{\beta}^*, \Lambda^*) = \mathbb{P}\ell(\mbf{\beta}_0, \Lambda_0)$.
By the property of the Kullback-Leibler divergence, $L(\mathring{\mbf{\beta}}, \mathring{\Lambda}) = L(\mbf{\beta}^*, \Lambda^*)=L(\mbf{\beta}_0, \Lambda_0)$ with probability one. 
We let $(Y, \Delta) = (t, 0)$ to obtain that 
\[
\exp\left[-G\{\int_0^t e^{\mathring{\mbf{\beta}}^{\trans}\mbf{X}(s)}d\mathring{\Lambda}(s)\}\right] = \exp\left[-G\{\int_0^t e^{{\mbf{\beta}^*}^{\trans}\mbf{X}(s)}d\Lambda^*(s)\}\right].
\] 
By Condition~\ref{transformation}, the above equality implies that $\int_0^t e^{\mathring{\mbf{\beta}}^{\trans}\mbf{X}(s)}d\mathring{\Lambda}(s) = \int_0^t e^{{\mbf{\beta}^*}^{\trans}\mbf{X}(s)}d\Lambda^*(s)$.
Differentiation with respect to $t$ and then taking the log transform on both sides yield
\[
(\mathring{\mbf{\beta}}-\mbf{\beta}^*)^{\trans}\mbf{X}(t)+\log\frac{d\mathring{\Lambda}(t)}{d\Lambda^*(t)} = 0, \quad \text{ for any } t\in[0,\tau].
\]
By Condition~\ref{X_cont_diff}, we obtain that $(\mathring{\mbf{\beta}}, \mathring{\Lambda}) = (\mbf{\beta}^*, \Lambda^*)$.
The desired conclusion follows by the continuity of $\Lambda^*$.

{\bf Step 4.} We establish an upper bound for $E\|\widehat{S}(t|\mbf{X})-S^*(t|\mbf{X})\|_{L_2}^2$.
Since $(\mbf{\beta}^*, \Lambda^*)$ maximizes $\mathbb{M}(\mbf{\beta}, \Lambda)$, by Taylor expansion at $(\mbf{\beta}^*, \Lambda^*)$ and Lemma \ref{info_lower_bound}, we have
\begin{align*}
& \mathbb{M}(\mbf{\beta}, \Lambda)-\mathbb{M}(\mbf{\beta}^*, \Lambda^*) \\
={} & 
\begin{multlined}[t]
-\mathbb{P}\dot{\ell}(\mbf{\beta}^*, \Lambda^*)[\mbf{\beta}-\mbf{\beta}^*, \Lambda-\Lambda^*]^2
-\xi \widetilde{\mathbb{P}}\dot{\psi}(\mbf{\beta}^*, \Lambda^*)[\mbf{\beta}-\mbf{\beta}^*, \Lambda-\Lambda^*]^2 \\
+o_p\left\{(1+\xi)(\|\mbf{\beta}-\mbf{\beta}^*\|^2+\|\Lambda-\Lambda^*\|_{L_2}^2)\right\} 
\end{multlined} \\
\lesssim{} & -(1+\xi)\left(\|\mbf{\beta}-\mbf{\beta}^*\|^2+\|\Lambda-\Lambda^*\|_{L_2}^2\right),
\end{align*}
where $x\lesssim y$ means that $x\le cy$ for a positive constant $c$.
Moreover, for sufficiently small $d$ and any $(\mbf{\beta}, \Lambda)\in \Omega_a$ satisfying $(1+\xi)(\|\mbf{\beta}-\mbf{\beta}^*\|^2+\|\Lambda-\Lambda^*\|_{L_2}^2)<d^2$, we have 
\begin{align*}
& \sup_{(\mbf{\beta}, \Lambda)}\left|(\mathbb{M}_{n,m}-\mathbb{M})(\mbf{\beta}, \Lambda)-(\mathbb{M}_{n,m}-\mathbb{M})(\mbf{\beta}^*, \Lambda^*)\right| \\
={} & \sup_{(\mbf{\beta}, \Lambda)} \left|(\mathbb{P}_n-\mathbb{P})\{\ell(\mbf{\beta}, \Lambda)-\ell(\mbf{\beta}^*, \Lambda^*)\}+
(\xi_n\widetilde{\mathbb{P}}_m-\xi\widetilde{\mathbb{P}})\{\psi(\mbf{\beta}, \Lambda)-\psi(\mbf{\beta}^*, \Lambda^*)\}\right| \\
\le{} & 
\begin{multlined}[t]
\sup_{(\mbf{\beta}, \Lambda)} \left|(\mathbb{P}_n-\mathbb{P})\{\ell(\mbf{\beta}, \Lambda)-\ell(\mbf{\beta}^*, \Lambda^*)\}\right| 
+\xi_n\sup_{(\mbf{\beta}, \Lambda)} \left|(\widetilde{\mathbb{P}}_m-\widetilde{\mathbb{P}})\{\psi(\mbf{\beta}, \Lambda)-\psi(\mbf{\beta}^*, \Lambda^*)\}\right| \\
+\sup_{(\mbf{\beta}, \Lambda)}\left|(\xi_n-\xi)\widetilde{\mathbb{P}}\{\psi(\mbf{\beta}, \Lambda)-\psi(\mbf{\beta}^*, \Lambda^*)\}\right|. 
\end{multlined}
\end{align*}
By the Donsker property shown in Lemma \ref{donsker} and the mean-value theorem, the first term on the right-hand side can be bounded from above by $O_p\{n^{-1/2}(1+\xi)^{-1/2}d\}$, and the third term can be bounded from above by $O_p\{|\xi_n-\xi|(1+\xi)^{-1/2}d\}$.
In addition, by Theorem 2.14.2 of \citet{vaart1996weak} and the arguments in the proof of Lemma \ref{donsker}, the second term can be bounded by 
\begin{align*}
& \xi_n\sup_{(\mbf{\beta}, \Lambda)} \left|(\widetilde{\mathbb{P}}_m-\widetilde{\mathbb{P}})\{\psi(\mbf{\beta}, \Lambda)-\psi(\mbf{\beta}^*, \Lambda^*)\}\right| \\
\lesssim{} & \xi_n m^{-1/2}\int_0^{(1+\xi)^{-1/2}d}\sqrt{1+\log N_{[]}(\epsilon, \mathcal{F}_2, L_2(\widetilde{\mathbb{P}}))}d\epsilon \\
\lesssim{} & \xi_n m^{-1/2}(1+\xi)^{-1/4}d^{1/2},
\end{align*}
where $\mathcal{F}_2 = \{\psi(\mbf{\beta}, \Lambda): (\mbf{\beta}, \Lambda)\in\Omega_a\}$, whose Donsker property is established in Lemma \ref{donsker}.
Combining the above results, we obtain an upper bound for the continuity modulus:
\[
\sup_{(\mbf{\beta}, \Lambda)}\left|(\mathbb{M}_{n,m}-\mathbb{M})(\mbf{\beta}, \Lambda)-(\mathbb{M}_{n,m}-\mathbb{M})(\mbf{\beta}^*, \Lambda^*)\right|
\lesssim \phi(d),
\]
where the supremum is taken over all $(\mbf{\beta}, \Lambda)$ satisfying $(1+\xi)(\|\mbf{\beta}-\mbf{\beta}^*\|^2+\|\Lambda-\Lambda^*\|_{L_2}^2)<d^2$ and 
\[
\phi(d) = n^{-1/2}(1+\xi)^{-1/2}d + \xi_n m^{-1/2}(1+\xi)^{-1/4}d^{1/2} + |\xi_n-\xi|(1+\xi)^{-1/2}d.
\]
We choose an $r$ such that $r^2\phi(r^{-1})\le O(1)$, which implies that 
\[
r\le O\{n^{1/2}(1+\xi)^{1/2} \land m^{1/3}\xi_n^{-2/3}(1+\xi)^{1/6} \land |\xi_n-\xi|^{-1}(1+\xi)^{1/2}\}.
\]
Thus, by Theorem 3.4.1 of \citet{vaart1996weak}, 
\[
(1+\xi)(\|\widehat{\mbf{\beta}}-\mbf{\beta}^*\|^2+\|\widehat{\Lambda}-\Lambda^*\|_{L_2}^2)=
\begin{multlined}[t]
O_p\left\{n^{-1}(1+\xi)^{-1} + m^{-2/3}\xi_n^{4/3}(1+\xi)^{-1/3}\right. \\
 \left.+ (\xi_n-\xi)^{2}(1+\xi)^{-1}\right\}.
\end{multlined}
\]
Since $\xi$ is finite and does not depend on $n$, the above equation further entails that 
\[
\|\widehat{\mbf{\beta}}-\mbf{\beta}^*\|^2+\|\widehat{\Lambda}-\Lambda^*\|_{L_2}^2 = O_p\left\{n^{-1} + m^{-2/3}\xi_n^{4/3}+(\xi_n-\xi)^{2}\right\}.
\]
Finally, note that for any $(\mbf{\beta}_1, \Lambda_1), (\mbf{\beta}_2, \Lambda_2)\in\Omega_a$, 
\begin{equation} \label{bound_S_L2norm}
\begin{aligned}
& \|S(t|\mbf{X};\mbf{\beta}_1, \Lambda_1)-S(t|\mbf{X};\mbf{\beta}_2, \Lambda_2)\|_{L_2}^2 \\
={} & \int_0^{\tau}\left|\exp\left[-G\left\{\int_0^t e^{\mbf{\beta}_1^{\trans}\mbf{X}(s)}d\Lambda_1(s)\right\}\right]-
\exp\left[-G\left\{\int_0^t e^{\mbf{\beta}_2^{\trans}\mbf{X}(s)}d\Lambda_2(s)\right\}\right]\right|^2dt \\
\lesssim{} & \|\mbf{\beta}_1-\mbf{\beta}_2\|^2 + \int_0^{\tau} \left|\int_0^t e^{\mbf{\beta}_2^{\trans}\mbf{X}(s)}d(\Lambda_1-\Lambda_2)(s)\right|^2dt \\
\lesssim{} & \|\mbf{\beta}_1-\mbf{\beta}_2\|^2 + \int_0^{\tau}\left\{|\Lambda_1(t)-\Lambda_2(t)|^2+\int_0^t|\Lambda_1(s)-\Lambda_2(s)|^2ds\right\}dt \\
\lesssim{} & \|\mbf{\beta}_1-\mbf{\beta}_2\|^2 + \|\Lambda_1-\Lambda_2\|_{L_2}^2.
\end{aligned}
\end{equation}
It then follows that  
\begin{equation} \label{bound_hat_star}
\begin{aligned}
E\|\widehat{S}(t|\mbf{X})-S^*(t|\mbf{X})\|_{L_2}^2 \le{} & O_p(\|\widehat{\mbf{\beta}}-\mbf{\beta}^*\|^2+\|\widehat{\Lambda}-\Lambda^*\|_{L_2}^2) \\
={} & O_p\left\{n^{-1} + m^{-2/3}\xi_n^{4/3}+(\xi_n-\xi)^{2}\right\}.
\end{aligned}
\end{equation}

{\bf Step 5.} We establish an upper bound for $E\|\widehat{S}(t|\mbf{X})-S_0(t|\mbf{X})\|_{L_2}^2$.
It suffices to bound $E\|S^*(t|\mbf{X})-S_0(t|\mbf{X})\|_{L_2}^2$.
By the definition of $(\mbf{\beta}^*, \Lambda^*)$, $\mathbb{M}(\mbf{\beta}^*, \Lambda^*)\ge \mathbb{M}(\mbf{\beta}_0, \Lambda_0)$, that is, 
\begin{equation} \label{diff_star_0}
\mathbb{P}\left\{\ell(\mbf{\beta}^*, \Lambda^*)-\ell(\mbf{\beta}_0, \Lambda_0)\right\} \ge \xi\widetilde{\mathbb{P}} \left\{\psi(\mbf{\beta}_0, \Lambda_0)-\psi(\mbf{\beta}^*, \Lambda^*)\right\}.
\end{equation}
By Taylor expansion, Lemma \ref{info_lower_bound}, and \eqref{bound_S_L2norm}, we can bound the left-hand side of \eqref{diff_star_0} by 
\begin{align*}
& \mathbb{P}\left\{\ell(\mbf{\beta}^*, \Lambda^*)-\ell(\mbf{\beta}_0, \Lambda_0)\right\} \\
={} & -\mathbb{P}\left\{\dot{\ell}(\mbf{\beta}_0, \Lambda_0)[\mbf{\beta}^*-\mbf{\beta}_0, \Lambda^*-\Lambda_0]^2\right\}+o(\|\mbf{\beta}^*-\mbf{\beta}_0\|^2+\|\Lambda^*-\Lambda_0\|_{L_2}^2) \\
\le{} & -C_3(\|\mbf{\beta}^*-\mbf{\beta}_0\|^2+\|\Lambda^*-\Lambda_0\|_{L_2}^2) \\
\le{} & -C_4E\|S^*(t|\mbf{X})-S_0(t|\mbf{X})\|_{L_2}^2,
\end{align*}
where $C_3$ and $C_4$ are positive constants. 
In addition, write $\psi(\mbf{\beta}, \Lambda)$ as $\psi\{S(\widetilde{Y}|\widetilde{\mbf{X}};\mbf{\beta}, \Lambda)\}$ and note that $\widecheck{S}$ maximizes $\psi(S)$ for every fixed $(\widetilde{Y}, \widetilde{\mbf{X}})$. 
We then perform a second-order Taylor expansion of $\psi(S)$ at $\widecheck{S}$ to obtain the following inequality for some constant $C>0$:
\begin{equation} \label{equiv_psi_L2}
1/CE\|\widecheck{S}(t|\mbf{X})-S(t|\mbf{X})\|_{L_2}^2 \le \widetilde{\mathbb{P}} \{\psi(\widecheck{S})-\psi(S)\} \le CE\|\widecheck{S}(t|\mbf{X})-S(t|\mbf{X})\|_{L_2}^2,
\end{equation}
where the existence of the constant $C$ is guaranteed by Condition~\ref{Scheck_bounded}.
Using the above result, together with the convergence rate of $\widecheck{S}$ and the dissimilarity between $S_0$ and $S_1$ stated in Section~\ref{m-sec:theory} of the main manuscript, we can bound the right-hand side of \eqref{diff_star_0} by
\begin{align*}
& \xi\widetilde{\mathbb{P}} \left\{\psi(\widecheck{S})-\psi(S^*)\right\}-\xi\widetilde{\mathbb{P}} \left\{\psi(\widecheck{S})-\psi(S_0)\right\} \\
\ge{} & C_5\xi E\|S^*(t|\mbf{X})-\widecheck{S}(t|\mbf{X})\|_{L_2}^2-C_6\xi E\|\widecheck{S}(t|\mbf{X})-S_0(t|\mbf{X})\|_{L_2}^2 \\
\ge{} & C_5\xi E\|S^*(t|\mbf{X})-S_0(t|\mbf{X})\|_{L_2}^2-(C_5+C_6)\xi E\|\widecheck{S}(t|\mbf{X})-S_0(t|\mbf{X})\|_{L_2}^2 \\
\ge{} & 
\begin{multlined}[t]
C_5\xi E\|S^*(t|\mbf{X})-S_0(t|\mbf{X})\|_{L_2}^2 \\
-C_7\xi \left\{E\|\widecheck{S}(t|\mbf{X})-S_1(t|\mbf{X})\|_{L_2}^2+E\|S_1(t|\mbf{X})-S_0(t|\mbf{X})\|_{L_2}^2\right\} 
\end{multlined} \\
\ge{} & C_5\xi E\|S^*(t|\mbf{X})-S_0(t|\mbf{X})\|_{L_2}^2-C_8\xi (q_N+\eta),
\end{align*}
where $C_i$ ($i=5, 6, 7, 8$) are positive constants. 
Replacing both sides of \eqref{diff_star_0} by their respective bounds derived above, we obtain
\[
E\|S^*(t|\mbf{X})-S_0(t|\mbf{X})\|_{L_2}^2\le O_p\left\{\xi(1+\xi)^{-1}(\eta+q_N)\right\}.
\]
Combining the above inequality with \eqref{bound_hat_star}, we obtain
\begin{align*}
& E\|\widehat{S}(t|\mbf{X})-S_0(t|\mbf{X})\|_{L_2}^2 \\
\lesssim{} & E\|\widehat{S}(t|\mbf{X})-S^*(t|\mbf{X})\|_{L_2}^2 + E\|S^*(t|\mbf{X})-S_0(t|\mbf{X})\|_{L_2}^2 \\ 
\le{} & O_p\bigl\{n^{-1} + m^{-2/3}\xi_n^{4/3}+(\xi_n-\xi)^{2}+\xi(\eta+q_N)\bigr\}.
\end{align*}

\medskip
\begin{flushleft}
{\bf \underline{Case 2: $\xi=\infty$}}
\end{flushleft}

{\bf Step 1.} We show that $\limsup_n \widehat{\Lambda}(\tau)<\infty$ almost surely. 
Let $\overline{\Lambda}$ be a step function which takes jumps at all generated time points $\widetilde{Y}_i$ and satisfies $\overline{\Lambda}(\widetilde{Y}_i) = \Lambda^*(\widetilde{Y}_i)$ ($i=1,\dots,m$). 
Clearly, $\overline{\Lambda}$ converges to $\Lambda^*$ uniformly over $[0,\tau]$.
By the definition of $(\widehat{\mbf{\beta}}, \widehat{\Lambda})$, $\mathbb{M}_{n,m}(\widehat{\mbf{\beta}}, \widehat{\Lambda})\ge \mathbb{M}_{n,m}(\mbf{\beta}^*, \overline{\Lambda})$, that is, 
\[
\xi_n^{-1}\mathbb{P}_n\ell(\widehat{\mbf{\beta}}, \widehat{\Lambda})+\widetilde{\mathbb{P}}_m \psi(\widehat{\mbf{\beta}}, \widehat{\Lambda}) \ge \xi_n^{-1}\mathbb{P}_n\ell(\mbf{\beta}^*, \overline{\Lambda})+\widetilde{\mathbb{P}}_m \psi(\mbf{\beta}^*, \overline{\Lambda}). 
\]
Using arguments similar to those in Step 2 of Case 1, we can eventually obtain that 
\[
O(1) \le 
\begin{multlined}[t]
-\xi_n^{-1}(2n)^{-1}\sum_{i=1}^n(\Delta_i+\rho)I(Y_i=\tau)\log \{1+\widehat{\Lambda}(\tau)\} - C_1\rho \log\{1+\widehat{\Lambda}(\tau)\} \\
-\xi_n^{-1}\Biggl\{(2n)^{-1}\sum_{i=1}^n(\Delta_i+\rho)I(Y_i=\tau)\log \{1+\widehat{\Lambda}(\tau)\}\Biggr. \\
-\Biggl.n^{-1}\sum_{i=1}^n \Delta_iI\{Y_i\in[s_{1}, s_{0}]\}\log\{1+\widehat{\Lambda}(\tau)\}\Biggr\} \\
-\xi_n^{-1}\sum_{r=1}^{R-1}\Biggl\{n^{-1}\sum_{i=1}^n(\Delta_i+\rho)I\{Y_i\in[s_r, s_{r-1})\}\log \{1+\widehat{\Lambda}(s_r)\}\Biggr. \\
\Biggl.-n^{-1}\sum_{i=1}^n \Delta_iI\{Y_i\in[s_{r+1}, s_{r})\}\log\{1+\widehat{\Lambda}(s_{r})\}\Biggr\}.
\end{multlined}
\]
We can choose $\tau = s_0>s_1>\cdots>s_R=0$ such that the first line on the right-hand side diverges to $-\infty$ as $\widehat{\Lambda}(\tau)$ diverges to $\infty$, and the coefficients in front of $\log\{1+\widehat{\Lambda}(s_{r})\}$ in the remaining terms are all non-positive when $n$ is large enough. This contradicts the fact that the right-hand side should be bounded. 
Thus, we conclude that $\limsup_n \widehat{\Lambda}(\tau)<\infty$ almost surely.

{\bf Step 2.} We show that $\|\widehat{\mbf{\beta}}-\mbf{\beta}^*\|+ \|\widehat{\Lambda}-\Lambda^*\|_{\infty}\overset{a.s.}{\rightarrow}0$.
As in Step 3 of Case 1, it follows from Helly's selection lemma that for any subsequence of $(\widehat{\mbf{\beta}}, \widehat{\Lambda})$, we can choose a further subsequence such that $\widehat{\mbf{\beta}}\rightarrow \mathring{\mbf{\beta}}$ and $\widehat{\Lambda}$ converges pointwise to some increasing function $\mathring{\Lambda}$ in $[0,\tau]$.
We want to show that $(\mathring{\mbf{\beta}}, \mathring{\Lambda}) = (\mbf{\beta}^*, \Lambda^*)$.

Note that 
\[
\xi_n^{-1}\mathbb{P}_n\ell(\widehat{\mbf{\beta}}, \widehat{\Lambda})+\widetilde{\mathbb{P}}_m \psi(\widehat{\mbf{\beta}}, \widehat{\Lambda}) \ge \xi_n^{-1}\mathbb{P}_n\ell(\mbf{\beta}^*, \overline{\Lambda})+\widetilde{\mathbb{P}}_m \psi(\mbf{\beta}^*, \overline{\Lambda}).
\] 
By the Donsker property shown in Lemma \ref{donsker} and the dominated convergence theorem, as $n,m\rightarrow\infty$, we have
\begin{gather*}
\xi_n^{-1}\mathbb{P}_n\ell(\widehat{\mbf{\beta}}, \widehat{\Lambda})+\widetilde{\mathbb{P}}_m \psi(\widehat{\mbf{\beta}}, \widehat{\Lambda}) \overset{a.s.}{\rightarrow} \widetilde{\mathbb{P}} \psi(\mathring{\mbf{\beta}}, \mathring{\Lambda}), \\
\xi_n^{-1}\mathbb{P}_n\ell(\mbf{\beta}^*, \overline{\Lambda})+\widetilde{\mathbb{P}}_m \psi(\mbf{\beta}^*, \overline{\Lambda}) \overset{a.s.}{\rightarrow} \widetilde{\mathbb{P}} \psi(\mbf{\beta}^*, \Lambda^*).
\end{gather*}
Thus, $\widetilde{\mathbb{P}} \psi(\mathring{\mbf{\beta}}, \mathring{\Lambda})\ge \widetilde{\mathbb{P}} \psi(\mbf{\beta}^*, \Lambda^*)$.
By Condition~\ref{identifiability}, we obtain that $\mathring{\mbf{\beta}} = \mbf{\beta}^*$ and $\mathring{\Lambda}(t) = \Lambda^*(t)$ for $t\in[0,\tau]$. 
By the continuity of $\Lambda^*$ stated in Condition~\ref{identifiability}, we conclude that $\|\widehat{\mbf{\beta}}-\mbf{\beta}^*\|+ \|\widehat{\Lambda}-\Lambda^*\|_{\infty}\overset{a.s.}{\rightarrow}0$.

{\bf Step 3.} We establish an upper bound for $E\|\widehat{S}(t|\mbf{X})-S^*(t|\mbf{X})\|_{L_2}^2$.
By Taylor expansion at $(\mbf{\beta}^*, \Lambda^*)$ and the arguments in the proof of Lemma \ref{info_lower_bound}, we have
\begin{align*}
& \mathbb{M}(\mbf{\beta}, \Lambda)-\mathbb{M}(\mbf{\beta}^*, \Lambda^*) \\
={} & \widetilde{\mathbb{P}}\{\psi(\mbf{\beta}, \Lambda)-\psi(\mbf{\beta}^*, \Lambda^*)\} \\
={} & 
\begin{multlined}[t]
-\widetilde{\mathbb{P}}\dot{\psi}(\mbf{\beta}^*, \Lambda^*)[\mbf{\beta}-\mbf{\beta}^*, \Lambda-\Lambda^*]^2 
+o_p\left(\|\mbf{\beta}-\mbf{\beta}^*\|^2+\|\Lambda-\Lambda^*\|_{L_2}^2\right)
\end{multlined} \\
\lesssim{} & -\left(\|\mbf{\beta}-\mbf{\beta}^*\|^2+\|\Lambda-\Lambda^*\|_{L_2}^2\right).
\end{align*}
Moreover, for sufficiently small $d$ and any $(\mbf{\beta}, \Lambda)\in \Omega_a$ satisfying $\|\mbf{\beta}-\mbf{\beta}^*\|^2+\|\Lambda-\Lambda^*\|_{L_2}^2<d^2$, we have 
\begin{align*}
& \sup_{(\mbf{\beta}, \Lambda)}\left|(\mathbb{M}_{n,m}-\mathbb{M})(\mbf{\beta}, \Lambda)-(\mathbb{M}_{n,m}-\mathbb{M})(\mbf{\beta}^*, \Lambda^*)\right| \\
={} & \sup_{(\mbf{\beta}, \Lambda)} \left|\xi_n^{-1}\mathbb{P}_n\{\ell(\mbf{\beta}, \Lambda)-\ell(\mbf{\beta}^*, \Lambda^*)\}+
(\widetilde{\mathbb{P}}_m-\widetilde{\mathbb{P}})\{\psi(\mbf{\beta}, \Lambda)-\psi(\mbf{\beta}^*, \Lambda^*)\}\right| \\
\le{} & 
\begin{multlined}[t]
\xi_n^{-1}\sup_{(\mbf{\beta}, \Lambda)} \left|(\mathbb{P}_n-\mathbb{P})\{\ell(\mbf{\beta}, \Lambda)-\ell(\mbf{\beta}^*, \Lambda^*)\}\right| 
+\xi_n^{-1}\sup_{(\mbf{\beta}, \Lambda)} \left|\mathbb{P}\{\ell(\mbf{\beta}, \Lambda)-\ell(\mbf{\beta}^*, \Lambda^*)\}\right| \\
+\sup_{(\mbf{\beta}, \Lambda)}\left|(\widetilde{\mathbb{P}}_m-\widetilde{\mathbb{P}})\{\psi(\mbf{\beta}, \Lambda)-\psi(\mbf{\beta}^*, \Lambda^*)\}\right|. 
\end{multlined}
\end{align*}
Using arguments similar to those in Step 4 of Case 1, we can bound the first term by $O_p(\xi_n^{-1}n^{-1/2}d)$, the second term by $O_p(\xi_n^{-1}d)$, and the third term by $O_p(m^{-1/2}d^{1/2})$.
Thus, the continuity modulus can be bounded by
\[
\sup_{(\mbf{\beta}, \Lambda)}\left|(\mathbb{M}_{n,m}-\mathbb{M})(\mbf{\beta}, \Lambda)-(\mathbb{M}_{n,m}-\mathbb{M})(\mbf{\beta}^*, \Lambda^*)\right|
\lesssim O_p(\xi_n^{-1}d + m^{-1/2}d^{1/2}).
\]
Therefore, by Theorem 3.4.1 of \citet{vaart1996weak}, 
\[
\|\widehat{\mbf{\beta}}-\mbf{\beta}^*\|^2+\|\widehat{\Lambda}-\Lambda^*\|_{L_2}^2=O_p\left(\xi_n^{-2} + m^{-2/3}\right),
\]
which further implies that  
\begin{equation} \label{bound_hat_star_case2}
E\|\widehat{S}(t|\mbf{X})-S^*(t|\mbf{X})\|_{L_2}^2 \le O_p\left(\xi_n^{-2} + m^{-2/3}\right).
\end{equation}

{\bf Step 4.} We derive an upper bound for $E\|S^*(t|\mbf{X})-S_0(t|\mbf{X})\|_{L_2}^2$ so as to establish the convergence rate for $\widehat{S}(t|\mbf{X})$.
By the definition of $(\mbf{\beta}^*, \Lambda^*)$ and the result in \eqref{equiv_psi_L2}, we have 
\begin{align*}
0 \le{} & \mathbb{M}(\mbf{\beta}^*, \Lambda^*) - \mathbb{M}(\mbf{\beta}_0, \Lambda_0) \\
={} & \widetilde{\mathbb{P}}\{\psi(\mbf{\beta}^*, \Lambda^*)-\psi(\mbf{\beta}_0, \Lambda_0)\} \\
={} & \widetilde{\mathbb{P}} \left\{\psi(\widecheck{S})-\psi(S_0)\right\}-\widetilde{\mathbb{P}} \left\{\psi(\widecheck{S})-\psi(S^*)\right\} \\
\le{} & C_9E\|\widecheck{S}(t|\mbf{X})-S_0(t|\mbf{X})\|_{L_2}^2-C_{10}E\|S^*(t|\mbf{X})-\widecheck{S}(t|\mbf{X})\|_{L_2}^2 \\
\le{} & (C_9+C_{10})E\|\widecheck{S}(t|\mbf{X})-S_0(t|\mbf{X})\|_{L_2}^2-C_{10}E\|S^*(t|\mbf{X})-S_0(t|\mbf{X})\|_{L_2}^2.
\end{align*}
Thus, 
\begin{align*}
E\|S^*(t|\mbf{X})-S_0(t|\mbf{X})\|_{L_2}^2\lesssim{} & E\|\widecheck{S}(t|\mbf{X})-S_0(t|\mbf{X})\|_{L_2}^2 \\
\lesssim{} & E\|\widecheck{S}(t|\mbf{X})-S_1(t|\mbf{X})\|_{L_2}^2 + E\|S_1(t|\mbf{X})-S_0(t|\mbf{X})\|_{L_2}^2 \\
\le{} & O_p(q_N+\eta).
\end{align*}
Combining the above result with \eqref{bound_hat_star_case2}, we obtain
\[
E\|\widehat{S}(t|\mbf{X})-S_0(t|\mbf{X})\|_{L_2}^2 \le O_p\bigl(\xi_n^{-2} + m^{-2/3} + \eta+q_N\bigr).
\]

Combining the results in Case 1 and Case 2, we conclude that 
\[
E\|\widehat{S}(t|\mbf{X})-S_0(t|\mbf{X})\|_{L_2}^2 \le 
\begin{cases}
O_p\bigl\{n^{-1} + m^{-2/3}\xi_n^{4/3}+(\xi_n-\xi)^{2}+\xi(\eta+q_N)\bigr\} & \text{ if } \xi<\infty, \\
O_p\bigl(\xi_n^{-2} + m^{-2/3} + \eta+q_N\bigr) & \text{ if } \xi=\infty.
\end{cases}
\]

\medskip
\begin{flushleft}
{\bf Proof of Corollary 1.}
\end{flushleft}
For the convergence rate established in Theorem 1, we can choose $m$ to be sufficiently large such that the terms involving $m^{-2/3}$ are negligible. 
Thus, to obtain the optimal convergence rate, it suffices to minimize the following upper bound with respect to $\xi_n$ and $\xi$:
\[
  I(\xi<\infty)O_p\bigl\{n^{-1} +(\xi_n-\xi)^{2}+\xi(\eta+q_N)\bigr\}+I(\xi=\infty)O_p\bigl(\xi_n^{-2} + \eta+q_N\bigr).
\]
If $\eta+q_N=o(n^{-1})$, we can choose $\xi=\infty$ and $\xi_n^{-2}\le O(\eta+q_N)$, such that the above bound attains the minimum $O_p(\eta+q_N)$.
Otherwise, we can simply set $\xi_n = \xi=0$ to achieve the minimum $O_p(n^{-1})$, which is the target-only convergence rate. 
Therefore, the optimal convergence rate for $\widehat{S}(t|\mbf{X})$ is given by 
\[
E\|\widehat{S}(t|\mbf{X})-S_0(t|\mbf{X})\|_{L_2}^2 \le O_p\left\{n^{-1} \land (\eta+q_N)\right\},
\]
which is guaranteed to be no slower than the target-only rate.

\medskip
\begin{flushleft}
{\bf Proof of Theorem 2.}
\end{flushleft}
By the arguments for semiparametric transformation models \citep{zeng2010general}, the target-only estimator $\widehat{S}^{\text{tar}}(t|\mbf{X})$ satisfies $E\|\widehat{S}^{\text{tar}}(t|\mbf{X})-S_0(t|\mbf{X})\|_{L_2}^2 = O_p(n^{-1})$.
By the central limit theorem and triangle inequality,
\begin{align*}
& \widetilde{\mathbb{P}}_m\|\widecheck{S}(t|\mbf{X})-\widehat{S}^{\text{tar}}(t|\mbf{X})\|_{L_2}^{2} \\
={} & (\widetilde{\mathbb{P}}_m-\widetilde{\mathbb{P}})\|\widecheck{S}(t|\mbf{X})-\widehat{S}^{\text{tar}}(t|\mbf{X})\|_{L_2}^{2}+E\|\widecheck{S}(t|\mbf{X})-\widehat{S}^{\text{tar}}(t|\mbf{X})\|_{L_2}^{2} \\
\le{} & 
\begin{multlined}[t]
O_p\bigl\{m^{-1/2}+E\|\widecheck{S}(t|\mbf{X})-S_1(t|\mbf{X})\|_{L_2}^2+E\|S_1(t|\mbf{X})-S_0(t|\mbf{X})\|_{L_2}^2 \\
+E\|S_0(t|\mbf{X})-\widehat{S}^{\text{tar}}(t|\mbf{X})\|_{L_2}^2\bigr\}
\end{multlined} \\
\le{} & O_p(m^{-1/2}+q_N+\eta+n^{-1}),
\end{align*}
where the term $m^{-1/2}$ can be omitted since $m$ can be arbitrarily large. 
We consider the following two scenarios separately. 
\begin{itemize}
  \item If $\eta+q_N\le O(n^{-1})$, then $\widetilde{\mathbb{P}}_m\|\widecheck{S}(t|\mbf{X})-\widehat{S}^{\text{tar}}(t|\mbf{X})\|_{L_2}^{2}+q_N\le cn^{-1}$ for some positive constant $c$, so that 
  \[
  \xi_n^{\text{opt}} = \left\{\widetilde{\mathbb{P}}_m\|\widecheck{S}(t|\mbf{X})-\widehat{S}^{\text{tar}}(t|\mbf{X})\|_{L_2}^{2}\right\}^{-\nu/2}
  \ge C_{11}(q_N+\eta+n^{-1})^{-\nu/2},
  \]
  for some positive constant $C_{11}$. 
  Thus, $\xi_n^{\text{opt}}\rightarrow \infty$.
  We plug $\xi_n^{\text{opt}}$ into the second upper bound in Theorem 1 and omit the term $m^{-2/3}$ to obtain that
  \begin{align*}
  E\|\widehat{S}(t|\mbf{X})-S_0(t|\mbf{X})\|_{L_2}^2\le{} & O_p\left[\left\{\widetilde{\mathbb{P}}_m\|\widecheck{S}(t|\mbf{X})-\widehat{S}^{\text{tar}}(t|\mbf{X})\|_{L_2}^{2}\right\}^{\nu}+\eta+q_N\right] \\
  \le{} & O_p\left\{(q_N+\eta+n^{-1})^{\nu}+\eta+q_N\right\} \\
  \le{} & O_p\left(n^{-\nu}+\eta+q_N\right) \\
  ={} & O_p\left(\eta+q_N\right),
  \end{align*}
  where the last inequality follows from the condition $n^{-\nu}\le O(q_N)$.

  \item If $n^{-1} = o(\eta+q_N)$, then 
  \begin{align*}
  n^{-1} ={} & o\left\{E\|S_1(t|\mbf{X})-S_0(t|\mbf{X})\|_{L_2}^2+q_N\right\} \\
  \le{} & 
  \begin{multlined}[t]
  o\bigl\{E\|S_1(t|\mbf{X})-\widecheck{S}(t|\mbf{X})\|_{L_2}^2+E\|\widecheck{S}(t|\mbf{X})-\widehat{S}^{\text{tar}}(t|\mbf{X})\|_{L_2}^{2} \\
  +E\|\widehat{S}^{\text{tar}}(t|\mbf{X})-S_0(t|\mbf{X})\|_{L_2}^2+q_N\bigr\}
  \end{multlined} \\
  \le{} & o\left\{q_N+n^{-1}-(\widetilde{\mathbb{P}}_m-\widetilde{\mathbb{P}})\|\widecheck{S}(t|\mbf{X})-\widehat{S}^{\text{tar}}(t|\mbf{X})\|_{L_2}^{2}+\widetilde{\mathbb{P}}_m\|\widecheck{S}(t|\mbf{X})-\widehat{S}^{\text{tar}}(t|\mbf{X})\|_{L_2}^{2}\right\} \\
  ={} & o\left\{q_N+n^{-1}+m^{-1/2}+\widetilde{\mathbb{P}}_m\|\widecheck{S}(t|\mbf{X})-\widehat{S}^{\text{tar}}(t|\mbf{X})\|_{L_2}^{2}\right\},
  \end{align*}
  which implies that the condition $\widetilde{\mathbb{P}}_m\|\widecheck{S}(t|\mbf{X})-\widehat{S}^{\text{tar}}(t|\mbf{X})\|_{L_2}^{2}+q_N\le cn^{-1}$ does not hold. 
  Thus, $\xi_n^{\text{opt}} = 0$ for every $n$. Therefore, $E\|\widehat{S}(t|\mbf{X})-S_0(t|\mbf{X})\|_{L_2}^2\le O_p(n^{-1})$. 
\end{itemize}
Hence, we conclude that $\xi_n^{\text{opt}}$ yields the optimal convergence rate $O_p\{n^{-1} \land (\eta+q_N)\}$.

\section{Useful Lemmas}
\label{sec:useful_lemmas}
\begin{lemma} \label{donsker}
Under Conditions~\ref{true_parameter}--\ref{identifiability}, the following function classes are Donsker.
\begin{align*}
\mathcal{F}_1 ={} & \left\{\log\Phi(\mbf{\beta}, \Lambda):\,(\mbf{\beta}, \Lambda)\in\Omega_a\right\}, \\
\mathcal{F}_2 ={} & \left\{\psi(\mbf{\beta}, \Lambda):\,(\mbf{\beta}, \Lambda)\in\Omega_a\right\}, \\
\mathcal{F}_{R_1} ={} & \left\{R_1(t;\mbf{\beta}, \Lambda): \, t\in[0,\tau], (\mbf{\beta}, \Lambda)\in\Omega_a\right\} \\
\mathcal{F}_{R_2} ={} & \left\{R_2(t;\mbf{\beta}, \Lambda): \, t\in[0,\tau], (\mbf{\beta}, \Lambda)\in\Omega_a\right\}.
\end{align*}
\end{lemma}

\begin{proof}
We first show that class $\mathcal{F}_1$ is Donsker. Recall the definition of $\Phi(\mbf{\beta}, \Lambda)$:
\[
\Phi(\mbf{\beta}, \Lambda) = \left[e^{\mbf{\beta}^{\trans}\mbf{X}(Y)}G^{\prime}\left\{\int_0^Ye^{\mbf{\beta}^{\trans}\mbf{X}(s)}d\Lambda(s)\right\}\right]^{\Delta}\exp\left[-G\left\{\int_0^Ye^{\mbf{\beta}^{\trans}\mbf{X}(s)}d\Lambda(s)\right\}\right].
\]
For any $(\mbf{\beta}_1, \Lambda_1), (\mbf{\beta}_2, \Lambda_2)\in\Omega_a$, by the mean-value theorem, 
\begin{equation} \label{diff_log_Phi}
\begin{aligned}
& |\log\Phi(\mbf{\beta}_1, \Lambda_1)-\log\Phi(\mbf{\beta}_2, \Lambda_2)| \\
={} & \frac{\dot{\Phi}(\mbf{\beta}_3, \Lambda_3)[\mbf{\beta}_1-\mbf{\beta}_2, \Lambda_1-\Lambda_2]}{\Phi(\mbf{\beta}_3, \Lambda_3)} \\
={} & 
\begin{multlined}[t]
\biggl|\Delta\mbf{X}(Y)^{\trans}(\mbf{\beta}_1-\mbf{\beta}_2)+\int_0^Y \zeta_1(t;\mbf{\beta}_3, \Lambda_3)\mbf{X}(t)^{\trans}(\mbf{\beta}_1-\mbf{\beta}_2)d\Lambda_3(t) \\
+\int_0^Y \zeta_1(t;\mbf{\beta}_3, \Lambda_3)d(\Lambda_1-\Lambda_2)(t)\biggr| 
\end{multlined} \\
\le{} & 
\begin{multlined}[t]
\left|\mbf{X}(Y)^{\trans}(\mbf{\beta}_1-\mbf{\beta}_2)\right|+\left|\int_0^Y \zeta_1(t;\mbf{\beta}_3, \Lambda_3)\mbf{X}(t)^{\trans}(\mbf{\beta}_1-\mbf{\beta}_2)d\Lambda_3(t)\right| \\
+\left|\int_0^Y \zeta_1(t;\mbf{\beta}_3, \Lambda_3)d(\Lambda_1-\Lambda_2)(t)\right|,
\end{multlined}
\end{aligned}
\end{equation}
where $(\mbf{\beta}_3, \Lambda_3)\in\Omega_a$ lies between $(\mbf{\beta}_1, \Lambda_1)$ and $(\mbf{\beta}_2, \Lambda_2)$, and $\zeta_1(t;\mbf{\beta}, \Lambda)$ is given by 
\[
\zeta_1(t;\mbf{\beta}, \Lambda) = \frac{\Delta G^{\prime\prime}\{\int_0^Y e^{\mbf{\beta}^{\trans}\mbf{X}(s)}d\Lambda(s)\}e^{\mbf{\beta}^{\trans}\mbf{X}(t)}}{G^{\prime}\{\int_0^Y e^{\mbf{\beta}^{\trans}\mbf{X}(s)}d\Lambda(s)\}}-G^{\prime}\{\int_0^Y e^{\mbf{\beta}^{\trans}\mbf{X}(s)}d\Lambda(s)\}e^{\mbf{\beta}^{\trans}\mbf{X}(t)}.
\]
By the smoothness of $\mbf{X}(\cdot)$ and $G(\cdot)$ stated in Conditions~\ref{X_cont_diff} and \ref{transformation}, the first and second terms on the right-hand side of \eqref{diff_log_Phi} can be bounded from above by $\|\mbf{\beta}_1-\mbf{\beta}_2\|$, up to a scaling factor.
In addition, by the integration by parts formula and the Cauchy–Schwarz inequality, the third term can be bounded from above by $|\Lambda_1(Y)-\Lambda_2(Y)|+\|\Lambda_1-\Lambda_2\|_{L_2}$, up to a scaling factor.
Thus, we have 
\[
\|\log\Phi(\mbf{\beta}_1, \Lambda_1)-\log\Phi(\mbf{\beta}_2, \Lambda_2)\|^2_{L_2(\mathbb{P})} \le O(1) \left\{\|\mbf{\beta}_1-\mbf{\beta}_2\|^2+\|\Lambda_1-\Lambda_2\|^2_{L_2(\mu)}\right\},
\]
for some finite positive measure $\mu$, where $\|f\|_{L_2(\mu)} = (\int f^2d\mu)^{1/2}$.

There are at most $O(\epsilon^{-p})$ $\epsilon$-brackets covering $\mathcal{B}$. In addition, by Theorem 2.7.5 of \citet{vaart1996weak}, the bracketing entropy of $\mathcal{L}_a$ satisfies $\log N_{[]}(\epsilon, \mathcal{L}_a, L_2(\mu))\lesssim \epsilon^{-1}$. 
Thus, there are a total of $O(\epsilon^{-p})\times \exp\{O(\epsilon^{-1})\}$ brackets $[\mbf{\beta}_1, \mbf{\beta}_2]\times [\Lambda_1, \Lambda_2]$ covering $\Omega_a$, such that $\|\log\Phi(\mbf{\beta}_1, \Lambda_1)-\log\Phi(\mbf{\beta}_2, \Lambda_2)\|^2_{L_2(\mathbb{P})} < O(\epsilon)$.
Therefore, the bracketing integral of $\mathcal{F}_1$ is finite, which entails that $\mathcal{F}_1$ is Donsker.  
By similar arguments, we can show that classes $\mathcal{F}_2$, $\mathcal{F}_{R_1}$, and $\mathcal{F}_{R_2}$ are Donsker. 
\end{proof}

\begin{lemma} \label{info_lower_bound}
Under Conditions~\ref{true_parameter}--\ref{identifiability}, for any $\xi\in[0,\infty)$, there exists some positive constant $c$ such that for any $\mbf{h} = (\mbf{h}_{\mbf{\beta}}, h_{\Lambda})\in \mathbb{R}^{p}\times L_2([0,\tau])$,  
\begin{gather*}
\mathbb{P}\dot{\ell}(\mbf{\beta}^*, \Lambda^*)[\mbf{h}_{\mbf{\beta}}, \int h_{\Lambda}d\Lambda^*]^2
+\xi \widetilde{\mathbb{P}}\dot{\psi}(\mbf{\beta}^*, \Lambda^*)[\mbf{h}_{\mbf{\beta}}, \int h_{\Lambda}d\Lambda^*]^2
\ge c(1+\xi)(\|\mbf{h}_{\mbf{\beta}}\|^2+\|h_{\Lambda}\|_{L_2}^2), \\
\mathbb{P}\dot{\ell}(\mbf{\beta}_0, \Lambda_0)[\mbf{h}_{\mbf{\beta}}, \int h_{\Lambda}d\Lambda_0]^2\ge c(\|\mbf{h}_{\mbf{\beta}}\|^2+\|h_{\Lambda}\|_{L_2}^2).
\end{gather*}
\end{lemma}

\begin{proof}
For simplicity of presentation, let $\dot{\ell}^*[\mbf{h}] = \dot{\ell}(\mbf{\beta}^*, \Lambda^*)[\mbf{h}_{\mbf{\beta}}, \int h_{\Lambda}d\Lambda^*]$ and $\dot{\psi}^*[\mbf{h}] = \dot{\psi}(\mbf{\beta}^*, \Lambda^*)[\mbf{h}_{\mbf{\beta}}, \int h_{\Lambda}d\Lambda^*]$.
Simple calculation yields the score statistics as follows:
\begin{equation} \label{score_func}
\begin{gathered}
\dot{\ell}^*[\mbf{h}]=\Delta\{\mbf{X}(Y)^{\trans}\mbf{h}_{\mbf{\beta}}+h_{\Lambda}(Y)\}+\int_0^{\tau}\zeta_1(t)\{\mbf{X}(t)^{\trans}\mbf{h}_{\mbf{\beta}}+h_{\Lambda}(t)\}d\Lambda^*(t), \\
\dot{\psi}^*[\mbf{h}]=\int_0^{\tau}\zeta_2(t)\{\widetilde{\mbf{X}}(t)^{\trans}\mbf{h}_{\mbf{\beta}}+h_{\Lambda}(t)\}d\Lambda^*(t),
\end{gathered}
\end{equation}
where 
\begin{gather*}
\zeta_1(t) = I(t\le Y)e^{{\mbf{\beta}^*}^{\trans}\mbf{X}(t)}\left[\frac{\Delta  G^{\prime\prime}\{\int_0^Y e^{{\mbf{\beta}^*}^{\trans}\mbf{X}(s)}d\Lambda^*(s)\}}{G^{\prime}\{\int_0^Y e^{{\mbf{\beta}^*}^{\trans}\mbf{X}(s)}d\Lambda^*(s)\}}-G^{\prime}\{\int_0^Y e^{{\mbf{\beta}^*}^{\trans}\mbf{X}(s)}d\Lambda^*(s)\}\right], \\
\zeta_2(t) = 
\begin{multlined}[t]
I(t\le \widetilde{Y})e^{{\mbf{\beta}^*}^{\trans}\widetilde{\mbf{X}}(t)}\Biggl[-\widecheck{S}(\widetilde{Y}|\widetilde{\mbf{X}})G^{\prime}\{\int_0^{\widetilde{Y}} e^{{\mbf{\beta}^*}^{\trans}\widetilde{\mbf{X}}(s)}d\Lambda^*(s)\} \\
+\{1-\widecheck{S}(\widetilde{Y}|\widetilde{\mbf{X}})\}\frac{G^{\prime}\{\int_0^{\widetilde{Y}} e^{{\mbf{\beta}^*}^{\trans}\widetilde{\mbf{X}}(s)}d\Lambda^*(s)\}}{\exp[G\{\int_0^{\widetilde{Y}} e^{{\mbf{\beta}^*}^{\trans}\widetilde{\mbf{X}}(s)}d\Lambda^*(s)\}]-1}\Biggr].
\end{multlined}
\end{gather*}
It is easy to observe that $\dot{\ell}^*[\mbf{h}]$ and $\dot{\psi}^*[\mbf{h}]$ can be reformulated as $\dot{\ell}^*[\mbf{h}] = \mbf{\zeta}_{3\mbf{\beta}}^{\trans}\mbf{h}_{\mbf{\beta}}+\int_0^{\tau}\zeta_{3\Lambda}(t)h_{\Lambda}(t)dt$ and $\dot{\psi}^*[\mbf{h}] = \mbf{\zeta}_{4\mbf{\beta}}^{\trans}\mbf{h}_{\mbf{\beta}}+\int_0^{\tau}\zeta_{4\Lambda}(t)h_{\Lambda}(t)dt$, respectively, where 
\begin{gather*}
\mbf{\zeta}_{3\mbf{\beta}} = \Delta\mbf{X}(Y)+\int_0^{\tau}\zeta_1(t)\mbf{X}(t)d\Lambda^*(t), \quad \zeta_{3\Lambda}(t) = \Delta \delta(t-Y)+\lambda^*(t)\zeta_1(t), \\
\mbf{\zeta}_{4\mbf{\beta}} = \int_0^{\tau}\zeta_2(t)\widetilde{\mbf{X}}(t)d\Lambda^*(t), \quad \zeta_{4\Lambda}(t) = \lambda^*(t)\zeta_2(t).
\end{gather*}
Thus, for any $\mbf{h}^{(1)}, \mbf{h}^{(2)}\in \mathbb{R}^{p}\times L_2([0,\tau])$, 
\begin{gather*}
\mathbb{P}\dot{\ell}^*[\mbf{h}^{(1)}]\dot{\ell}^*[\mbf{h}^{(2)}] = \mbf{\zeta}_{5\mbf{\beta}}(\mbf{h}^{(2)})^{\trans}\mbf{h}^{(1)}_{\mbf{\beta}}+\int_0^{\tau}\zeta_{5\Lambda}(t;\mbf{h}^{(2)})h^{(1)}_{\Lambda}(t)dt, \\
\widetilde{\mathbb{P}}\dot{\psi}^*[\mbf{h}^{(1)}]\dot{\psi}^*[\mbf{h}^{(2)}] = \mbf{\zeta}_{6\mbf{\beta}}(\mbf{h}^{(2)})^{\trans}\mbf{h}^{(1)}_{\mbf{\beta}}+\int_0^{\tau}\zeta_{6\Lambda}(t;\mbf{h}^{(2)})h^{(1)}_{\Lambda}(t)dt,
\end{gather*}
where 
\begin{gather*}
\mbf{\zeta}_{5\mbf{\beta}}(\mbf{h}) = \mathbb{P}\mbf{\zeta}_{3\mbf{\beta}}\left\{\mbf{\zeta}_{3\mbf{\beta}}^{\trans}\mbf{h}_{\mbf{\beta}}+\int_0^{\tau}\zeta_{3\Lambda}(t)h_{\Lambda}(t)dt\right\}, \\
\zeta_{5\Lambda}(t;\mbf{h}) = \mathbb{P}\zeta_{3\Lambda}(t)\left\{\mbf{\zeta}_{3\mbf{\beta}}^{\trans}\mbf{h}_{\mbf{\beta}}+\int_0^{\tau}\zeta_{3\Lambda}(s)h_{\Lambda}(s)ds\right\}, \\
\mbf{\zeta}_{6\mbf{\beta}}(\mbf{h}) = \widetilde{\mathbb{P}}\mbf{\zeta}_{4\mbf{\beta}}\left\{\mbf{\zeta}_{4\mbf{\beta}}^{\trans}\mbf{h}_{\mbf{\beta}}+\int_0^{\tau}\zeta_{4\Lambda}(t)h_{\Lambda}(t)dt\right\}, \\
\zeta_{6\Lambda}(t;\mbf{h}) = \widetilde{\mathbb{P}}\zeta_{4\Lambda}(t)\left\{\mbf{\zeta}_{4\mbf{\beta}}^{\trans}\mbf{h}_{\mbf{\beta}}+\int_0^{\tau}\zeta_{4\Lambda}(s)h_{\Lambda}(s)ds\right\}.
\end{gather*} 

Let $\mathbb{H} = \mathbb{R}^{p}\times L_2([0,\tau])$. We equip $\mathbb{H}$ with the inner product 
\[
\langle\mbf{h}^{(1)}, \mbf{h}^{(2)}\rangle = {\mbf{h}_{\mbf{\beta}}^{(1)}}^{\trans}{\mbf{h}_{\mbf{\beta}}^{(2)}}+\int_0^{\tau} h_{\Lambda}^{(1)}(t)h_{\Lambda}^{(2)}(t)dt.
\]
Clearly, $(\mathbb{H}, \|\cdot\|)$ is a Banach space, where $\|\cdot\|$ is the norm induced by the above inner product. 
Define the seminorm $\|\mbf{h}\|_{\ell} = \{\mathbb{P}\dot{\ell}^*[\mbf{h}]^2\}^{1/2}$.
We first show that $\|\cdot\|_{\ell}$ is a norm on the space $\mathbb{H}$. If $\|\mbf{h}\|_{\ell}=0$, then $\dot{\ell}^*[\mbf{h}]=0$ almost surely. 
By the expression of $\dot{\ell}^*[\mbf{h}]$ in \eqref{score_func}, we can set $(Y, \Delta)=(t,0)$ to obtain 
\[
\int_0^{t}G^{\prime}\{\int_0^t e^{{\mbf{\beta}^*}^{\trans}\mbf{X}(s)}d\Lambda^*(s)\}e^{{\mbf{\beta}^*}^{\trans}\mbf{X}(s)}\{\mbf{X}(s)^{\trans}\mbf{h}_{\mbf{\beta}}+h_{\Lambda}(s)\}d\Lambda^*(s) = 0,
\]
which further entails $\int_0^{t}e^{{\mbf{\beta}^*}^{\trans}\mbf{X}(s)}\{\mbf{X}(s)^{\trans}\mbf{h}_{\mbf{\beta}}+h_{\Lambda}(s)\}d\Lambda^*(s) = 0$ by Condition~\ref{transformation}.
Since $\lambda^*(t)$ is positive over $[0,\tau]$ under Condition~\ref{identifiability}, differentiating both sides with respect to $t$ yields $\mbf{X}(t)^{\trans}\mbf{h}_{\mbf{\beta}}+h_{\Lambda}(t)=0$, which holds for every $t\in[0,\tau]$.
By Condition~\ref{X_cont_diff}, $\mbf{h}_{\mbf{\beta}}=\mbf{0}$ and $h_{\Lambda}\equiv0$. 
Thus, $\|\cdot\|_{\ell}$ is a norm on $\mathbb{H}$.

Next, we show that $(\mathbb{H}, \|\cdot\|_{\ell})$ is a Banach space. It suffices to show the completeness of the space. 
Let $\{\mbf{h}^{(n)}\}_{n=1}^{\infty}$ be any arbitrary Cauchy sequence in $(\mathbb{H}, \|\cdot\|_{\ell})$. Then $\|\mbf{h}^{(n)}-\mbf{h}^{(m)}\|_{\ell}\overset{a.s.}{\rightarrow}0$, which implies that $\dot{\ell}^*[\mbf{h}^{(n)}-\mbf{h}^{(m)}]\overset{a.s.}{\rightarrow}0$.
Letting $(Y, \Delta)=(t,0)$, we have 
\[
\int_0^{t}\zeta_1(s)\left[\mbf{X}(s)^{\trans}\{\mbf{h}^{(n)}_{\mbf{\beta}}-\mbf{h}^{(m)}_{\mbf{\beta}}\}+\{h^{(n)}_{\Lambda}(s)-h^{(m)}_{\Lambda}(s)\}\right]d\Lambda^*(s)\overset{a.s.}{\rightarrow}0,
\]
for any $t\in[0,\tau]$.
It is clear that $\mbf{h}^{(n)}_{\mbf{\beta}}$ converges to some $\mbf{h}_{\mbf{\beta}}\in\mathbb{R}^p$, so we only need to show that $h^{(n)}_{\Lambda}$ converges to some $h_{\Lambda}\in L_2([0,\tau])$.
Define $F_n(t) = \int_0^t\zeta_1(s)h^{(n)}_{\Lambda}(s)d\Lambda^*(s)$, which is a compact operator for $h^{(n)}_{\Lambda}$ and thus converges to some function in $L_2([0,\tau])$.
Note that
\[
F_n(t) = \frac{\zeta_1(t)^2}{\zeta^{\prime}_1(t)}\frac{d}{dt}\frac{\int_0^t\{\zeta_1(t)-\zeta_1(s)\}h^{(n)}_{\Lambda}(s)d\Lambda^*(s)}{\zeta_1(t)},
\]
and $\zeta_1(t)$ and $\zeta^{\prime}_1(t)$ are uniformly bounded in $[0,\tau]$ under Conditions~\ref{X_cont_diff} and \ref{transformation}.
Thus, $h^{(n)}_{\Lambda}$ converges to some $h_{\Lambda}\in L_2([0,\tau])$.
Therefore, $(\mathbb{H}, \|\cdot\|_{\ell})$ is a Banach space.

It is easy to observe that $\|\mbf{h}\|_{\ell}\le c_1\|\mbf{h}\|$ for some positive constant $c_1$. By the open mapping theorem in Banach space, there exists some positive constant $c_2$ such that $\|\mbf{h}\|^2_{\ell}\ge c_2\|\mbf{h}\|^2$.
Thus, $\mathbb{P}\dot{\ell}^*[\mbf{h}]^2\ge c_2(\|\mbf{h}_{\mbf{\beta}}\|^2+\|h_{\Lambda}\|_{L_2}^2)$.
Since all the above arguments are still valid when replacing $(\mbf{\beta}^*, \Lambda^*)$ by $(\mbf{\beta}_0, \Lambda_0)$, the second inequality of the lemma follows. 
In addition, by similar arguments, we can show that $\widetilde{\mathbb{P}}\dot{\psi}^*[\mbf{h}]^2\ge c_3(\|\mbf{h}_{\mbf{\beta}}\|^2+\|h_{\Lambda}\|_{L_2}^2)$ for some positive constant $c_3$.
Combining the above results, we conclude that 
\[
\mathbb{P}\dot{\ell}^*[\mbf{h}]^2+\xi\widetilde{\mathbb{P}}\dot{\psi}^*[\mbf{h}]^2\ge c_4(1+\xi)(\|\mbf{h}_{\mbf{\beta}}\|^2+\|h_{\Lambda}\|_{L_2}^2)
\]
for some positive constant $c_4$.
\end{proof}

\section{Justification for the EM algorithm}
\label{sec:justification_for_the_em_algorithm}
We show that in the presence of individual weights, the observed-data likelihood still increases over each iteration in the EM algorithm. 
Let $O_i$, $Z_i$, and $w_i$ denote the observed data, missing data, and weight for the $i$th subject ($i=1,\dots,n$), respectively.
The weighted log-likelihood function for the unknown parameter $\theta$ is 
\begin{equation} \label{log-likelihood}
\ell(\theta) = \sum_{i=1}^n w_i\log p(O_i| \theta)
=\sum_{i=1}^n w_i\bigl\{\log p(O_i, Z_i| \theta)-\log p(Z_i| O_i, \theta)\bigr\},
\end{equation}
where $p(O_i| \theta)$, $p(O_i, Z_i| \theta)$, and $p(Z_i| O_i, \theta)$ denote the observed-data, complete-data, and conditional likelihood functions, respectively, for the $i$th subject.
Let $\theta^{(t)}$ denote the $\theta$ estimator at the $t$th iteration. 
Then at the $(t+1)$th iteration, we take the conditional expectation over $Z_i$ given $O_i$ and $\theta^{(t)}$ on both sides of \eqref{log-likelihood} to obtain
\begin{align*}
\ell(\theta) ={} & \sum_{i=1}^n w_i \bigl[E_{Z|O,\theta^{(t)}}\{\log p(O_i, Z_i| \theta)\}-E_{Z|O,\theta^{(t)}}\{\log p(Z_i| O_i, \theta)\}\bigr] \\
={} & E_{Z|O,\theta^{(t)}}\{\ell_c(\theta)\}-\sum_{i=1}^n w_iE_{Z|O,\theta^{(t)}}\{\log p(Z_i| O_i, \theta)\},
\end{align*}
where $\ell_c(\theta)$ is the weighted complete-data log-likelihood. 
The above equation holds for every $\theta$.
Thus, we can plug in $\theta^{(t)}$ and $\theta^{(t+1)}$; taking the difference between the resulting two equations yields
\[
\ell(\theta^{(t+1)})-\ell(\theta^{(t)}) = 
\begin{multlined}[t]
E_{Z|O,\theta^{(t)}}\{\ell_c(\theta^{(t+1)})\}-E_{Z|O,\theta^{(t)}}\{\ell_c(\theta^{(t)})\} \\
-\sum_{i=1}^n w_iE_{Z|O,\theta^{(t)}}\left\{\log \frac{p(Z_i| O_i, \theta^{(t+1)})}{p(Z_i| O_i, \theta^{(t)})}\right\}.
\end{multlined}
\]
Since $\log$ is a concave function, by Jensen's inequality,
\begin{align*}
& -\sum_{i=1}^n w_iE_{Z|O,\theta^{(t)}}\left\{\log \frac{p(Z_i| O_i, \theta^{(t+1)})}{p(Z_i| O_i, \theta^{(t)})}\right\} \\
\ge{} & -\sum_{i=1}^n w_i \log E_{Z|O,\theta^{(t)}}\left\{\frac{p(Z_i| O_i, \theta^{(t+1)})}{p(Z_i| O_i, \theta^{(t)})}\right\} \\
={} & -\sum_{i=1}^n w_i\log \left\{\int_{Z_i}\frac{p(Z_i| O_i, \theta^{(t+1)})}{p(Z_i| O_i, \theta^{(t)})} \times p(Z_i| O_i, \theta^{(t)})dZ_i\right\} \\
={} & 0.
\end{align*}
Thus, 
\[
\ell(\theta^{(t+1)})-\ell(\theta^{(t)})\ge E_{Z|O,\theta^{(t)}}\{\ell_c(\theta^{(t+1)})\}-E_{Z|O,\theta^{(t)}}\{\ell_c(\theta^{(t)})\} \ge 0,
\]
where the second inequality follows by the fact that $\theta^{(t+1)}$ maximizes the $E_{Z|O,\theta^{(t)}}\{\ell_c(\theta)\}$.
Therefore, the weighted observed-data likelihood is guaranteed to increase over each iteration of the EM algorithm.

\section{Additional Simulation Studies}
\label{sec:additional_simulation_studies}
\subsection{Covariate Shift}
\label{sub:covariate_shift}
\begin{figure}
\centerline{%
\includegraphics[width=\textwidth]{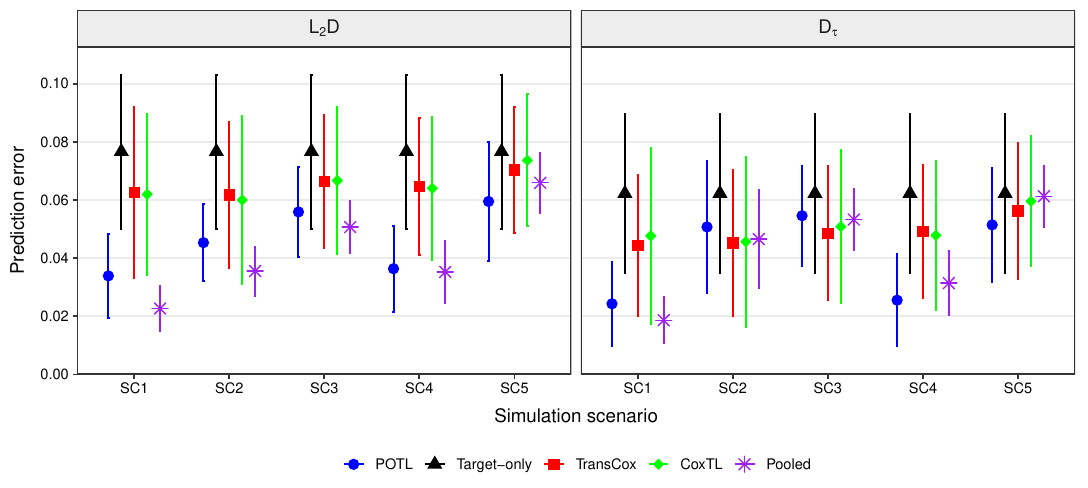}}
\caption{\revise{Simulation results for the L$_2$D and D$_\tau$ metrics under covariate shift between target and source data. Points represent median over 200 replicates, and vertical bars represent median $\pm$ median absolute deviation.}}
\label{fig:sim_pred_error_source_shift}
\end{figure}

\begin{figure}
\centerline{%
\includegraphics[width=\textwidth]{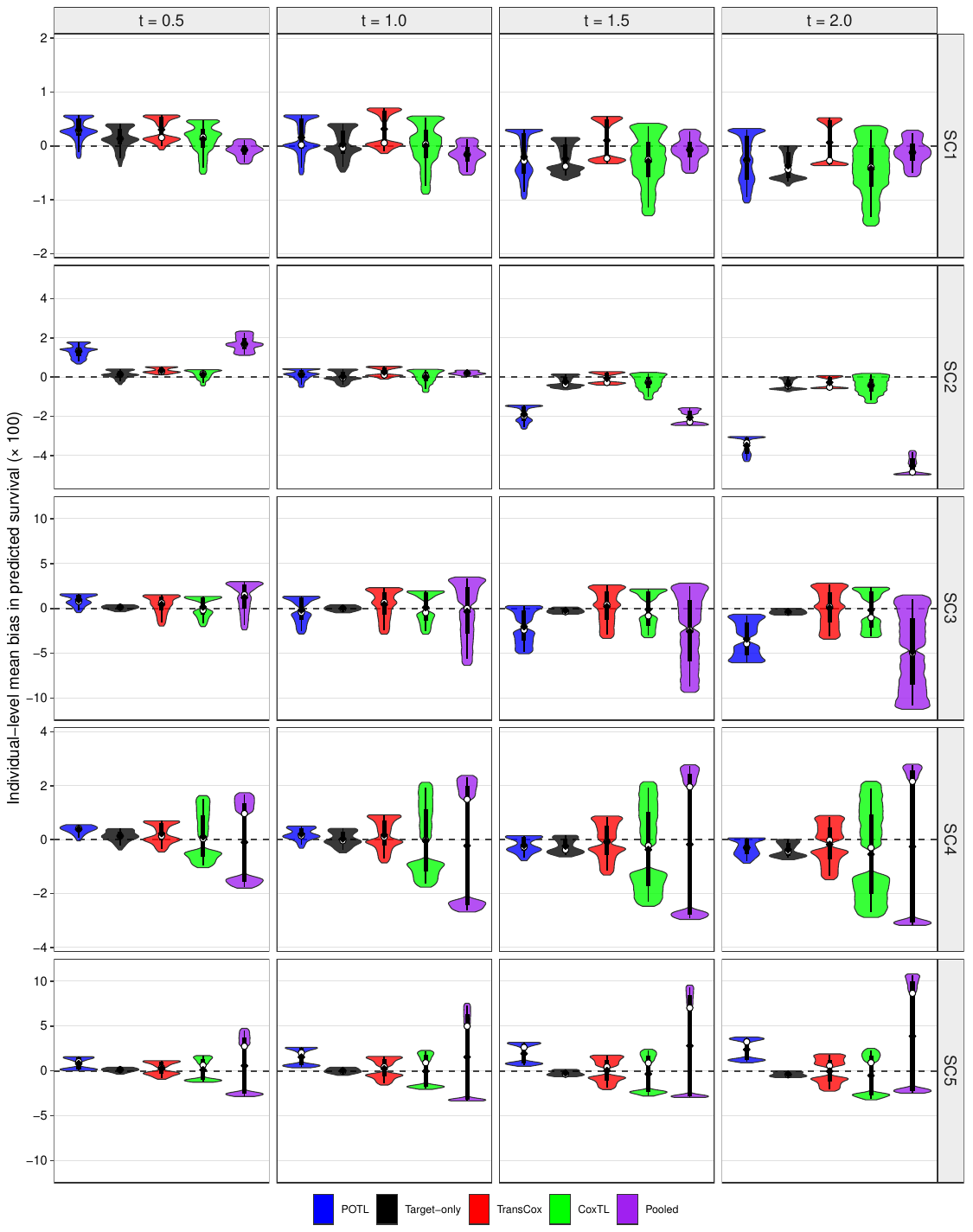}}
\caption{\revise{Violin plots of individual-level mean bias in predicted survival probabilities under covariate shift between target and source data. For each of 10,000 validation subjects, the mean signed bias is calculated across 200 replicates at prediction times $t=0.5$, $1.0$, $1.5$, and $2.0$. Biases are multiplied by 100 for display.}}
\label{fig:sim_indiv_bias_source_shift}
\end{figure}

\begin{figure}
\centerline{%
\includegraphics[width=\textwidth]{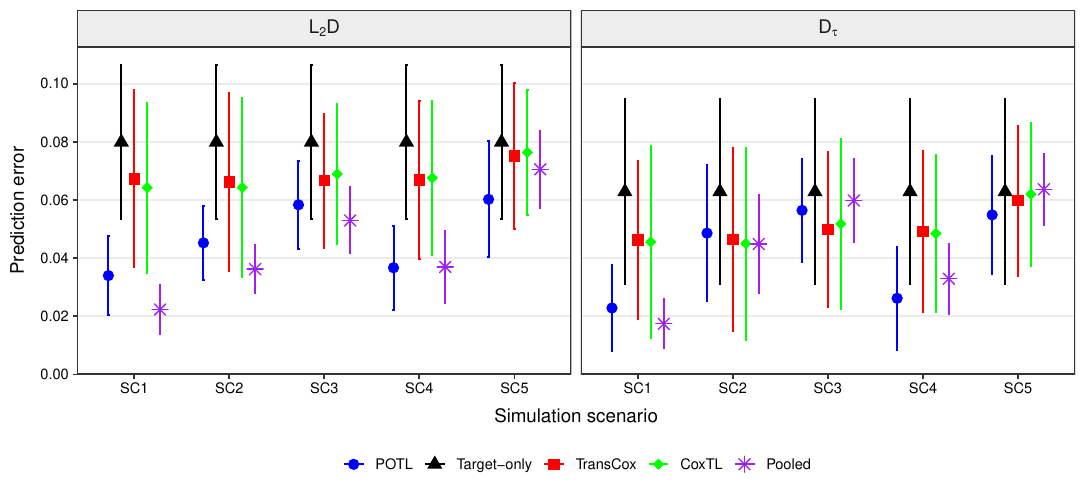}}
\caption{\revise{Simulation results for the L$_2$D and D$_\tau$ metrics under covariate shift between training and validation data. Points represent median over 200 replicates, and vertical bars represent median $\pm$ median absolute deviation.}}
\label{fig:sim_pred_error_validation_shift}
\end{figure}

\begin{figure}
\centerline{%
\includegraphics[width=\textwidth]{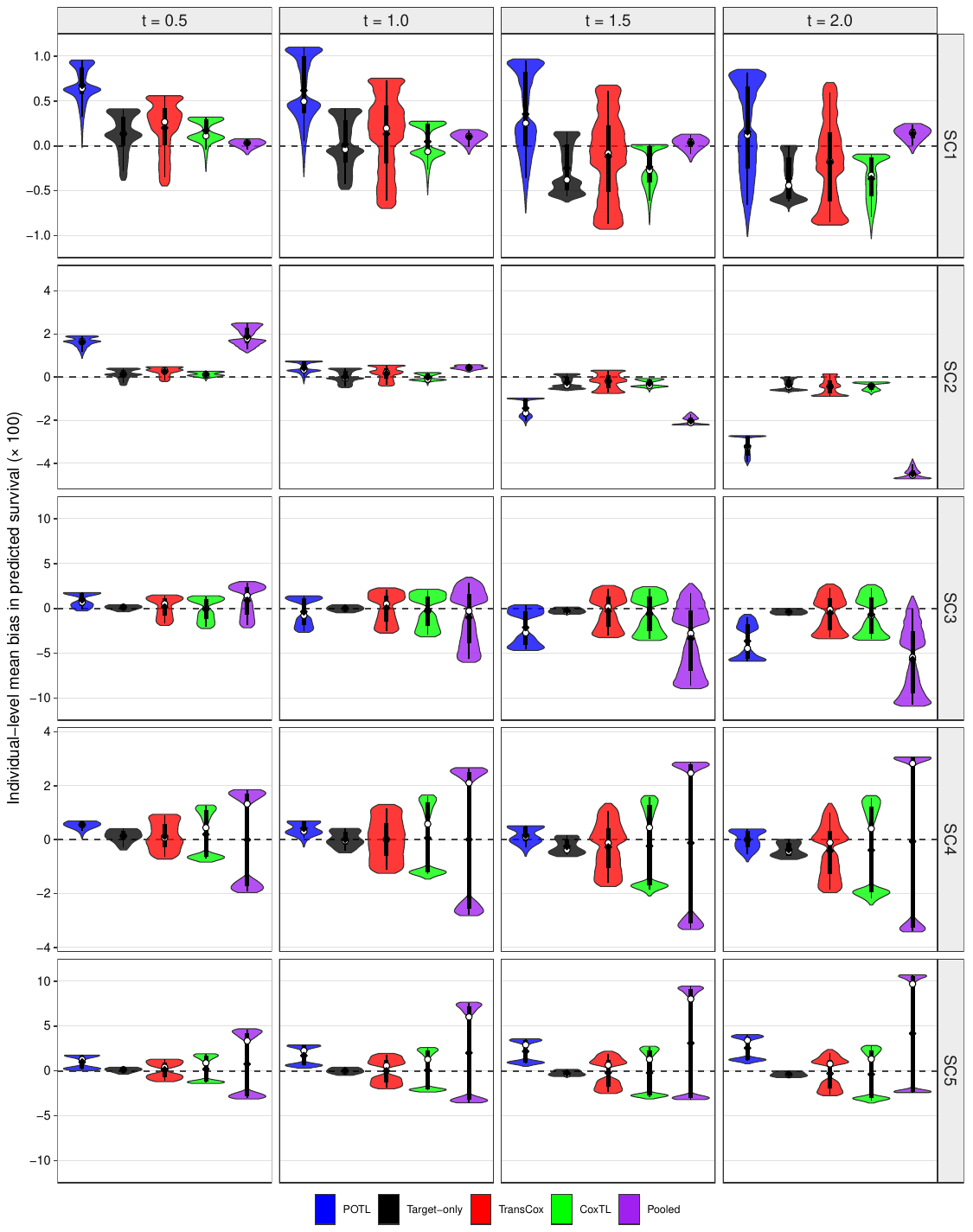}}
\caption{\revise{Violin plots of individual-level mean bias in predicted survival probabilities under covariate shift between training and validation data. For each of 10,000 validation subjects, the mean signed bias is calculated across 200 replicates at prediction times $t=0.5$, $1.0$, $1.5$, and $2.0$. Biases are multiplied by 100 for display.}}
\label{fig:sim_indiv_bias_validation_shift}
\end{figure}

Covariate shift often arises in practice due to heterogeneity in demographic and clinical characteristics (e.g., age and disease stage) across studies.
We conduct additional simulation studies to evaluate the performance of all five methods when covariate shift is present between the target and source data, as well as between the training and validation data.
Specifically, in separate simulations we generate the covariate $X_2$ from $\text{Beta}(1,2)$ for the source data only or for the validation data only while keeping all other settings unchanged.
\revise{Web Figures~\ref{fig:sim_pred_error_source_shift} and \ref{fig:sim_pred_error_validation_shift} display the simulation results for the L$_2$D and D$_\tau$ metrics, while Web Tables~\ref{tb:covar_shift_source} and \ref{tb:covar_shift_valid} provide the full results for all metrics. The distributions of the individual-level mean signed biases are shown in Web Figures~\ref{fig:sim_indiv_bias_source_shift} and \ref{fig:sim_indiv_bias_validation_shift}.}
The conclusions remain unchanged, demonstrating that POTL is robust to covariate shift.

\subsection{Different Covariate Sets}
\begin{figure}
\centerline{%
\includegraphics[width=\textwidth]{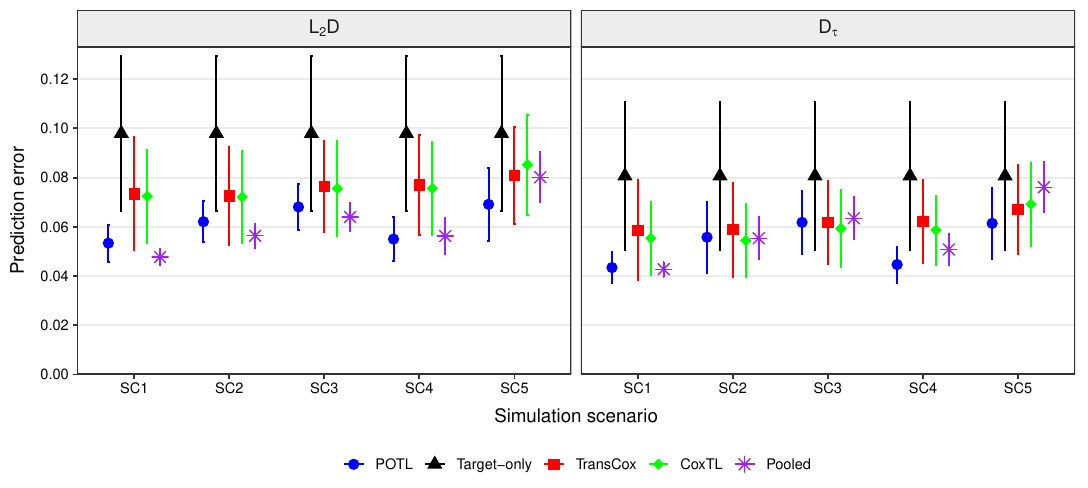}}
\caption{\revise{Simulation results for the L$_2$D and D$_\tau$ metrics when covariates differ between target and source studies. Points represent median over 200 replicates, and vertical bars represent median $\pm$ median absolute deviation.}}
\label{fig:sim_pred_error_diffcovar}
\end{figure}

\begin{figure}
\centerline{%
\includegraphics[width=\textwidth]{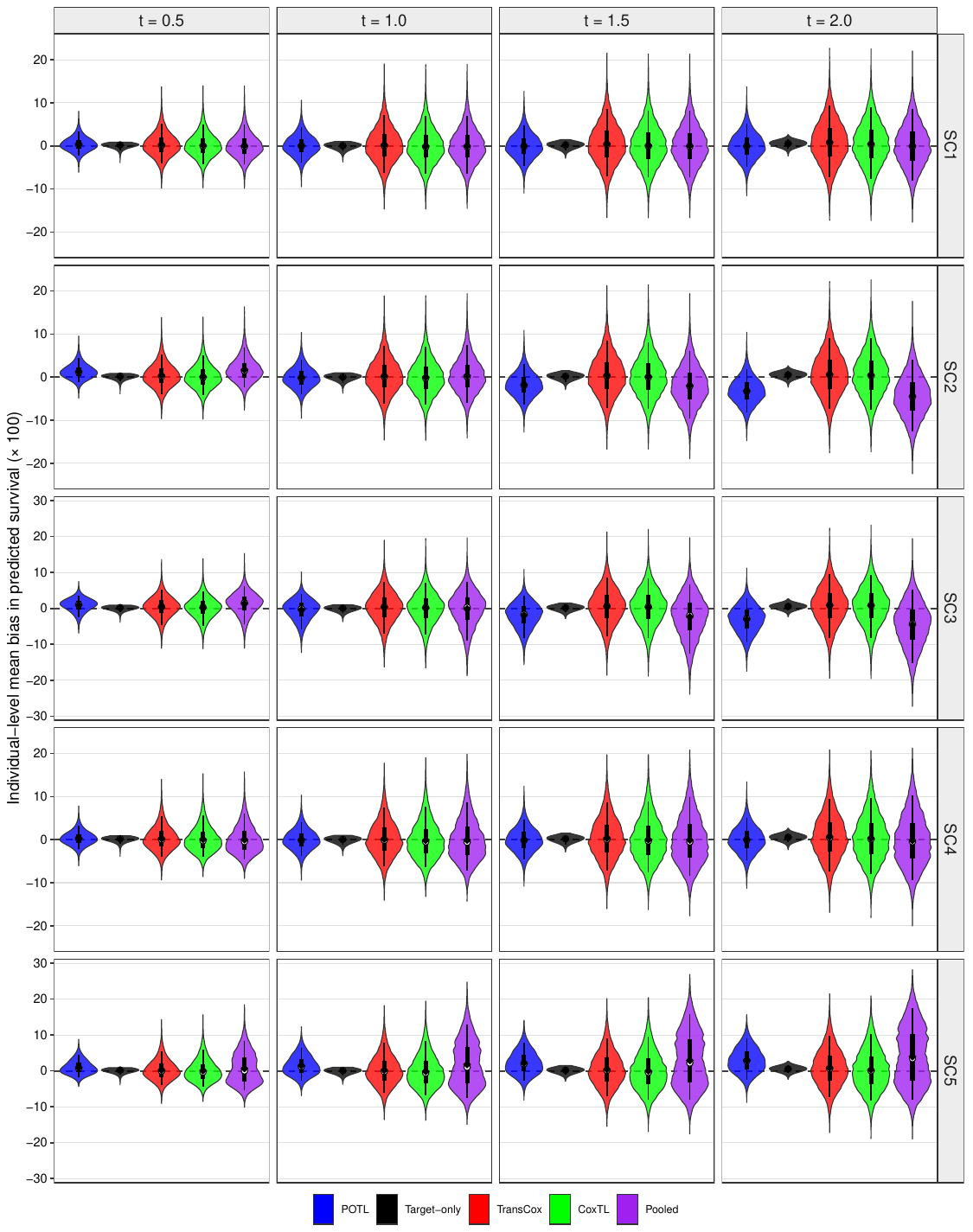}}
\caption{\revise{Violin plots of individual-level mean bias in predicted survival probabilities when covariates differ between target and source studies. For each of 10,000 validation subjects, the mean signed bias is calculated across 200 replicates at prediction times $t=0.5$, $1.0$, $1.5$, and $2.0$. Biases are multiplied by 100 for display.}}
\label{fig:sim_indiv_bias_diffcovar}
\end{figure}

We consider another common situation where the covariate sets differ between the target and source studies. 
For the target population, we generate four independent covariates: $X_1\sim\text{Ber}(0.5)$ and $X_2, X_3, X_4\sim\text{Unif}(0, 1)$, and consider an expanded Cox model:
\[
  \Lambda(t|X_1, X_2, X_3, X_4) = \Lambda(t)\exp(\beta_1X_1+\beta_2X_2+\beta_3X_3+\beta_4X_4),
\]
where $\Lambda(t) = \log(1+0.5t)$, and $(\beta_1,\beta_2,\beta_3,\beta_4)  = (0.5, -0.5, -0.5, 0.5)$.
For the POTL and target-only methods, we include all covariates available in each study when fitting the corresponding target and source models. 
For TransCox, CoxTL, and pooled analysis, which require the target and source studies to share the same covariate set, we include only $X_1$ and $X_2$ in the models. 
\revise{Web Figure~\ref{fig:sim_pred_error_diffcovar} and Web Table~\ref{tb:diff_covar} present the results for the five population-level metrics. POTL performs comparably to pooled analysis in SC1--SC3 and achieves the best performance in SC4--SC5. Web Figure~\ref{fig:sim_indiv_bias_diffcovar} displays the distributions of the individual-level mean signed biases. Compared to TransCox, CoxTL, and pooled analysis, POTL generally has substantially narrower bias distributions across scenarios and prediction times, indicating more stable individual-level calibration.}
These results further demonstrate the broad applicability of the proposed method to real-world studies with heterogeneous covariate sets.

\subsection{Multiple Sources}
\label{sub:multiple_sources}
\begin{figure}
\centerline{%
\includegraphics[width=\textwidth]{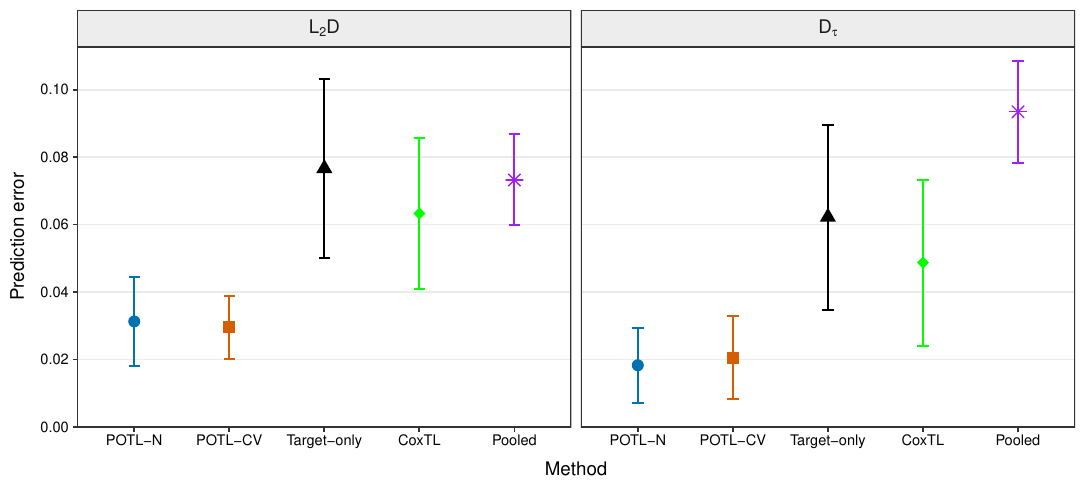}}
\caption{\revise{Multi-source simulation results for the L$_2$D and D$_\tau$ metrics. Points represent median over 200 replicates, and vertical bars represent median $\pm$ median absolute deviation. For POTL-N, source weights are set proportional to the sample size of each source study; for POTL-CV, source weights are selected using 5-fold cross-validation.}}
\label{fig:sim_pred_error_multi_source}
\end{figure}

\begin{figure}
\centerline{%
\includegraphics[width=\textwidth]{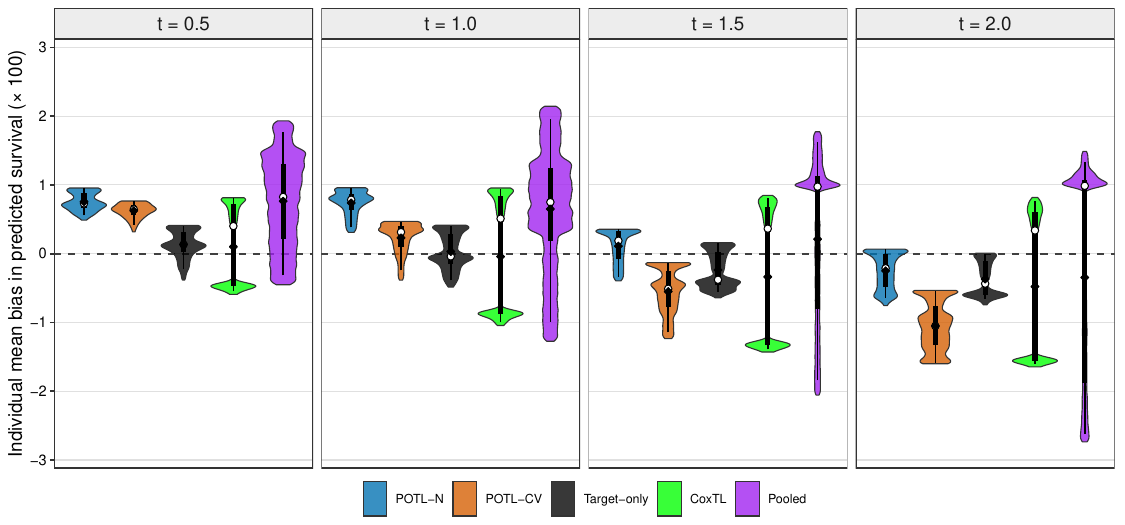}}
\caption{\revise{Multi-source violin plots of individual-level mean bias in predicted survival probabilities. For each of 10,000 validation subjects, the mean signed bias is calculated across 200 replicates at prediction times $t=0.5$, $1.0$, $1.5$, and $2.0$. Biases are multiplied by 100 for display. For POTL-N, source weights are set proportional to the sample size of each source study; for POTL-CV, source weights are selected using 5-fold cross-validation.}}
\label{fig:sim_indiv_bias_multi_source}
\end{figure}

We assess the performance of each method in the presence of multiple source studies. Specifically, we consider five source studies corresponding to SC1--SC5 in Section~\ref{m-sec:simulation} of the main manuscript, with sample sizes $(200,300,400,500,600)$ and study durations $(4,4,5,6,6)$, respectively. For POTL, we consider two weighting strategies: prespecified source weights proportional to the sample size of each source study, and data-driven source weights selected by 5-fold cross-validation. Because TransCox does not accommodate multiple source studies, it is excluded from this comparison. 
\revise{Web Figure~\ref{fig:sim_pred_error_multi_source} and Web Table~\ref{tb:sim_ms} present the results for the five population-level metrics. The two source weighting strategies for POTL yield similar results, and both substantially outperform all competing methods in terms of the L$_2$D and D$_\tau$ metrics. Web Figure~\ref{fig:sim_indiv_bias_multi_source} displays the distributions of the individual-level mean signed biases. Both weighting strategies for POTL yield small individual-level biases and substantially narrower bias distributions than CoxTL and pooled analysis. Together, these results demonstrate the advantages of POTL in multi-source settings.}

\begin{table}
\caption{\revise{Simulation results for all methods based on various metrics. Median among 200 replicates is reported, with median absolute deviation shown in parentheses. SC4$^{\prime}$--SC5$^{\prime}$ are identical to SC4--SC5, respectively, except that POTL uses a standard Cox model for source prediction.}}%
\begin{tabular}{ll*{5}{r}}
\Hline
SC & Metric & POTL & Target-only & TransCox & CoxTL & Pooled \\
\hline
1 & L$_2$D & 0.031 (0.014) & 0.077 (0.026) & 0.061 (0.028) & 0.064 (0.030) & 0.021 (0.007) \\ 
   & D$_\tau$ & 0.022 (0.015) & 0.062 (0.027) & 0.046 (0.027) & 0.048 (0.032) & 0.016 (0.007) \\ 
   & C-index & 0.581 (0.004) & 0.578 (0.007) & 0.581 (0.004) & 0.581 (0.004) & 0.581 (0.003) \\ 
   & IBS & 0.190 (0.002) & 0.193 (0.003) & 0.192 (0.002) & 0.192 (0.003) & 0.190 (0.001) \\ 
   & RMST & 0.650 (0.007) & 0.652 (0.012) & 0.651 (0.013) & 0.652 (0.012) & 0.651 (0.005) \\ 
  \hline
  2 & L$_2$D & 0.043 (0.012) & 0.077 (0.026) & 0.061 (0.025) & 0.063 (0.031) & 0.035 (0.008) \\ 
   & D$_\tau$ & 0.046 (0.017) & 0.062 (0.027) & 0.043 (0.028) & 0.047 (0.033) & 0.046 (0.013) \\ 
   & C-index & 0.581 (0.004) & 0.578 (0.007) & 0.581 (0.004) & 0.581 (0.004) & 0.581 (0.003) \\ 
   & IBS & 0.191 (0.002) & 0.193 (0.003) & 0.192 (0.002) & 0.192 (0.002) & 0.190 (0.001) \\ 
   & RMST & 0.652 (0.007) & 0.652 (0.012) & 0.652 (0.011) & 0.651 (0.012) & 0.653 (0.005) \\ 
  \hline
  3 & L$_2$D & 0.055 (0.014) & 0.077 (0.026) & 0.065 (0.023) & 0.067 (0.026) & 0.049 (0.009) \\ 
   & D$_\tau$ & 0.052 (0.016) & 0.062 (0.027) & 0.048 (0.023) & 0.050 (0.026) & 0.052 (0.010) \\ 
   & C-index & 0.581 (0.004) & 0.578 (0.007) & 0.581 (0.004) & 0.581 (0.004) & 0.581 (0.003) \\ 
   & IBS & 0.191 (0.002) & 0.193 (0.003) & 0.192 (0.002) & 0.192 (0.003) & 0.191 (0.001) \\ 
   & RMST & 0.649 (0.007) & 0.652 (0.012) & 0.648 (0.010) & 0.648 (0.013) & 0.649 (0.005) \\ 
  \hline
  4 & L$_2$D & 0.034 (0.014) & 0.077 (0.026) & 0.069 (0.024) & 0.065 (0.025) & 0.037 (0.012) \\ 
   & D$_\tau$ & 0.025 (0.015) & 0.062 (0.027) & 0.048 (0.022) & 0.049 (0.024) & 0.033 (0.012) \\ 
   & C-index & 0.581 (0.004) & 0.578 (0.007) & 0.579 (0.006) & 0.579 (0.005) & 0.579 (0.005) \\ 
   & IBS & 0.190 (0.002) & 0.193 (0.003) & 0.192 (0.002) & 0.192 (0.002) & 0.190 (0.001) \\ 
   & RMST & 0.651 (0.007) & 0.652 (0.012) & 0.652 (0.011) & 0.655 (0.013) & 0.657 (0.005) \\ 
  \hline
  5 & L$_2$D & 0.056 (0.018) & 0.077 (0.026) & 0.071 (0.023) & 0.075 (0.020) & 0.069 (0.013) \\ 
   & D$_\tau$ & 0.051 (0.016) & 0.062 (0.027) & 0.057 (0.022) & 0.060 (0.021) & 0.064 (0.012) \\ 
   & C-index & 0.578 (0.006) & 0.578 (0.007) & 0.578 (0.008) & 0.575 (0.011) & 0.566 (0.012) \\ 
   & IBS & 0.192 (0.002) & 0.193 (0.003) & 0.193 (0.002) & 0.193 (0.002) & 0.192 (0.002) \\ 
   & RMST & 0.650 (0.007) & 0.652 (0.012) & 0.654 (0.013) & 0.657 (0.014) & 0.657 (0.005) \\ 
   \hline
   4$^{\prime}$ & L$_2$D & 0.048 (0.018) & 0.077 (0.026) & 0.069 (0.024) & 0.065 (0.025) & 0.037 (0.012) \\ 
   & D$_\tau$ & 0.039 (0.017) & 0.062 (0.027) & 0.048 (0.022) & 0.049 (0.024) & 0.033 (0.012) \\ 
   & C-index & 0.578 (0.006) & 0.578 (0.007) & 0.579 (0.006) & 0.579 (0.005) & 0.579 (0.005) \\ 
   & IBS & 0.191 (0.002) & 0.193 (0.003) & 0.192 (0.002) & 0.192 (0.002) & 0.190 (0.001) \\ 
   & RMST & 0.654 (0.007) & 0.652 (0.012) & 0.652 (0.011) & 0.655 (0.013) & 0.657 (0.005) \\
   \hline 
  5$^{\prime}$ & L$_2$D & 0.072 (0.019) & 0.077 (0.026) & 0.071 (0.023) & 0.075 (0.020) & 0.069 (0.013) \\ 
   & D$_\tau$ & 0.062 (0.021) & 0.062 (0.027) & 0.057 (0.022) & 0.060 (0.021) & 0.064 (0.012) \\ 
   & C-index & 0.573 (0.013) & 0.578 (0.007) & 0.578 (0.008) & 0.575 (0.011) & 0.566 (0.012) \\ 
   & IBS & 0.193 (0.002) & 0.193 (0.003) & 0.193 (0.002) & 0.193 (0.002) & 0.192 (0.002) \\ 
   & RMST & 0.653 (0.008) & 0.652 (0.012) & 0.654 (0.013) & 0.657 (0.014) & 0.657 (0.005) \\ 
   \hline
\end{tabular}
\label{tb:main_sim}
\end{table}

\begin{table} 
\caption{\revise{Simulation results with covariate shift between target and source data. Median among 200 replicates is reported, with median absolute deviation shown in parentheses.}}
\label{tb:covar_shift_source}
\begin{tabular}{ll*{5}{r}}
\Hline
SC & Metric & POTL & Target-only & TransCox & CoxTL & Pooled \\
\hline
1 & L$_2$D & 0.034 (0.015) & 0.077 (0.026) & 0.063 (0.029) & 0.062 (0.028) & 0.023 (0.008) \\ 
   & D$_\tau$ & 0.024 (0.014) & 0.062 (0.027) & 0.044 (0.024) & 0.048 (0.030) & 0.019 (0.008) \\ 
   & C-index & 0.581 (0.004) & 0.578 (0.007) & 0.580 (0.004) & 0.581 (0.004) & 0.581 (0.003) \\ 
   & IBS & 0.191 (0.002) & 0.193 (0.003) & 0.192 (0.002) & 0.192 (0.002) & 0.190 (0.001) \\ 
   & RMST & 0.651 (0.007) & 0.652 (0.012) & 0.650 (0.012) & 0.652 (0.012) & 0.652 (0.005) \\ 
  \hline
  2 & L$_2$D & 0.045 (0.013) & 0.077 (0.026) & 0.062 (0.025) & 0.060 (0.029) & 0.036 (0.008) \\ 
   & D$_\tau$ & 0.051 (0.023) & 0.062 (0.027) & 0.045 (0.025) & 0.046 (0.029) & 0.047 (0.017) \\ 
   & C-index & 0.581 (0.004) & 0.578 (0.007) & 0.580 (0.004) & 0.581 (0.004) & 0.581 (0.003) \\ 
   & IBS & 0.191 (0.002) & 0.193 (0.003) & 0.192 (0.002) & 0.192 (0.002) & 0.190 (0.001) \\ 
   & RMST & 0.653 (0.008) & 0.652 (0.012) & 0.651 (0.012) & 0.653 (0.012) & 0.653 (0.005) \\ 
  \hline
  3 & L$_2$D & 0.056 (0.015) & 0.077 (0.026) & 0.066 (0.023) & 0.067 (0.025) & 0.051 (0.009) \\ 
   & D$_\tau$ & 0.055 (0.017) & 0.062 (0.027) & 0.049 (0.023) & 0.051 (0.026) & 0.053 (0.011) \\ 
   & C-index & 0.580 (0.005) & 0.578 (0.007) & 0.581 (0.004) & 0.581 (0.004) & 0.581 (0.003) \\ 
   & IBS & 0.192 (0.002) & 0.193 (0.003) & 0.192 (0.002) & 0.192 (0.002) & 0.191 (0.001) \\ 
   & RMST & 0.649 (0.008) & 0.652 (0.012) & 0.647 (0.012) & 0.648 (0.013) & 0.649 (0.005) \\ 
  \hline
  4 & L$_2$D & 0.036 (0.015) & 0.077 (0.026) & 0.065 (0.023) & 0.064 (0.025) & 0.035 (0.011) \\ 
   & D$_\tau$ & 0.026 (0.016) & 0.062 (0.027) & 0.049 (0.023) & 0.048 (0.026) & 0.031 (0.011) \\ 
   & C-index & 0.580 (0.004) & 0.578 (0.007) & 0.580 (0.004) & 0.580 (0.005) & 0.580 (0.005) \\ 
   & IBS & 0.191 (0.002) & 0.193 (0.003) & 0.192 (0.002) & 0.192 (0.002) & 0.190 (0.001) \\ 
   & RMST & 0.651 (0.007) & 0.652 (0.012) & 0.653 (0.013) & 0.656 (0.012) & 0.657 (0.006) \\ 
  \hline
  5 & L$_2$D & 0.059 (0.021) & 0.077 (0.026) & 0.070 (0.022) & 0.074 (0.023) & 0.066 (0.010) \\ 
   & D$_\tau$ & 0.051 (0.020) & 0.062 (0.027) & 0.056 (0.023) & 0.060 (0.022) & 0.061 (0.011) \\ 
   & C-index & 0.579 (0.005) & 0.578 (0.007) & 0.578 (0.006) & 0.576 (0.010) & 0.569 (0.012) \\ 
   & IBS & 0.192 (0.002) & 0.193 (0.003) & 0.193 (0.002) & 0.193 (0.002) & 0.192 (0.002) \\ 
   & RMST & 0.650 (0.008) & 0.652 (0.012) & 0.653 (0.012) & 0.657 (0.014) & 0.657 (0.005) \\ 
   \hline
\end{tabular}
\end{table}

\begin{table} 
\caption{\revise{Simulation results with covariate shift between training and validation data. Median among 200 replicates is reported, with median absolute deviation shown in parentheses.}}
\label{tb:covar_shift_valid}
\begin{tabular}{ll*{5}{r}}
\Hline
SC & Metric & POTL & Target-only & TransCox & CoxTL & Pooled \\
\hline
1 & L$_2$D & 0.034 (0.014) & 0.080 (0.027) & 0.067 (0.030) & 0.064 (0.029) & 0.022 (0.009) \\ 
   & D$_\tau$ & 0.023 (0.015) & 0.063 (0.032) & 0.046 (0.027) & 0.046 (0.033) & 0.017 (0.008) \\ 
   & C-index & 0.577 (0.003) & 0.575 (0.006) & 0.577 (0.004) & 0.577 (0.003) & 0.577 (0.003) \\ 
   & IBS & 0.195 (0.002) & 0.198 (0.003) & 0.197 (0.003) & 0.197 (0.002) & 0.195 (0.001) \\ 
   & RMST & 0.662 (0.006) & 0.665 (0.011) & 0.663 (0.010) & 0.664 (0.010) & 0.663 (0.005) \\ 
  \hline
  2 & L$_2$D & 0.045 (0.013) & 0.080 (0.027) & 0.066 (0.031) & 0.064 (0.031) & 0.036 (0.008) \\ 
   & D$_\tau$ & 0.049 (0.023) & 0.063 (0.032) & 0.046 (0.032) & 0.045 (0.033) & 0.045 (0.017) \\ 
   & C-index & 0.577 (0.003) & 0.575 (0.006) & 0.577 (0.004) & 0.577 (0.003) & 0.577 (0.003) \\ 
   & IBS & 0.196 (0.002) & 0.198 (0.003) & 0.197 (0.002) & 0.197 (0.002) & 0.195 (0.001) \\ 
   & RMST & 0.664 (0.005) & 0.665 (0.011) & 0.663 (0.009) & 0.664 (0.010) & 0.664 (0.004) \\ 
  \hline
  3 & L$_2$D & 0.058 (0.015) & 0.080 (0.027) & 0.067 (0.023) & 0.069 (0.024) & 0.053 (0.011) \\ 
   & D$_\tau$ & 0.056 (0.018) & 0.063 (0.032) & 0.050 (0.027) & 0.052 (0.029) & 0.060 (0.014) \\ 
   & C-index & 0.577 (0.004) & 0.575 (0.006) & 0.577 (0.003) & 0.577 (0.003) & 0.578 (0.003) \\ 
   & IBS & 0.196 (0.002) & 0.198 (0.003) & 0.197 (0.002) & 0.197 (0.002) & 0.196 (0.001) \\ 
   & RMST & 0.662 (0.006) & 0.665 (0.011) & 0.660 (0.010) & 0.660 (0.011) & 0.662 (0.005) \\ 
  \hline
  4 & L$_2$D & 0.037 (0.014) & 0.080 (0.027) & 0.067 (0.027) & 0.068 (0.026) & 0.037 (0.012) \\ 
   & D$_\tau$ & 0.026 (0.018) & 0.063 (0.032) & 0.049 (0.028) & 0.048 (0.027) & 0.033 (0.012) \\ 
   & C-index & 0.577 (0.003) & 0.575 (0.006) & 0.576 (0.004) & 0.576 (0.004) & 0.576 (0.004) \\ 
   & IBS & 0.195 (0.002) & 0.198 (0.003) & 0.197 (0.003) & 0.197 (0.002) & 0.195 (0.001) \\ 
   & RMST & 0.663 (0.006) & 0.665 (0.011) & 0.665 (0.010) & 0.668 (0.011) & 0.669 (0.005) \\ 
  \hline
  5 & L$_2$D & 0.060 (0.020) & 0.080 (0.027) & 0.075 (0.025) & 0.076 (0.021) & 0.071 (0.013) \\ 
   & D$_\tau$ & 0.055 (0.020) & 0.063 (0.032) & 0.060 (0.026) & 0.062 (0.025) & 0.064 (0.012) \\ 
   & C-index & 0.576 (0.004) & 0.575 (0.006) & 0.575 (0.005) & 0.573 (0.008) & 0.565 (0.011) \\ 
   & IBS & 0.197 (0.002) & 0.198 (0.003) & 0.198 (0.003) & 0.198 (0.002) & 0.198 (0.002) \\ 
   & RMST & 0.663 (0.006) & 0.665 (0.011) & 0.668 (0.010) & 0.670 (0.011) & 0.670 (0.005) \\
   \hline
\end{tabular}
\end{table}

\begin{table} 
\caption{\revise{Simulation results with different covariates between target and source data. Median among 200 replicates is reported, with median absolute deviation shown in parentheses.}}
\label{tb:diff_covar}
\begin{tabular}{ll*{5}{r}}
\Hline
SC & Metric & POTL & Target-only & TransCox & CoxTL & Pooled \\
\hline
1 & L$_2$D & 0.053 (0.008) & 0.098 (0.032) & 0.073 (0.023) & 0.072 (0.019) & 0.048 (0.003) \\ 
   & D$_\tau$ & 0.043 (0.006) & 0.081 (0.030) & 0.059 (0.020) & 0.055 (0.015) & 0.043 (0.003) \\ 
   & C-index & 0.580 (0.005) & 0.575 (0.014) & 0.579 (0.004) & 0.579 (0.005) & 0.580 (0.004) \\ 
   & IBS & 0.191 (0.002) & 0.195 (0.004) & 0.192 (0.002) & 0.192 (0.002) & 0.190 (0.001) \\ 
   & RMST & 0.651 (0.006) & 0.651 (0.014) & 0.653 (0.013) & 0.654 (0.013) & 0.653 (0.005) \\ 
  \hline
  2 & L$_2$D & 0.062 (0.008) & 0.098 (0.032) & 0.073 (0.020) & 0.072 (0.019) & 0.056 (0.005) \\ 
   & D$_\tau$ & 0.056 (0.014) & 0.081 (0.030) & 0.059 (0.019) & 0.054 (0.015) & 0.055 (0.009) \\ 
   & C-index & 0.580 (0.005) & 0.575 (0.014) & 0.579 (0.005) & 0.579 (0.005) & 0.580 (0.004) \\ 
   & IBS & 0.191 (0.002) & 0.195 (0.004) & 0.192 (0.002) & 0.192 (0.002) & 0.191 (0.001) \\ 
   & RMST & 0.654 (0.006) & 0.651 (0.014) & 0.652 (0.012) & 0.654 (0.013) & 0.654 (0.005) \\ 
  \hline
  3 & L$_2$D & 0.068 (0.009) & 0.098 (0.032) & 0.076 (0.019) & 0.076 (0.019) & 0.064 (0.006) \\ 
   & D$_\tau$ & 0.062 (0.013) & 0.081 (0.030) & 0.062 (0.017) & 0.059 (0.016) & 0.063 (0.009) \\ 
   & C-index & 0.580 (0.005) & 0.575 (0.014) & 0.579 (0.004) & 0.579 (0.005) & 0.580 (0.004) \\ 
   & IBS & 0.192 (0.002) & 0.195 (0.004) & 0.192 (0.003) & 0.192 (0.002) & 0.191 (0.001) \\ 
   & RMST & 0.650 (0.006) & 0.651 (0.014) & 0.649 (0.014) & 0.649 (0.013) & 0.650 (0.005) \\ 
  \hline
  4 & L$_2$D & 0.055 (0.009) & 0.098 (0.032) & 0.077 (0.020) & 0.076 (0.019) & 0.056 (0.007) \\ 
   & D$_\tau$ & 0.045 (0.007) & 0.081 (0.030) & 0.062 (0.017) & 0.059 (0.014) & 0.051 (0.006) \\ 
   & C-index & 0.580 (0.005) & 0.575 (0.014) & 0.578 (0.005) & 0.578 (0.005) & 0.577 (0.004) \\ 
   & IBS & 0.191 (0.002) & 0.195 (0.004) & 0.192 (0.002) & 0.192 (0.002) & 0.191 (0.001) \\ 
   & RMST & 0.651 (0.006) & 0.651 (0.014) & 0.654 (0.013) & 0.657 (0.011) & 0.658 (0.005) \\ 
  \hline
  5 & L$_2$D & 0.069 (0.015) & 0.098 (0.032) & 0.081 (0.020) & 0.085 (0.020) & 0.080 (0.010) \\ 
   & D$_\tau$ & 0.061 (0.014) & 0.081 (0.030) & 0.067 (0.018) & 0.069 (0.017) & 0.076 (0.010) \\ 
   & C-index & 0.578 (0.007) & 0.575 (0.014) & 0.577 (0.007) & 0.573 (0.011) & 0.564 (0.011) \\ 
   & IBS & 0.192 (0.002) & 0.195 (0.004) & 0.193 (0.002) & 0.193 (0.002) & 0.193 (0.002) \\ 
   & RMST & 0.651 (0.006) & 0.651 (0.014) & 0.655 (0.013) & 0.659 (0.013) & 0.658 (0.005) \\
   \hline
\end{tabular}
\end{table}

\begin{table}
\caption{\revise{Multi-source simulation results for all methods based on various metrics. Median among 200 replicates is reported, with median absolute deviation shown in parentheses. For POTL-N, source weights are set proportional to the sample size of each source study; for POTL-CV, source weights are selected using 5-fold cross-validation.}}%
\begin{tabular}{lccccc}
\Hline
Metric & POTL-N & POTL-CV & Target-only & CoxTL & Pooled \\
\hline
L$_2$D & 0.031 (0.013) & 0.030 (0.009) & 0.077 (0.026) & 0.063 (0.022) & 0.073 (0.013) \\ 
  D$_\tau$ & 0.018 (0.011) & 0.021 (0.012) & 0.062 (0.027) & 0.049 (0.025) & 0.093 (0.015) \\ 
  C-index & 0.581 (0.004) & 0.581 (0.004) & 0.578 (0.007) & 0.580 (0.005) & 0.581 (0.003) \\ 
  IBS & 0.191 (0.002) & 0.190 (0.002) & 0.193 (0.003) & 0.192 (0.002) & 0.190 (0.001) \\ 
  RMST & 0.649 (0.006) & 0.650 (0.006) & 0.652 (0.012) & 0.654 (0.013) & 0.651 (0.004) \\  
\hline
\end{tabular}
\label{tb:sim_ms}
\end{table}

\begin{table} 
\caption{Summary of patients' clinical characteristics in the TCGA--BRCA and METABRIC studies. For continuous variables, median (Q1, Q3) are reported. For categorical variables, number (proportion) of cases are reported.}
\label{tb:summary_profile}
\begin{tabular}{l*{2}{r}}
\Hline
Characteristic & TCGA--BRCA ($n=762$) & METABRIC ($N=1393$) \\
\hline
{\bf Age} & 58 (48, 66) & 61 (51, 70) \\
{\bf Lymph nodes} & 0 (0, 2) & 0 (0, 2) \\
{\bf Tumor stage} \\
\multicolumn{1}{r}{Early} & 580 (76\%) & 1269 (91\%) \\
\multicolumn{1}{r}{Advanced} & 182 (24\%) & 124 (9\%) \\
{\bf Histological type} \\
\multicolumn{1}{r}{IDC} & 532 (70\%) & 1087 (78\%) \\
\multicolumn{1}{r}{Others} & 230 (30\%) & 306 (22\%) \\
{\bf ER status} \\
\multicolumn{1}{r}{Positive} & 591 (78\%) & 1071 (77\%) \\
\multicolumn{1}{r}{Negative} & 171 (22\%) & 322 (23\%) \\
{\bf PR status} \\
\multicolumn{1}{r}{Positive} & 514 (67\%) & 732 (53\%) \\ 
\multicolumn{1}{r}{Negative} & 248 (33\%) & 661 (47\%) \\
{\bf HER2 status} \\
\multicolumn{1}{r}{Positive} & 122 (16\%) & 173 (12\%) \\
\multicolumn{1}{r}{Negative} & 640 (84\%) & 1220 (88\%) \\
\hline
\end{tabular}
\end{table}

\backmatter



%
 \bibliographystyle{biom} 
\bibliography{paper-ref}






\label{lastpage}